\documentclass[journal,transmag]{IEEEtran}

\usepackage[pdftex]{graphicx}
\usepackage{amsmath,amssymb}

\usepackage{multirow}
\usepackage[ruled,vlined]{algorithm2e}
\usepackage{bm}
\graphicspath{{images/}}

\usepackage{xcolor}
\newcommand{\vc}[3]{\overset{#2}{\underset{#3}{#1}}}
\begin{document}
\title{EmoGen: Quantifiable Emotion Generation and Analysis for Experimental Psychology}

\author{\IEEEauthorblockN{Nadejda Roubtsova\IEEEauthorrefmark{1},
Martin Parsons\IEEEauthorrefmark{1}, 
Nicola Binetti\IEEEauthorrefmark{2},
Isabelle Mareschal\IEEEauthorrefmark{2},
Essi Viding\IEEEauthorrefmark{3} and
Darren Cosker\IEEEauthorrefmark{1}
\IEEEauthorblockA{
\IEEEauthorrefmark{1} CAMERA, Department of Computer Science, 
University of Bath, Bath, BA2 7AY, United Kingdom \\
\IEEEauthorrefmark{2} Psychology Department, Queen Mary University of London, London, E1 4NS, United Kingdom \\
\IEEEauthorrefmark{3} Division of Psychology and Language Sciences, University College London, London, WC1E 7HB, United Kingdom 
}}
}

\IEEEtitleabstractindextext{
\begin{abstract}
3D facial modelling and animation in computer vision and graphics traditionally require either digital artist's skill or complex pipelines with objective-function-based solvers to fit models to motion capture. This inaccessibility of quality modelling to a non-expert is an impediment to effective quantitative study of facial stimuli in experimental psychology. The EmoGen methodology we present in this paper solves the issue democratising facial modelling technology. EmoGen is a robust and configurable framework letting anyone author arbitrary quantifiable facial expressions in 3D through a user-guided genetic algorithm search. Beyond sample generation, our methodology is made complete with techniques to analyse distributions of these expressions in a principled way. This paper covers the technical aspects of expression generation, specifically our production-quality facial blendshape model, automatic corrective mechanisms of implausible facial configurations in the absence of artist's supervision and the genetic algorithm implementation employed in the model space search. Further, we provide a comparative evaluation of ways to quantify generated facial expressions in the blendshape and geometric domains and compare them theoretically and empirically. The purpose of this analysis is 1.~to define a similarity cost function to simulate model space search for convergence and parameter dependence assessment of the genetic algorithm and 2.~to inform the best practices in the data distribution analysis for experimental psychology.
\end{abstract}
}

\maketitle
\IEEEdisplaynontitleabstractindextext
\IEEEpeerreviewmaketitle
\section{Introduction}
Facial animation has traditionally been a task requiring either technical know-how to transfer captured motion from an actor's performance or artistic skill to manipulate the model controls manually. Consequently, for a long time facial modelling has been mainly confined to applications in the entertainment industries enjoying tailored specialist support. However, if the animation control could be made accessible to an non-expert, the technology enabling generation of arbitrary facial expressions easily finds applications in other fields, such as psychological research, plastic surgery and forensics. In each field, specific goals and task parameters result in different demands on the non-expert animation control. In this paper, we concentrate on the facial modelling requirements for experimental psychology research.
\par Human facial expression dynamics is instrumental in non-verbal communication \cite{Barret2019}. This basic fact generates many research questions in experimental psychology. Many of these in some way involve an attempt to gauge a characterisation or a classification of individual human perception of facial stimuli within some system. One such topic of interest is the individual perception of a facial stimulus in terms of its classification as one of the basic emotions. The ability to characterise and classify depends on the individual's expected learnt understanding of the reference system. In the case of emotion classification, the knowledge of the system can be described as the individual's set of \textit{internalised facial representations} embodying happiness, sadness, anger, fear etc. 
\par There are several difficulties in trying to measure these internalised representations by non-expert facial modelling that we shall address in formulating our approach to the problem. Firstly, the internalised representation is a latent target without a quantitative description. Secondly, the individual target can be ill-defined. Thirdly, as the target may be intricate, the non-expert should be given access to sufficiently accurate photorealistic modelling tools that are nonetheless controlled in an intuitive manner. Lastly, any experimental methodology for internalised representation estimation must have a solid system for data quantification and comparison to enable analysis. We address all these difficulties in formulating EmoGen  - a methodology for quantifiable emotion generation and analysis in experimental psychology.
\par The proposed \textit{EmoGen methodology} democratises facial modelling allowing anyone to control a production-quality facial blendshape model via a user-friendly tool with the end goal of generating quantifiable target emotion representations. The \textit{generation} part of our experimental methodology rests on two pillars. Firstly, the non-expert user control of blendshape animation for arbitrary expression generation is realised as an iterative space search via a genetic algorithm (GA). We argue that the GA search is uniquely suited for optimisation  towards a latent, ill-defined and \textit{a priori} unquantifiable internalised representation target, while also lending itself well to intuitive user guidance. Secondly, as illustrated in Figure~\ref{corrected_meshes}, in EmoGen, detection and correction of implausible facial configurations are automated by mesh analysis and optimisation in 3D prior to rendering while also retaining expression quantification. Distorted expressions are possible in facial animation by bringing together incompatible elements and activation levels of facial dynamics. Our contribution of automated corrective mechanisms is important as it replaces the otherwise necessary artist's intervention, thus sustaining the requirement of purely non-expert control of the process.
\par  In addition to EmoGen's practical and configurable \textit{generation} applicable to a wide range of user studies, the methodology is made complete with guidelines for collected data \textit{analysis} in the given context. The optimal practices, informed by a systematic study of the best \textit{quantitative representation} and sample \textit{similarity metrics}, are also presented in this paper. Finally, an evaluation of the convergence behaviour and parameter sensitivity of our GA-based generation results in justified recommendations for parameter settings in human testing and significance thresholds in data analysis within the proposed methodology. 
\par In summary,  with our EmoGen methodology a psychologist can configure and run user studies to
generate consistent quantitative estimates of individual internalised targets (assuming their sufficiently stable latent definition) and then analyse the resultant estimate distributions for robust human behaviour trend extraction with metrics tailored to the quantitative data representation. This paper is organised as follows. In Section~\ref{related_work}, we sketch the fields of facial dynamics modelling and genetic algorithm applications. We focus on comparing the proposed EmoGen framework to its prototype variant \cite{ReedCGF2019}, highlighting the significant improvements in the static expression generation with respect to robustness, freedom of model space exploration and visual sample quality, as well as EmoGen's substantive extension in terms of data analysis tools and performance assessment.  Then, in Section~\ref{methodology}, we discuss the principles of blendshape modelling and mesh correction automation, the chosen GA-approach and its application to facial modelling. Our built-in process customisation options enabling diverse controllable  studies are also presented. In Section~\ref{evaluation}, we present a comparative study of expression quantification and similarity metrics in different domains. Further in this section we provide tool performance analysis in terms of convergence and sensitivity to configurable parameters. 

\section{Related Work}
\label{related_work}
\textbf{Facial modelling} is a complex, interesting and widely applicable area of research. In psychology, the facial coding system (FACS) \cite{FACS:1978} constituted the first attempt to standardise units of human facial deformation space. In computer science, the interest in faces has been fuelled by security applications (e.g. facial recognition) and the entertainment industries i.e. film, gaming, virtual reality and social media. Due to the high variability along with the complex non-linear facial deformation and dynamics, efforts have mainly been invested in learning the model space from data. With roots in the Point Distribution Models (PDMs) and Active Shape Models (ASMs) \cite{CootesCVIU1995}, Active Appearance Models (AAMs)~\cite{CootesPAMI2001} were designed to fit facial shape and appearance in 2D. These statistical models are linear and built by PCA decomposition of training image data. Application of higher order singular value decomposition (HOSVD) opened the door to multilinear analysis in the so-called TensorFaces~\cite{10.5555/645315.649173} modelling facial geometries, expressions, head poses, and lighting conditions. In face recognition, independent component analysis (ICA) \cite{COMON1994287} has been used, as a generalisation of PCA, to extract higher-order relationships from pixel data \cite{Bartlett2002}. ICA is in essence the blind source separation problem, which has an extensively researched space of solution approaches \cite{JuttenKarhunen}, including neural networks. Hence, ICA became an early herald of the subsequently discovered potency of deep learning for feature extraction in facial modelling tasks amongst a myriad of other applications.
\par 3D Morphable Models~\cite{Blanz_CVPR_2003,Booth_CVPR_2016} are an extension of the principle behind the AAM to 3D. 3DMMs are denser in representation and, being three dimensional, unlike AAMs, can handle occlusion \cite{1315210}. 3DMMs focus on modelling the base identity variation in shape and texture. 
More complete multi-linear models \cite{10.1145/1186822.1073209,Thies_CVPR_2016, BolkartCVPR2016, FLAME:2017} extending to expression dynamics also subsequently appeared. Further information on the state-of-the-art, as well as historical and future development, of statistical human face models for 3D data can be found in the detailed surveys \cite{BRUNTON2014, 10.1145/3395208}.
\par As a side note, let us mention the recent interesting trend of training facial generative models as neural variational autoencoders (VAEs) \cite{8354111,COMA:ECCV2018}. An important direction in the field is enforcing a semantically meaningful latent variable layer \cite{MOFA2017}, as the direct control of the output significantly expands the range of applications.
\par Blendshape modelling \cite{LewisetalEurographics2014} emerged as the dominant industrial method for character animation. Conventionally, for high quality, personalised blendshape model rigging required manual work of a professional digital artist. At the expense of reduced quality, deformation transfer \cite{10.1145/1015706.1015736} would often be used for automatic personalisation from a template blendshape set. However, procedural blendshape model personalisation, as an active research area, has seen significant recent advances. Hao Li and coleagues \cite{li2010example} pioneer example-based rigging from a handful of training poses of the new identity. The most impressive advance in the field is the work of Jiaman Li and colleagues \cite{10.1145/3414685.3417817} who generate personalised blendshapes with the corresponding dynamic textures from a single neutral scan. Their approach is based on training two cascaded neural networks, specifically one for the personalised blendshape generation followed by another for the generation of dynamic texture maps (albedo, specular intensity, displacement), tailored to the expression geometries.  Their results are of near-production quality and are compatible with the professional pipelines to enable use and any necessary asset polishing. However, as a limitation, the training process of their neural network cascade requires a database of over 4000 scans with pore-level detail and high variability in terms of expression and identity. Also the approach shows predictable generalisation limitations to subjects under-represented in the training data e.g. children and bearded men. Along with the academic publications on automated rigging, there are now also commercial solutions \cite{MetaHuman} offering to a non-artist customisable photorealistic human character generation with blendshape model personalisation to model facial dynamics. 
\par The blendshape face model offers a number of unique practical advantages. Firstly, the rig consists of shapes loosely corresponding to facial action units from FACS, which immediately translates into intuitive control for a trained professional. The model is linear, which can generate implausible facial configurations. However, with artistic skill dedicated to sculpting an arsenal of corrective shapes, the lacking non-linearity can be compensated for. Further, the artist (and increasingly also automated pipelines) can customise the model to any new identity in terms of both base geometry and individual facial dynamics. With statistical models on the other hand one is limited to variation within learnt spaces of shape and expression. Often the expression control in the statistical models is much less intuitive than the blendshape weights. 
\par The blendshape model is animated using blendshape weights and/or higher level rig controls. In other words, the vector of blendshape weights offers a unique representation to every facial expression in the model space. In the EmoGen methodology we propose, this quantification facilitates a way to traverse the facial manifold using a genetic algorithm for novel expression generation and perform comparative analysis of facial expression samples in data analysis. Note where traditionally the artist is involved in both creation and animation of the blendshape model, with EmoGen the non-expert, supported by suitable corrective and space traversal algorithms, is able to control the professionally designed model independently.
\par In the context of user-friendly facial modelling in 3D, FaceGen~\cite{FaceGen} and FACSGEN~\cite{FACSGEN} must be mentioned. Unlike our modelling tool  aimed for open-source distribution to non-commercial users, both FaceGen and FACSGEN are only available free of charge in their limited demo versions. Further, the applications do not offer the same functionality as EmoGen. FaceGen offers a practical interface to generate custom identities of specified age, gender, race etc. and to fit models to existing photographs. However, there is no way to generate custom expressions without a template photo. FACSGEN offers a set of FACS Action Unit sliders to generate static expressions and an interface for authoring animation curves via control points. Although it is clearly made more user-friendly than the standard artist's interface e.g. by employing more intuitive action units, objectively it still seems unlikely that a user can work with FACSGEN without some prior training. Hence, for psychology research into internalised expression representations, the proposed EmoGen is better suited than both FaceGen and FACSGEN for several reasons. Firstly, it is inherently applicable to the \textit{latent} (i.e. without a template photo) target expression search via genetic evolution. Secondly, EmoGen's trivial control through visual sample selection is unique in truly requiring no training and thus opening up the way for faster testing and inclusion of subjects of diverse abilities (e.g. children). Finally, in FaceGen and FACSGEN, the focus is not on the generated sample quantification and subsequent analysis. In contrast, EmoGen provides complete quantitative data from all stages of the generation process and offers validated methods for analysis.
\par \textbf{Evolutionary algorithms (EA) and their applications.} Inspired by natural selection, EA is an umbrella term for a family of population-based metaheuristic approaches including genetic algorithms (GA), evolutionary strategies (ES), differential evolution (DE) and estimation of distribution algorithms (EDA) \cite{DBLP:reference/sp/CorneL18}. Multi-objective evolutionary algorithms (MOEAs) have also been proposed \cite{996017}. The EA approach is attractive having no strict mathematical requirements beyond a way to assess relative fitness of samples. It is applicable to non-linear problems, constrained or unconstrained with both discrete and continuous search spaces \cite{SivanandamDeepaSpringerBook2010}. Despite being parametric, EA algorithms have not been found overly sensitive to optimal settings. The main drawback of EA is the lack of guarantees on the \textit{absolute} optimality of the found solution; the approach ensures only relative improvement with respect to discarded samples. The EA metaheuristic is most suitable for problems where the absolute optimum is not well defined and hence untestable. This reflects the original evolution problem in the natural world where there is no \textit{a priori} definition of the fittest species, only an iterative search towards it using the relative survival potential as the objective function. 
\par Prototyping of internalised facial emotion representation, addressed in this paper and earlier by Reed and Cosker~\cite{ReedCGF2019}, is a illustrative example of a problem inherently suitable for the EA approach. Since the internalised target, our absolute optimum, is latent and perhaps also ill-defined, no analytical cost function can be defined and minimised conventionally. For this problem, the EA offers a way to still explore the space using user selections as a latent cost function but leaves no way to test the final result optimality beyond subjective user satisfaction scores. However, it is possible to assess convergence of the face prototyping evolutionary search to the absolute optimum  by \textit{simulation}. For that, the iterative stochastic framework of the EA approach is kept the same while the target is quantitatively defined and the user selections are replaced by a consistent similarity metric seeking samples most similar to the target. Convergence to the absolute optimum is then gauged by the similarity score of the final solution to the defined target. Thus the key evolutionary mechanisms are adapted to both the intended human-centred use-case and for validation of the generation part of the proposed EmoGen methodology.    
\par For further examples of EA application to problems in other branches of engineering the reader is referred to the survey by Slowik and Kwasnicka \cite{Slowik2020EvolutionaryAA}.
\par \textbf{EA for facial modelling: state-of-the-art.} Reed and Cosker~\cite{ReedCGF2019} first demonstrate efficacy of an EA framework to generate customisable shape and dynamics in facial modelling. In their framework, static expressions are represented by arbitrary blendshape rig controls, while sampled dynamics is encoded in non-linear animation curves procedurally, by constrained keyframing of linear source-to-target interpolations. In both static and dynamic case, subsequent user-guided evolution is realised through cross-breeding and mutation processes. Reed and Cosker present a range of applications including  non-expert prototyping of static expressions towards an internalised or observed target and authoring non-linear dynamics for face regions.
\par The strength and focus of the pioneering work consists in the demonstration of a wide ranging applicability of the EA approach to facial modelling tasks.
However, in the specific task of static expression generation by a non-expert user their work has some important limitations. Firstly, the face blendshape model they make publicly available has only a crude 2.5D untextured geometry. The model is automatically generated by deformation transfer from a template rig with visible shape distortion and misalignment artefacts. Further, their geometry lacks mouth cavity and proper integration of the eyes and teeth into the blendshapes. The latter means that extreme facial expressions, e.g. jaw fully open, cannot be sampled as there is no way to facilitate corresponding teeth motion for sample realism. 
\par Generally speaking, their facial modelling framework suffers from unrealistic expression samples with no corrective mechanisms beyond evolution constraints. Specifically, Reed and Cosker employ a series of EA process constraints (e.g. select feature variation, local sampling, rig control ranking etc.) with different scenarios arranged into sets of configurations (``control schemas''). In their paper, the EA convergence is said to depend on informed task-dependent selection of the right configuration, which requires familiarity with their framework unlikely for an non-expert user.  Further, the necessity of their default coarse-to-fine evolution approach and mutation rate limitations are not experimentally validated and can equally be hypothesised to act to the detriment in limiting variability of generated samples.  Further, Reed and Cosker provide a limited study of the framework's convergence properties and effects of parameters. Finally, the implementation in Maya with limited quantitative data output hinders both accessibility and usefulness as a diagnostic tool in psychology research.
\par Building on the work of Reed and Cosker~\cite{ReedCGF2019}, in this paper we present EmoGen as an advanced methodology, specifically tailored to static facial analysis research in experimental psychology. The methodology is complete offering both improved \textit{data generation} and a principled approach to \textit{data analysis} to inform trend extraction in psychology research.   
\par Implemented as a standlone C++ program with OpenGL rendering and a user-friendly interface,  \textit{EmoGen} first of all addresses the limitations of \cite{ReedCGF2019} in  expression generation. Specifically, we make available a production-quality full 3D textured blendshape model with proper sub-geometry (teeth, eyes and mouth cavity) integration. The model is currently available in two identities (male and female) and we intend to diversify in the future specifically to different ages and ethnicities. Further, unlike~\cite{ReedCGF2019}, we implement a mechanism for automatic detection and correction of unrealistic facial configurations by application of two types of corrective blendshapes. As our model is more robust against straying off the plausible facial dynamics manifold, we have much weaker constraints on the evolution process (e.g. unlimited mutation rate), which allows better space exploration. We show convergence accuracy and consistency \textit{without} many of the limiting constraints of \cite{ReedCGF2019} (e.g. facial region selection), while using analogously defined primary GA mechanisms of cross-breeding and mutation. Further, we devise and evaluate initialisation procedures tailored specifically to the nature of psychology experiments and generate exhaustive data from all stages of the evolution process for analysis. 
\par Beyond better data generation, EmoGen methodology includes a principled \textit{data analysis} approach resting on a comparative study of both sample representations in multiple domains and metrics  for best expression similarity assessment. A series of convergence studies via simulation in different configurations is presented to assess performance, inform parameter settings and provide significance thresholds for user studies.

\section{EmoGen Methodology}
\label{methodology}
\subsection{EmoGen: facial modelling}
\label{blnd_model}
EmoGen generates samples in 3D, subsequently rendering in 2D. The knowledge of geometric properties (vertex 3D position, normals, topological interconnectivity) affords the freedom to render in an arbitrary way in terms of camera projection matrix and illumination as appearance can be recomputed accordingly. Facial 3D mesh $\mathcal{F} =(\mathcal{V}, \mathcal{T})$ is described by a set of vertices in $\mathcal{V}$ and triangular faces in $\mathcal{T}$. For a fixed topology model, faces, each defined by a vertex index triplet, remain constant. Facial 3D mesh is deformed by altering Cartesian coordinates $(x,y,z)$ of the vertices from $\mathcal{V}$. In the case of fixed topology 3D mesh $\mathcal{F}$ is defined by $\mathcal{V}$ only.
\par EmoGen accommodates facial dynamics spanned by a model created by a professional digital artist. Specifically, we use a facial \textit{blendshape model}~\cite{LewisetalEurographics2014}, which, in its simplest form, is a set of aligned fixed topology meshes of the neutral expression mesh $\mathbf{B}_{0}$ and K of its deformations defining an non-orthogonal basis for facial deformation. Facial expression $\mathcal{F}$ is a linear combination of $\mathbf{B}_{0}$ and weighted deviations from $\mathbf{B}_{0}$:
\begin{equation}
\mathcal{F} = \mathbf{B}_{0} + \vc{\sum}{K}{k=1}\alpha_k(\mathbf{B}_k - \mathbf{B}_{0}).
\end{equation}
where $\boldsymbol{\alpha} =(\alpha_1, ..., \alpha_{K})$ is the blendshape weight vector defining the facial expression with all $\alpha_{k}=[0,1]$. A professional blendshape model typically consists of \textit{core} blendshapes and \textit{correctives}.
\par \textbf{Core blendshapes} correspond to facial deformation modes identified by an artist and are loosely inspired by specific muscle group activations. While all blendshape weights are normalised to the $[0,1]$ range, 
blendshape offsets in the vertex space are arbitrary. When applicable, core blendshapes typically come as left/right pairs for independent control of facial sides. EmoGen enforces facial symmetry activating left and right shapes equally. The key problem with unconstrained traversal of the blendshape model space lies in the mutual non-orthogonality of blendshape deformation, which can lead to unrealistic facial expressions. We address this problem by employing \textit{corrective} blendshapes also sculpted by the artist.
\par \textbf{Corrective blendshapes} come in two types: combinational and collision.
\par \textit{Combinational} correctives are designed to neutralise the unrealistic effect brought about by a certain set of core shapes being activated together. The common practice is to activate the corrective by the product of the combined core shape weights. The rationale behind the practice is two-fold. Firstly, it is a masking mechanism ensuring that a non-activation of any one blendshape in the combination blocks activation of the corrective. Secondly, such non-linear activation process, in a targeted way, dampens corrective contribution for core shapes combined with low weights. The effect is justified as the distortion at this intensity level of the core shapes is likely to be unnoticeable.
\par
\begin{figure}[h!]
\centering
\begin{tabular}{ccc}
\rotatebox{90}{\parbox{4cm}{deformation imperfection} } &
\includegraphics[scale=0.18]{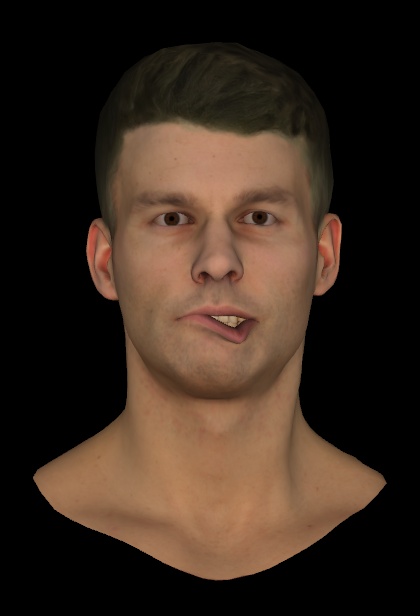}&
\includegraphics[scale=0.18]{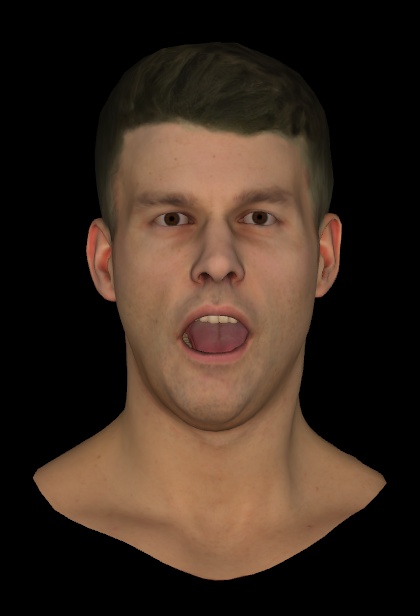}\\
\rotatebox{90}{\parbox{4cm}{automatic corrections}}  &
\includegraphics[scale=0.18]{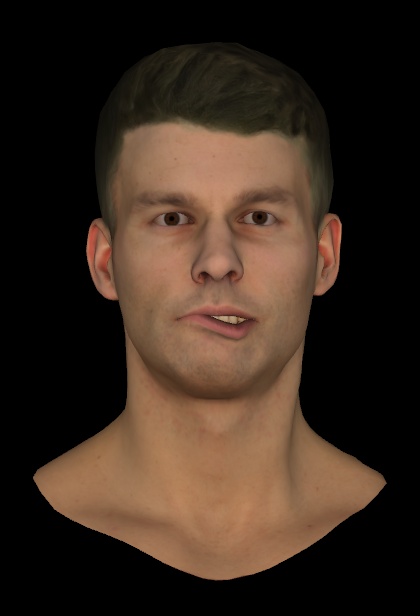}&
\includegraphics[scale=0.18]{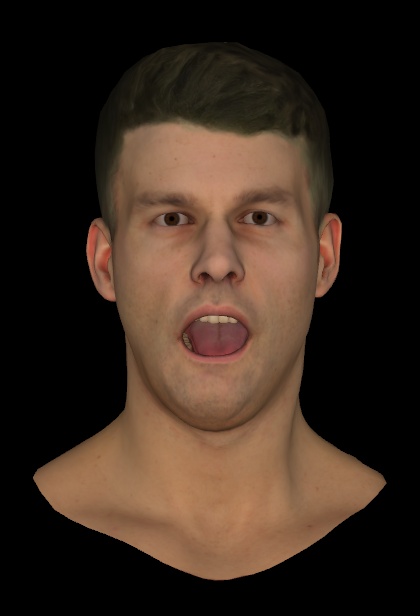}\\
\end{tabular}
\caption{Performance of combinational correctives in enhancing facial geometries. \textbf{Row 1:} examples of imperfect facial deformation in given configurations; \textbf{Row 2:} corresponding automatic corrections by application of combinational correctives.}
\label{combinational}
\end{figure}

\begin{figure*}[h!]
\begin{tabular}{c@{\hspace{0.2mm}}c@{\hspace{0.0mm}}c@{\hspace{0.0mm}}c@{\hspace{0.0mm}}c@{\hspace{0.0mm}}c@{\hspace{0.0mm}}c}
\rotatebox{90}{\parbox{4cm}{sub-geometry collisions}}  &
\includegraphics[scale=0.18]{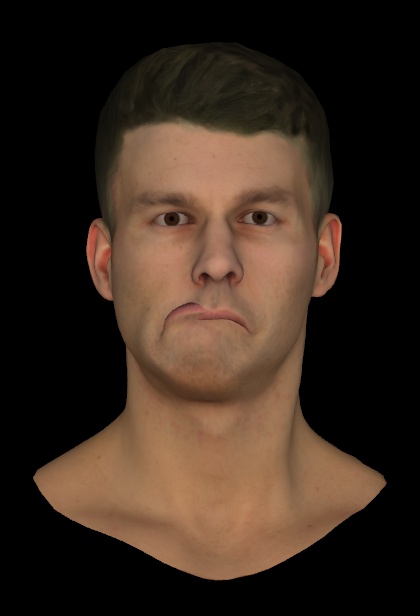}&
\includegraphics[scale=0.18]{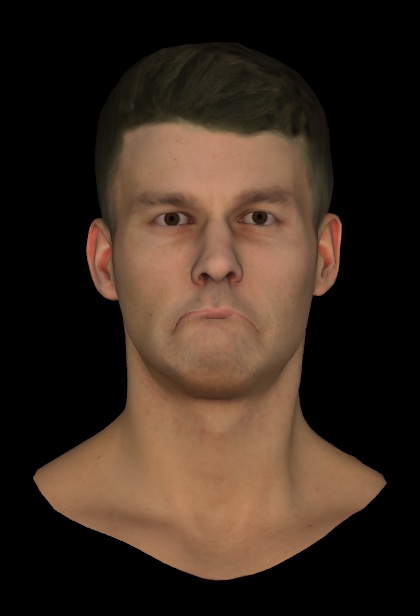}&
\includegraphics[scale=0.18]{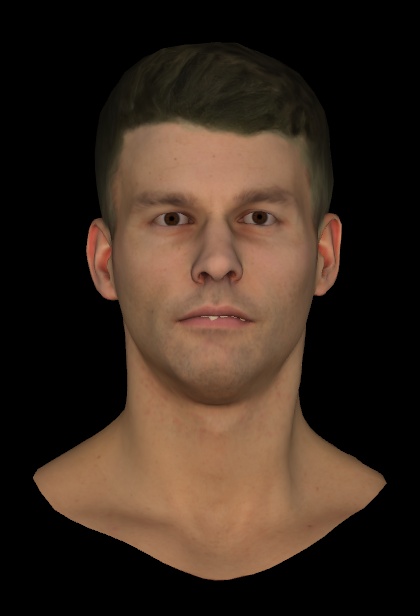}&
\includegraphics[scale=0.18]{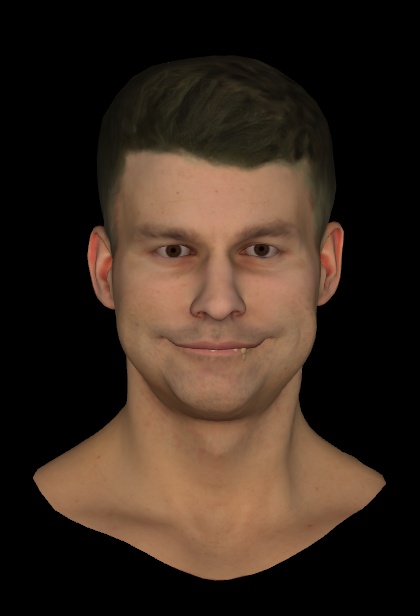}&
\includegraphics[scale=0.18]{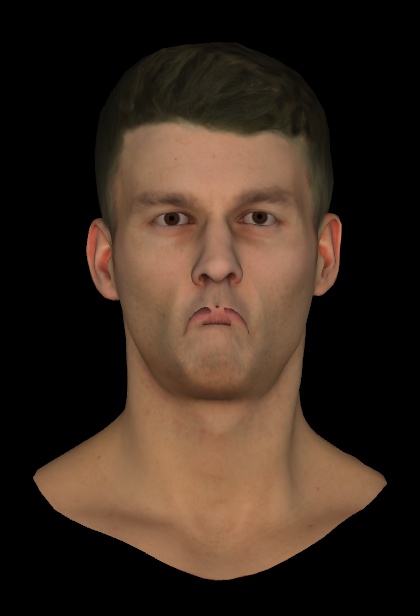}&
\includegraphics[scale=0.18]{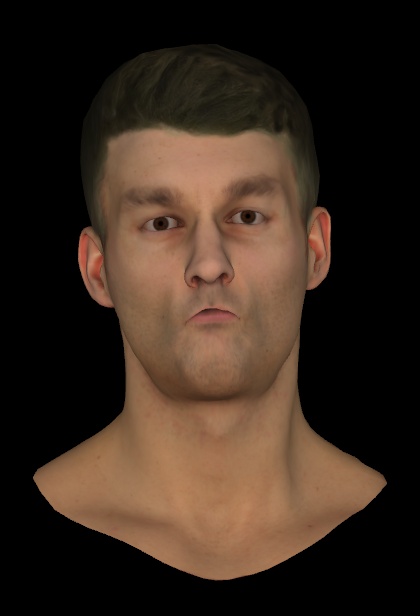} \\
\rotatebox{90}{\parbox{4cm}{automatic corrections}}  &
\includegraphics[scale=0.18]{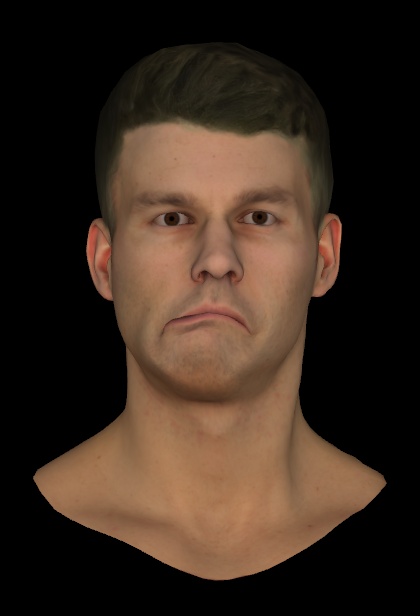}&
\includegraphics[scale=0.18]{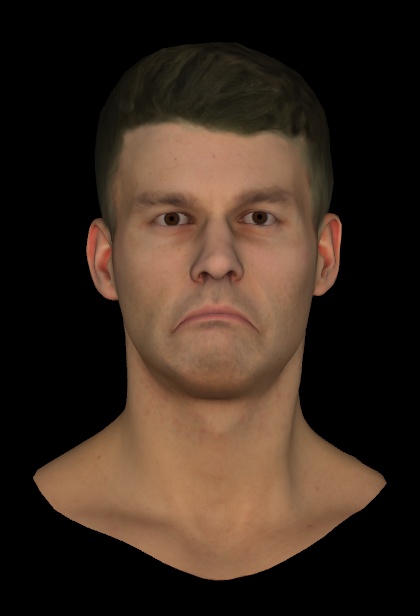}&
\includegraphics[scale=0.18]{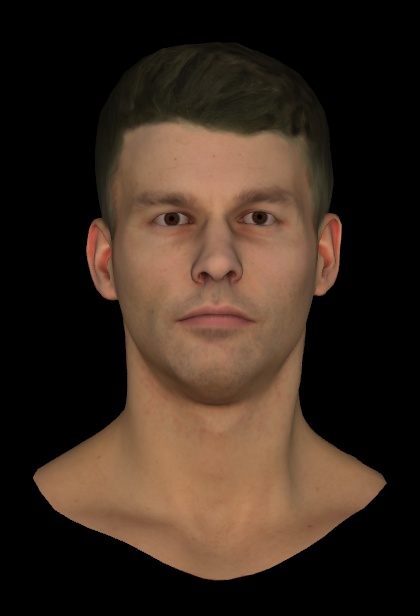}&
\includegraphics[scale=0.18]{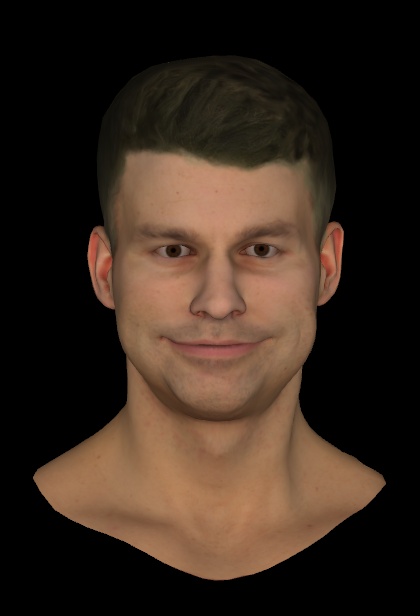}&
\includegraphics[scale=0.18]{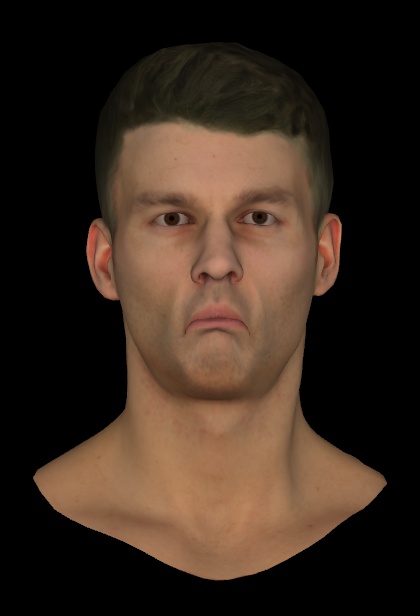}&
\includegraphics[scale=0.18]{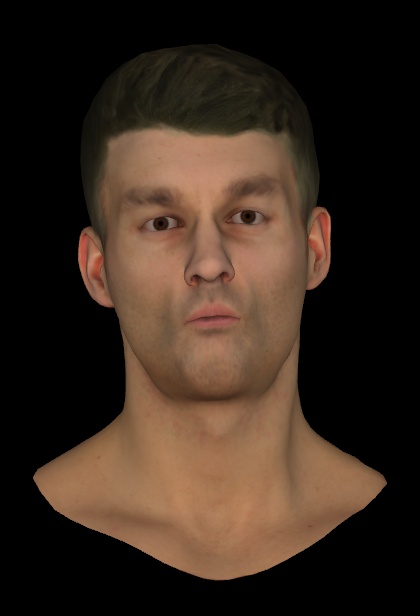} \\
\end{tabular}
\caption{Performance of the automatic collision detection and correction in sample generation of EmoGen. \textbf{Row 1:} examples of unrealistic facial configurations characterised by the interpenetration of sub-geometries (i.e. collisions); \textbf{Row 2:} corresponding automatic corrections by application of the anti-collision mechanisms described in Section~\ref{blnd_model}. }
\label{corrected_meshes}
\end{figure*}

 Figure~\ref{combinational} shows a couple of examples of combinational correctives in action. Their effect is visually subtler than that of the collision correctives (see below and Figure~\ref{corrected_meshes}). For example, combinational correctives act to bring about more geometric smoothness into the deformation (e.g. the lower lip in the left-hand-side example) or to realise a more plausible local behaviour, given a certain expression configuration (e.g. thinning of the lower lip at activation of the \textit{lower-lip-suck} blendshape in the right-hand-side example).
\par
\begin{figure}
\centering
\begin{tabular}{c@{\hspace{0mm}}|c@{\hspace{0.0mm}}}
lip collision correction & teeth collision correction\\ \hline
\includegraphics[scale=0.22, trim = 60mm 110mm 60mm 110mm, clip]{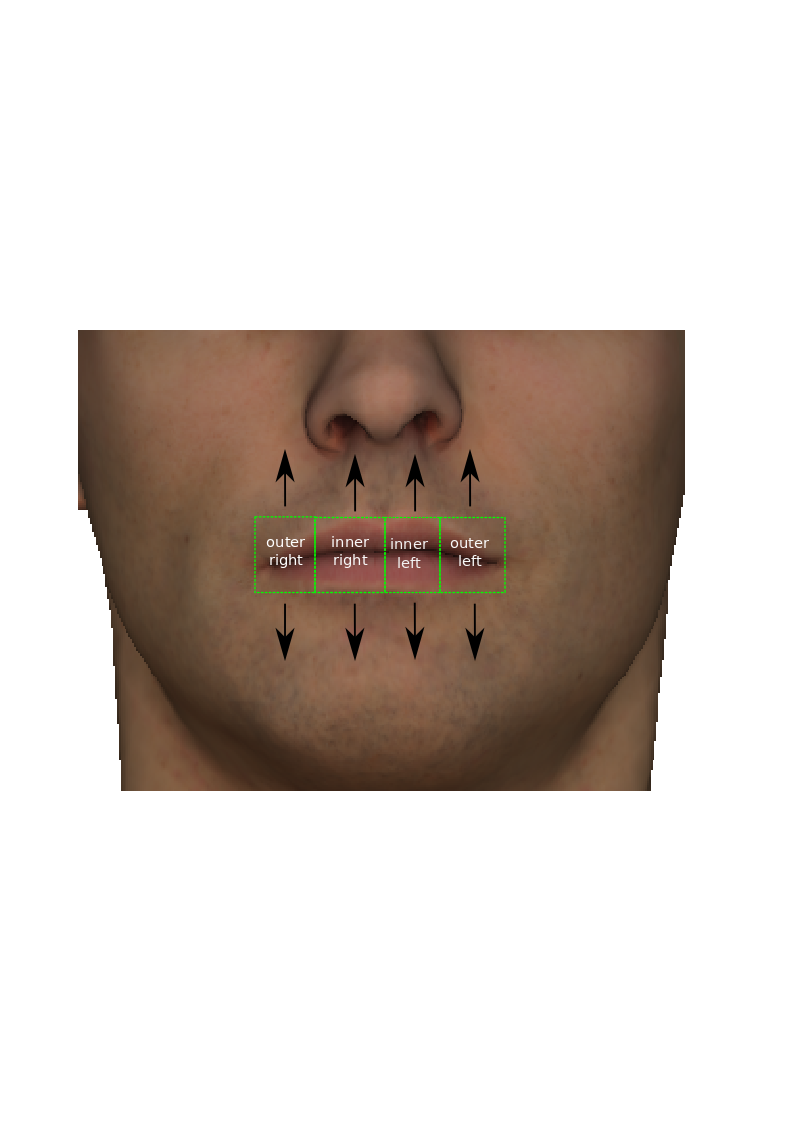} &
\includegraphics[scale=0.22, trim = 60mm 110mm 60mm 110mm, clip]{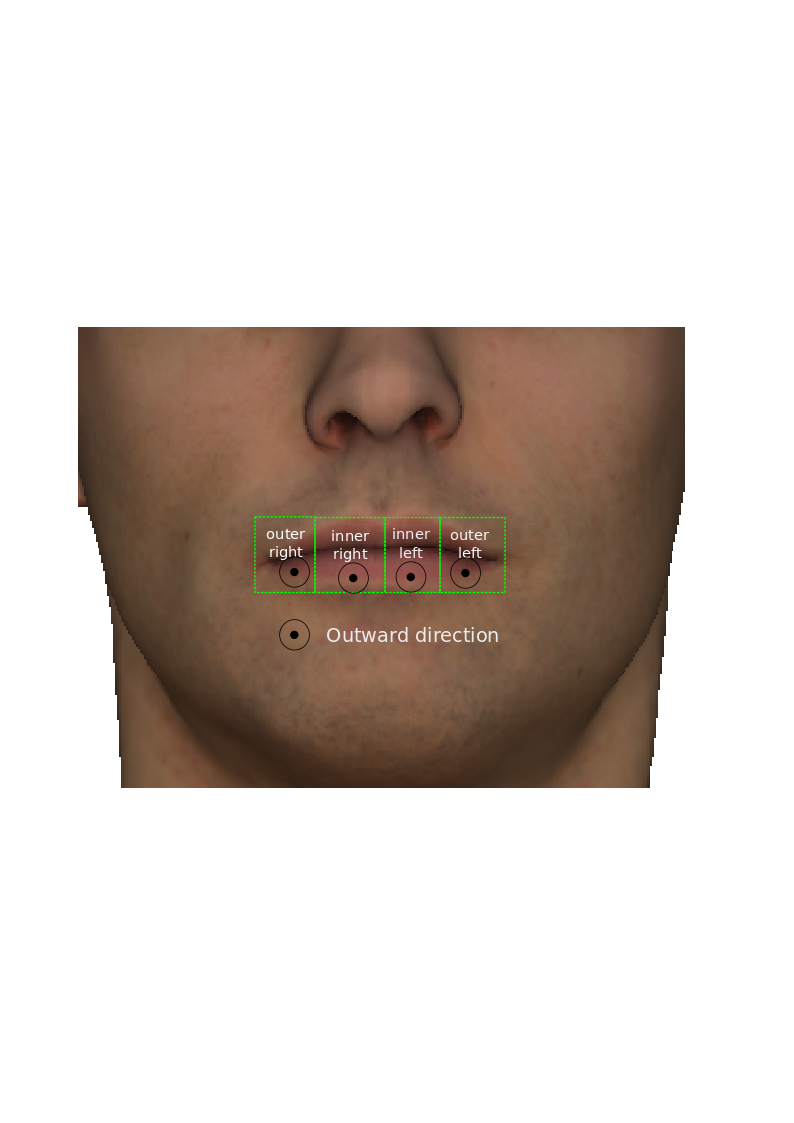} \\
\end{tabular}
\caption{Schematic representation of the collision correction mechanisms for respectively lips and teeth  hard-coded as corrective collision blendshapes in the facial model within EmoGen.}
\label{correction}
\end{figure}

 \textit{Collision} correctives are designed to deal with the unrealistic interpenetration of facial sub-geometries such as in the examples of Figure~\ref{corrected_meshes}~(row 1). The assignment of weights to collision correctives is not straightforward as the activation amount would depend on the intersection depth in the specific facial configuration. In animation these weights are set manually by the artist. In order to automate collision correction in the context of EmoGen's sample generation, we have devised an optimisation procedure detailed and formalised in the following paragraphs. In our model collision correctives are split into teeth and lip types, respectively corresponding to the \textit{upper teeth-to-lower lip} and \textit{upper-to-lower lip} collisions. Further, we split the collision region into four zones (outer left, inner left, inner right and outer right) resulting in eight collision correctives in total i.e. four per collision type. Figure~\ref{correction} illustrates the anti-collision mechanisms hard-coded in the model per type. Broadly speaking, the lip collision is corrected by a vertical corrective offset of both upper and lower lips, whereas the teeth collision correction is addressed by the outward lower lip motion. The results of our automatic application of the anti-collision mechanisms are shown in Figure~\ref{corrected_meshes}~(row 2) for some representative examples.
\par \textbf{Collision corrective offset formalisation.} Given the identified collision modes typical of the specific blendshape model, three regions of interest are selected on the topology: upper front teeth, upper lip and lower lip. Each region is represented by a set of barycentric anchor point coordinates, typically the centres of selected faces. Let us denote the sets $P_\text{teeth}=\{(t_{i}, c_{1,i}, c_{2,i})\}_{N_\text{teeth}}$, $P_\text{upr\_lip}=\{(t_{i}, c_{1,i}, c_{2,i})\}_{N_\text{upr\_lip}}$ and
$P_\text{lwr\_lip}=\{(t_{i}, c_{1,i}, c_{2,i})\}_{N_\text{lwr\_lip}}$, where $t_i\in\mathcal{T}$ of the model's topology and $(c_{1,i}, c_{2,i})$ are the barycentric coordinates defining the point's position within face $t_i$. We will now formalise the corrective blendshape offsets for the anchor sets using $P_\text{lwr\_lip}$ as an example. The definition is analogous for $P_\text{teeth}$ and $P_\text{upr\_lip}$. 
\par Cartesian coordinates corresponding to $N_\text{lwr\_lip}$ anchor points can be extracted from any mesh $\mathcal{F}$ in our topology by application of barycentric coordinates: $\mathcal{F}(P_\text{lwr\_lip}) = \mathcal{V}_\text{lwr\_lip} = \{(x_i, y_i, z_i)\}_{N_\text{lwr\_lip}}$. For each collision corrective $\mathbf{B}_\text{clsn}$, we define offsets $\Delta\mathbf{B}_\text{clsn}$ from the neutral  $\mathbf{B}_0$. These offsets can be sampled specifically for the set of anchor points $P_\text{lwr\_lip}$ :
\begin{equation}
\Delta\mathbf{B}_\text{clsn}(P_\text{lwr\_lip}) =  \mathbf{B}_\text{clsn}(P_\text{lwr\_lip}) - \mathbf{B}_0(P_\text{lwr\_lip})
\end{equation}

\par In the lip collision blendshapes, upper and lower lip anchor points move away from each other. In the teeth collision blendshapes, the lip point motion dominates and any teeth motion relative to the neutral is expected to be negligible although can be factored in for generality of formalisation. As  will be discussed further on, the simultaneous movement in the opposite directions needs to be taken into account when computing collision depth eliminated per unit corrective blendshape weight activation.
\par \textbf{Collision detection.} Anchor points of the upper lip or teeth found to be within the lower lip or having passed through it are indicative of sub-geometry interpenetration. For each such anchor point we count the number of intersections with the lower lip faces when sampled with the rays in the collision directions of Figure~\ref{correction}. Without loss of generality, let us denote the direction of lower lip motion in the corrective mechanism as positive. There are two intersection scenarios possible:
\par \textit{Within.} A tested anchor point \textit{within} the lower lip would have two intersections, one along each positive and negative sampling ray (for robustness, detection of this intersection scenario in either one of the collision directions in Figure~\ref{correction} is sufficient). 
\par \textit{Through.} Two intersections with oppositely oriented lower lip faces along the negative ray indicate the anchor point having possibly passed \textit{through} the lower lip. \par Both are intersection scenarios for which interpenetration depth can be quantified as will be explained further on. However, collision detection itself is based strictly on the point count \textit{within} the lower lip to avoid misclassification in ambiguous facial configurations e.g. lower lip tucked behind the upper teeth. 
\par
\begin{figure*}
\centering
\begin{tabular}{cc}
upper-to-lower lip intersection & upper teeth-to-lower lip intersection \\ \hline
\includegraphics{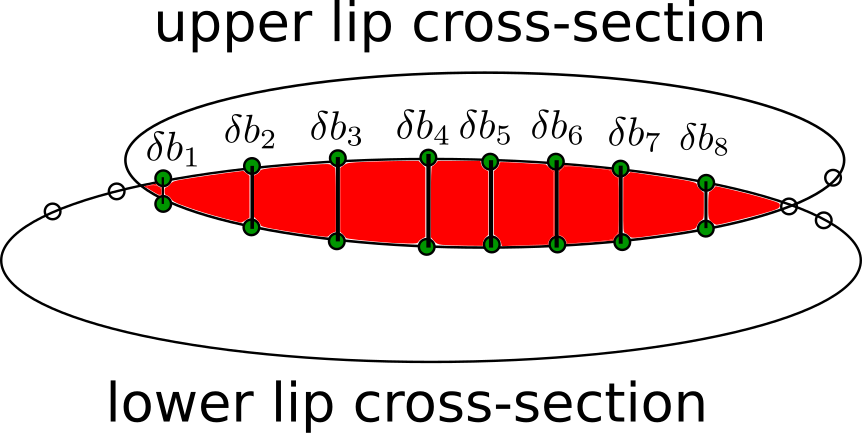} & \includegraphics{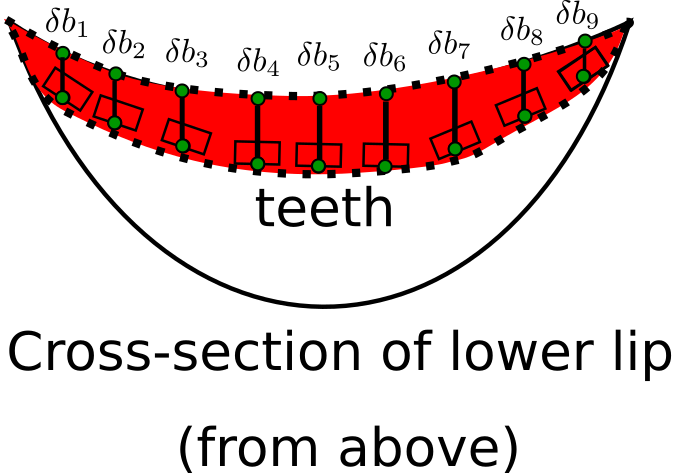}\\
\end{tabular}
\caption{Schematic representation of interpenetration depth quantification for upper-to-lower lip and upper teeth-to-lower lip collision types as a set of distance offsets $\delta b$ in the directions pre-defined by collision corrective blendshapes. Interpenetration region is shown in red.}
\label{quant_schematic}
\end{figure*}

 \textbf{Collision depth quantification} is based on identifying \textit{intersection point pairs} that need to be separated in space along the direction pre-defined by the collision blendshapes (Figure~\ref{correction}) to eliminate the sub-geometry interpenetration. The anchor points  on the upper lip and teeth, previously defined in the context of collision corrective offsets formalisation, initialise the pair. The second point in each pair is the intersection with the lower lip when sampling with a ray from the anchor point in the collision-neutralising direction. Figure~\ref{quant_schematic} aims to clarify the principle schematically for each collision type. For the upper-to-lower lip collision, the lips are shown shifted laterally to illustrate a realistic expression induced case which results in the points paired up differently than they would in the neutral lip placement. The interpenetration depth quantifying offsets $\delta b$ are in the vertical and outward directions for the two collision types. Only the points within the interpenetration segment (in red) generate constraints. 
\par \textbf{Collision correction optimisation.} In quantification of collision depth, we identify the lower lip intersection point, both its absolute 3D position and the topological coordinate i.e. the face number within the region of interest for collision detection. Recall from formalisation of collision corrective offsets that each face centroid in the regions of interest has an associated maximum corrective blendshape offset from the neutral. The maxima are used to define the \textit{eliminated interpenetration depth per unit corrective blendshape weight} tailored to each intersection point pair involved in quantifying collision.
\par Let us formalise this on the upper-to-lower lip collision example. Let a point on the upper lip $p_{i, \text{upr}} \in P_\text{upr\_lip}$ form an intersection pair with a lower lip point $p_{i, \text{lwr}} \in P_\text{lwr\_lip}$ to result in constraint $\delta{b_{i}}$. For each collision corrective blendshape $\mathbf{B}_\text{clsn}$, the eliminated collision depth $\delta a_{i}$  per unit blendshape weight is:
\begin{equation}
\delta a_{i} = (\Delta\mathbf{B}_\text{clsn}(p_{i,\text{upr}}))_{z} - (\Delta \mathbf{B}_\text{clsn}(p_{i,\text{lwr}}))_{z},
\end{equation}
where $z$ denotes the vertical dimension of the $\Delta \mathbf{B}$ vector. Lip collision correctives are split into zones as indicated in Figure~\ref{correction} but with some mutual overlap necessitating four $\delta a_{i}$ coefficient values. Furthermore, teeth collision correctives, solved for jointly, may also contribute to the eliminated lip collision depth resulting in four additional coefficients per constraint. We only allow teeth collision corrective contribution to the lip correcting dimension (and vice versa) if there are detected teeth collisions to be solved as well, otherwise the corresponding $\delta a_{i}$ values  are disabled by setting to zero.  The restriction constrains corrective mechanisms to the primary artist-defined modes e.g. lip interpenetration primarily solved by lip correctives. 
\par The analogous formulation of the eliminated collision depth offset $\delta a_{i}$  per unit corrective blendshape weight for the upper teeth-to-lower lip type is:
\begin{equation}
\delta a_{i} = (\Delta \mathbf{B}_\text{clsn}(p_{i,\text{lwr}}))_y -(\Delta \mathbf{B}_\text{clsn}(p_{i,\text{teeth}}))_y ,
\end{equation}
where  $p_{i,\text{teeth}} \in P_\text{teeth}$ and $p_{i, \text{lwr}} \in P_\text{lwr\_lip}$ is its intersection point on the lower lip. The correction is in the ``outward'' dimension of $\Delta \mathbf{B}$ denoted by $y$. Also analogously, the lip  blendshape contribution to corrective action for these points and in this dimension is only enabled if lip collisions are being solved simultaneously.
\par To jointly solve for the eight-dimensional (four correctives per collision type) collision blendshape weight vector $\mathbf{w}_\text{clsn}$, we optimise the cost function:
\begin{equation}
\mathcal{C} =  w_1|| A_{M\times8} \mathbf{w}_\text{clsn} - \mathbf{b} ||^2_2 + (1 - w_1)M||\mathbf{w}_\text{clsn}||^2_2,
\label{cost_function}
\end{equation}
where $M$ is the total number of constraints $\delta{b_i}$ in $\mathbf{b}$ across both collision types
and $w_{1}$ is the data term weight. Based on some exploratory experimentation, we set $w_{1} = 0.98$.
\par If both teeth and lip collision constraints are being optimised in (\ref{cost_function}) we want to enforce the intended corrective mechanism for each type of constraint while taking into account the inadvertent contribution from the other type of corrective blendshapes. To this end, when defining constraints for one collision type, the jacobians of the other type are artificially set to zero. Thus the contribution of the unintended corrective blendshapes is treated as a constant in the constraint. The constant is updated only through optimisation of constraints of the other type where the relevant jacobians are not disabled.

\subsection{EmoGen: genetic algorithm (GA)}
Data generation within the proposed EmoGen methodology is based on a \textit{genetic algorithm}: facial expressions encoded by the blendshape weight vectors $\boldsymbol{\alpha}$ (i.e.~\textit{chromosomes}) are generated and refined through a gradual user-guided evolution of individual blendshape weights $\alpha_i$ (i.e.~\textit{genes}). Unlike the binary strings in traditional GAs, in our framework the genes take \textit{continuous} values in $[0,1)$. At each generation $g$, we present to the user a population $\mathbf{P}_g=(\boldsymbol{\alpha}_1,...,\boldsymbol{\alpha}_{10})$ of ten expression samples.  Generally speaking, the GA behaviour in space exploration is determined by the strategies adopted for the tasks of \textit{selection}, \textit{cross-breeding}, \textit{mutation}, \textit{replacement} and \textit{termination} shared by all members of this algorithm family~\cite{SivanandamDeepaSpringerBook2010}. In this section, we present and justify the choices made for each task to define the specific GA adopted within our EmoGen methodology.
\par \textbf{Selection.} The success of genetic optimisation directly depends on the chromosomes chosen to breed and propagate in each generation. Unlike textbook GAs, there is no continuous measure of fitness of each member of population $\mathbf{P}_g$ in our implementation. Instead, the user selects one best (\textit{elite}) facial expression and any number of additional approximations considered mutually equally fit by the algorithm.
Let us mathematically define our selection operator applied to population $\mathbf{P}_g$ at a given generation $g$ as:
\begin{equation}
[\boldsymbol{\alpha}_{g,\text{elite}}, \mathbf{p}_g]= \mathcal{S}(\mathbf{P}_g),
\end{equation}
where $\mathbf{p}_g \subset \mathbf{P}_g $ and the elite sample $\boldsymbol{\alpha}_{g,\text{elite}} \notin \mathbf{p}_g$. In selection, EmoGen employs \textit{elitism}  i.e. the propagation guarantee of the elite face in the original selected form and cross-bred by averaging with other options. The elite is kept separate from other selections ($\boldsymbol{\alpha}_{g,\text{elite}} \notin \mathbf{p}_g$) to enable these elitist mechanisms. We choose to avoid any perceptual ranking of complexity greater than a single elite selection as they may not be accurate and are likely to cause user fatigue. Propagation of all non-elite selections $\mathbf{p}_g$ is equally likely, drawing from a uniform random distribution with replacement. The mixed approach of random selection for propagation and elitism helps balance \textit{selection pressure} maintaining the diversity of the population, while also preventing loss of favourable genes.
\par \textbf{Cross-breeding and mutation.} Our GA employs \textit{uniform crossover}, which is a cross-breeding approach whereby the genetic information swap is per gene (blendshape weight $\alpha_i$), rather than a chromosome segment. As the blendshape order does not encode any information, segment swapping is not meaningful. 
\par We define the cross-breeding operator acting on any pair of non-identical samples 
$\boldsymbol{\alpha}_s$ and $\boldsymbol{\alpha}_{s'}$ from the current user selection $\{\boldsymbol{\alpha}_{g,\text{elite}}, \mathbf{p}_g\}$:
\begin{equation}
\boldsymbol{\alpha}' = \mathcal{B}(\boldsymbol{\alpha}_s,\boldsymbol{\alpha}_{s'}),
\label{breed}
\end{equation}
where for each gene $\alpha'_i = \alpha_{s',i}$ if the uniform continuous distribution $\mathcal{U}(0,1) < \frac{1}{2}$ and $\alpha'_i = \alpha_{s,i}$ otherwise. The chance of cross-breeding is 50\% (i.e. a coin flip) for each gene as each sampling of $\mathcal{U}(0,1)$ is independent.
\par The subsequent mutation is allowed for a fixed number $m$ of genes $\alpha_i$ (in our case $m=2$). The new weight is drawn from the uniform continuous distribution $\mathcal{U}(0,1)$. Unlike \cite{ReedCGF2019}, where the mutation rate is typically limited, our mutation always covers the \textit{entire} range of blendweight variation. Mutation operates on $m$ randomly selected genes defined by the index set $\{i\}_m$. The random index selection is realised by independent sampling of a \textit{discrete} uniform distribution $\mathcal{U}\{0,K_{\text{core}}\}$ defined over $K_{\text{core}}$ core shapes of the blendshape model. The mutation operator is:
\begin{equation}
\boldsymbol{\alpha}'' = \mathcal{M}(\boldsymbol{\alpha}', \{i\}_m),
\end{equation}
where $\alpha''_{i} = \mathcal{U}(0,1), \;\; \forall i \in \{i\}_m $ and $\alpha''_{i} = \alpha'_{i}, \;\; \forall i \notin \{i\}_m$. 
\par In addition to the uniform crossover, we employ \textit{whole arithmetic recombination}, which is also a type of cross-breeding. In our implementation, this cross-breeding type constitutes weight averaging of two or more selections. Let us define the averaging operator on any chromosome subset $\{\boldsymbol{\alpha}_s\}_S$ of some cardinality $S \leq|\boldsymbol{p}_g| + 1$:
\begin{equation}
\boldsymbol{\alpha'} = \mathcal{A}(\{\boldsymbol{\alpha}_s\}_S),
\end{equation}
where $\alpha'_i = \frac{1}{S}\sum\limits_{s=1}^{S}\alpha_{i,s} $. In our GA, the averaging operator is applied in two contexts: 1.~ $\{\boldsymbol{\alpha}_{g,\text{elite}},\boldsymbol{\alpha}_s\}_2$ i.e. to average the elite and another randomly selected sample from the user-defined $\boldsymbol{p}_g$; 2~$\{\boldsymbol{\alpha}_1, ...,\boldsymbol{\alpha}_{|\boldsymbol{p}_g|+1}\}$ i.e. to average all user selections. Note that such averaging is also a manifestation of elitism in our framework that ensures propagation of favourable genes.
\par In both mutation and cross-breeding, facial symmetry is enforced by assigning weight updates consistently in the left/right blendshape pairs. Further corrective blendshapes are never evolved by $\mathcal{B}(\boldsymbol{\alpha}_1,\boldsymbol{\alpha}_2)$, 
$\mathcal{M}(\boldsymbol{\alpha}', \{i\}_m)$ or $\mathcal{A}(\{\boldsymbol{\alpha}_s\}_S)$. Instead the values of corrective shapes are computed \textit{post factum} based the generated core expressions as described in Section~\ref{blnd_model}. 
\par \textbf{Replacement.} Typically, in the GAs, the decision to replace one or both parents by their children is determined by their fitness ranking in the population. In our implementation, both parents are replaced in the next generation but not necessarily by the children of the particular cross-breeding/mutation round. Specifically, as is clear from Equation \ref{breed}, each randomly chosen pair of parents produces one child to be propagated. The parents can then be drawn from the selected population $\{\boldsymbol{\alpha}_{g,\text{elite}}, \mathbf{p}_g\}$ again for the next pairing. It is hard to predict how our approach compares to the classical one where parent pairs are formed strictly based on fitness, breeding always produces two children to replace both parents and the same pair of parents cannot breed more than once. We shall simulate the classical approach for comparison.
\par \textbf{Termination.} We terminate the process at a configurable fixed number of generations $G$ for practical considerations. The choice of a specific number is informed by observation of the algorithm's typical point of subjectively sufficient convergence, signalled by reduced population diversity. Termination after a fixed number of generations also helps to standardise the process for participant testing in psychology experiments.
\par \textbf{Population diversity boosting.} Apart from the absence of a continuous fitness ranking, the GA of our EmoGen methodology  follows the unusual practice of always inserting new randomly generated population members unrelated to prior generations. The practice further lowers selection pressure to avoid pre-mature convergence. The population is thus boosted substantially, by 40\%  or four out of ten samples. Each new population sample is generated by first randomly selecting an index set of $x$ core shapes $\{i\}_{x}$, sampling $\mathcal{U}\{0,K_{\text{core}}\}$, and then assigning the gene weights, such that $\alpha_{i} = \mathcal{U}(0,1)\;\; \forall i \in \{i\}_{x}$ and $\alpha_{i}=0  \;\; \forall i \notin \{i\}_{x}$. In our implementation, $x=6$. This random generation operator is defined as:
\begin{equation}
\boldsymbol{\alpha_{\text{new}}} = \mathcal{R}(\{i\}_{x})
\end{equation}
As before, symmetry is always enforced  and implausible configurations are corrected afterwards.
\par \textbf{GA summary.} Algorithm~\ref{algorithm} summarises the protocol for facial expression sample generation and evolution in pseudo-code using the formalisation and operator definitions above.
\begin{algorithm}
\SetAlgoLined
\SetKwInOut{Input}{input}\SetKwInOut{Output}{output}
 \Input{G, $\mathbf{P}_0$, $g = 0$ }
 \Output{$\boldsymbol{\alpha}_{G,\text{elite}}$}
 \While{$g \leq G$}{
 $[\boldsymbol{\alpha}_{g,\text{elite}}, \mathbf{p}_g]= \mathcal{S}(\mathbf{P}_g)$
  
 \For{$\text{fc\_nr} = 1;\ \text{fc\_nr} \leq 10;\ \text{fc\_nr}=\text{fc\_nr} + 1$}{
 
       \uIf{$\text{fc\_nr} = 1$}{
           $\boldsymbol{\alpha}_{g+1,\text{fc\_nr}} =\boldsymbol{\alpha}_{g,\text{elite}}$
       }\uElseIf{ $\text{fc\_nr} = 2$ \textbf{and} $(|\boldsymbol{p}_g| + 1) > 2 $ } {
          $\boldsymbol{\alpha}_{g+1,\text{fc\_nr}} = \mathcal{A}(\{\boldsymbol{\alpha}_1, ...,\boldsymbol{\alpha}_{|\boldsymbol{p}_g|+1}\})$  
       } \uElseIf{ $\text{fc\_nr} = 3$ \textbf{and} $g \neq 1 $ \textbf{and} $g \neq 2 $  } {
         \uIf {$(|\boldsymbol{p}_g| + 1) \geq 2 $} {
         $\boldsymbol{\alpha}_{g+1,\text{fc\_nr}} = \mathcal{A}(\{\boldsymbol{\alpha}_{g,\text{elite}},\boldsymbol{\alpha}_s\}_2)$
         } \Else{
         
             $\boldsymbol{\alpha}_{g+1,\text{fc\_nr}} = \boldsymbol{\alpha}_{g,\text{fc\_nr}}$
         }
         
       } \uElseIf{ $\text{fc\_nr} < 7$  } {
          $\boldsymbol{\alpha}' = \mathcal{B}(\boldsymbol{\alpha}_s,\boldsymbol{\alpha}_{s'})$ \\
          $\boldsymbol{\alpha}'' = \mathcal{M}(\boldsymbol{\alpha}', \{i\}_2)$ \\
          $\boldsymbol{\alpha}_{g+1,\text{fc\_nr}} = \boldsymbol{\alpha}''$
       } \Else{
          $\boldsymbol{\alpha_{\text{new}}} = \mathcal{R}(\{i\}_{6})$ \\
          $\boldsymbol{\alpha}_{g+1,\text{fc\_nr}} = \boldsymbol{\alpha_{\text{new}}} $
       }
  }
  $g = g + 1$
  
 }
 \caption{EmoGen: the genetic algorithm}
 \label{algorithm}
\end{algorithm}

\subsection{EmoGen: configurability}
\label{config}
The methodology allows one flexible configuration options for expression sample generation to suit various user studies.
\begin{figure}
\centering
\begin{tabular}{cc}
\includegraphics[scale=0.18]{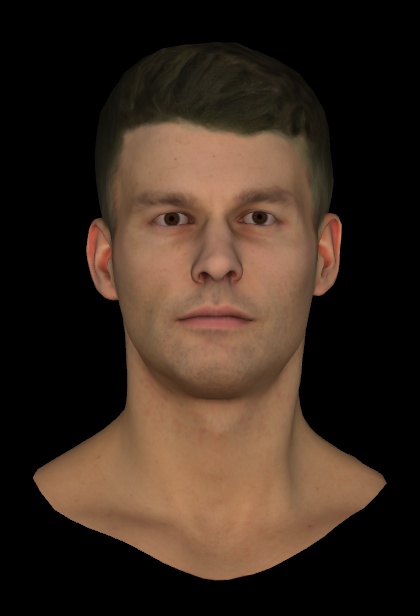}
\includegraphics[scale=0.18]{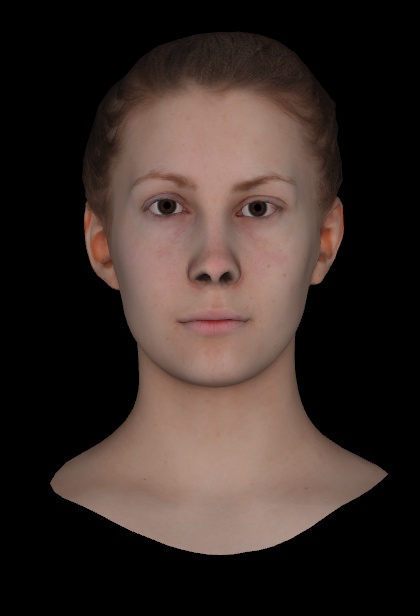}
\end{tabular}
\caption{Male and female identities available within EmoGen.}
\label{ID}
\end{figure}
\par \textbf{Model identity.} Currently, the user can choose either male or female identity: the corresponding neutral expressions are shown in Figure~\ref{ID}. EmoGen's sample generation will work consistently for any identity provided our template blendshape model is personalised to it. We aim to expand the pool of available identities in the future to include different ages and ethnicities.
\par \textbf{Maximum number of generations.} The parameter sets the search termination point of the GA. 
\par \textbf{Initialisation.} There are several scenarios that can be configured for initialisation of the GA search within EmoGen. Firstly, the user can choose between \textit{fixed} and \textit{protocol-generated} initialisations. The so-called fixed initialisation is pre-determined i.e. one of the options read from file, with the option number either specified or randomly picked. Protocol-based initialisation on the other hand prescribes random generation of two expressions of each emotion type (happy, sad, angry and fearful), one arbitrary expression and the neutral expression to compose the initial ten faces. Automated generation of specific emotion type samples for initialisation is achieved by randomly drawing from subsets of blendshapes defined by an artist to be key building blocks for particular emotions. 
The user can also specify whether to reinitialise or keep the initialisation set if the GA is reset.
\par \textbf{Minimum and maximum number of selections.} These parameters can either control how the user guides the algorithm and leave it unconstrained by setting the minimum at one and maximum at ten choices.
\par \textbf{Eye and pupil motion} can be optionally disabled if one wishes to avoid closed eyelids or gaze direction shifts in the generated expressions. 
\begin{figure}
\begin{tabular}{c@{\hspace{0.2mm}}c@{\hspace{0.2mm}}c@{\hspace{0mm}}}
\centering
$Target~1$ & $Target~2$ & $Target~3$  \\
\includegraphics[scale=0.18]{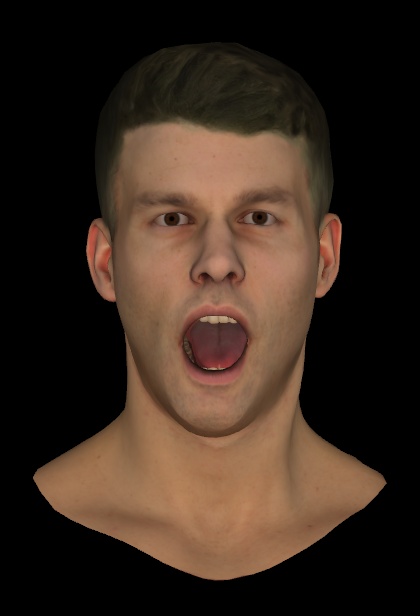} &
\includegraphics[scale=0.18]{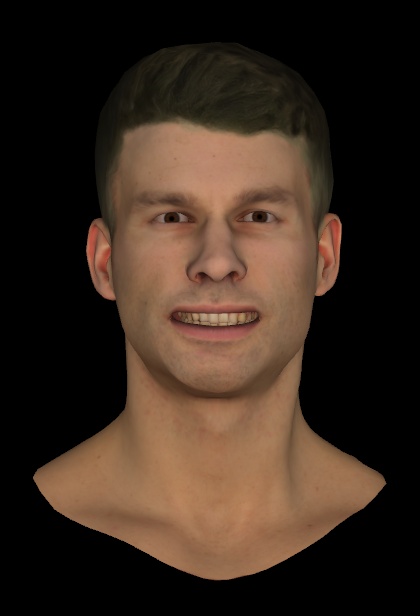} &
\includegraphics[scale=0.18]{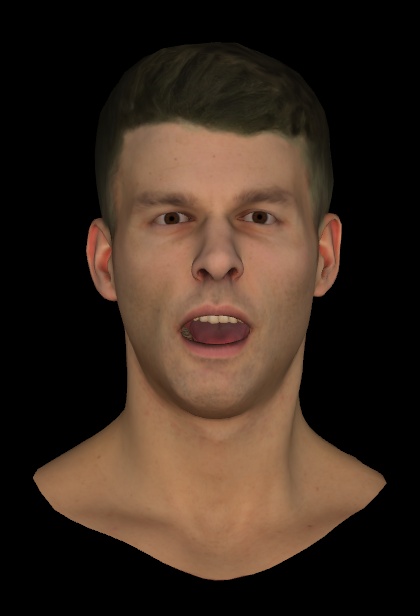} \\
\end{tabular}
\caption{Facial expression targets to evolve by EmoGen simulation.}
\label{targets}
\end{figure}

\section{Evaluation}
\label{evaluation}
The key EmoGen performance characteristic in sample generation is the ability to converge to the desired target face. The only way to quantitatively measure convergence to the absolute optimum is by simulation.
As discussed previouly in Section~\ref{related_work}, with human-driven selection we cannot assess the algorithm this way objectively as the GA target is latent, ill-defined and may be prone to advoc changes.   
In order to simulate consistent selection, a meaningful sample similarity metric needs to be formulated. The metric is meant to emulate some key factors in human perception of facial expression similarity although it is unlikely to model it fully. In this section we propose and evaluate a number of such metrics. Beyond justifying a mechanism for simulation, the study of these metrics leads to the formulation of the quantitative data analysis approach of the EmoGen methodology for psychology experiments. 
\par The studied metrics can be categorised as either \textit{empirical} i.e. trained on a distribution of user-generated expression samples or \textit{theoretical} i.e. derived from absolute notions of similarity in the representation space. We consider two representations of the facial mesh: 1. as a set of vertices $\mathcal{V}$ (\textit{vertex space}) and 2. as a vector of blendshape weights $\boldsymbol{\alpha}$ (\textit{blendshape space}). Derived spaces, such as the Principle Component Analysis (PCA) representation, are also considered.
\begin{figure}
\begin{tabular}{c@{\hspace{0mm}}c@{\hspace{0mm}}c@{\hspace{0mm}}c@{\hspace{0mm}}}
$0.3\boldsymbol{\alpha}_{tgt_2}$ & $0.5\boldsymbol{\alpha}_{tgt_2}$ & $0.7\boldsymbol{\alpha}_{tgt_2}$ & $\boldsymbol{\alpha}_{tgt_2}$ \\
\includegraphics[scale=0.15]{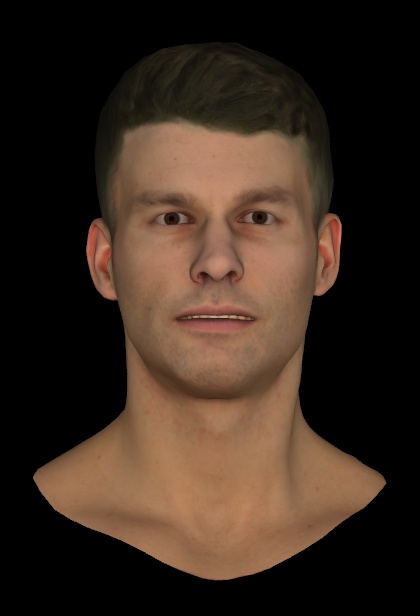} &
\includegraphics[scale=0.15]{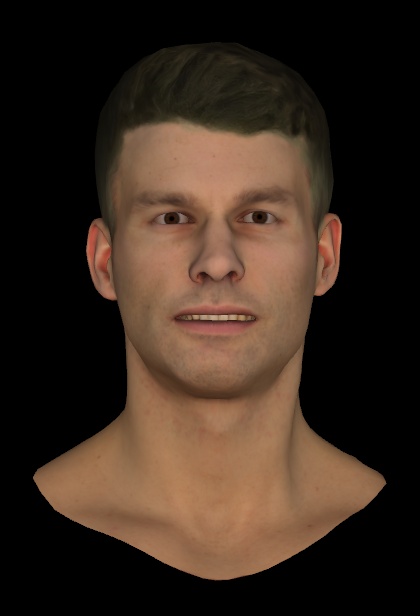} &
\includegraphics[scale=0.15]{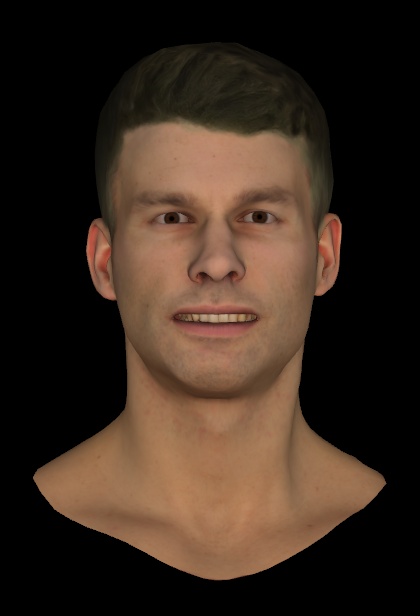} &
\includegraphics[scale=0.15]{target2.jpg} \\
\end{tabular}
\caption{Intensity variations along blendshape vector $\boldsymbol{\alpha}_{tgt_2}$ of target 2. These samples cannot be distinguished using the cosine distance metric $CD(\boldsymbol{\alpha}_{1}, \boldsymbol{\alpha}_{2})$ in Equation~\ref{cd} due to its agnosticism to the absolute intensity given equal blendshape composition.}
\label{intensity}
\end{figure}

\subsection{Default simulation settings}
\label{settings}
In the following sections, simulation results are presented for evaluation of similarity metrics and EmoGen convergence properties. As stated above, tool simulation consists in automation of the facial expression selection process with the goal of converging to a pre-defined target using the GA with consistent choices. While the similarity metric varies between experiments, the following \textit{simulation settings} are kept constant by default unless explicitly stated otherwise.
\begin{itemize}
\item  \textbf{Targets:} as in Figure~\ref{targets}, note the purposefully introduced strong similarity between Targets 1 and 3. Target 2 is complex in terms of triggering more blendshapes.
\item \textbf{Initialisation:} always a new protocol-generated expression set generated to initialise the GA.
\item \textbf{Number of repetitions:} to build reliable distributions each simulation is repeated 500 times by default.
\item \textbf{Number of generations:} each repetition of the GA search consists of ten generations excluding initialisation (generation 0).
\item \textbf{Number of choices per generation} is based on the average counts shown by human tool users. Specifically, that means two selections at initialisation and four and five selections for generations 1-5 and 6-10 respectively. As the process terminates after generation 10 only the elite selection from the last generation is used.
\end{itemize}

\subsection{Similarity metrics: definitions and assessment}
\subsubsection{Theoretical metrics}
\label{def_theoretical}
\textbf{Euclidean distance} (ED) of two blendshape weight vectors representing facial expressions is the na\"{i}ve metric presented as the baseline. We reduce vector dimensionality and sparsity by only comparing the unique core blendshapes. The reduction to the unique set of 45-55 blendshapes from the full set of 149 shapes is achieved by:
\begin{itemize}
	\item only including one blendshape in symmetrical (left/right) pairs; since facial symmetry is enforced, inclusion of both is redundant; 
	\item removing combinational correctives as these weights are computed deterministically based on core shape values;
	\item removal of unused model shapes such as lip seals (only used in 4D animation), head movement and any other shapes disabled through tool configuration parameters by the user (the number of core shapes varies due to this disabling discretion).
\end{itemize}
The metric is defined as:
\begin{equation}
ED(\boldsymbol{\alpha}{'}_{1}, \boldsymbol{\alpha}{'}_{2}) =  \|\boldsymbol{\alpha}{'}_{1} - \boldsymbol{\alpha}{'}_{2} \|,
\label{ed_blnd}
\end{equation}
where $\boldsymbol{\alpha}{'}_{1}$ and $\boldsymbol{\alpha}{'}_{2}$ denote the \textit{unique core} blendshape vectors of compared faces.
\begin{figure*}[!htbp]
\begin{tabular}{ccccc|ccc}
\hline
\multicolumn{8}{c}{$ED(\boldsymbol{\alpha}{'}_{1}, \boldsymbol{\alpha}{'}_{2})$} \\
\hline
&& \multicolumn{3}{c}{Cluster component mean of target:} & \multicolumn{3}{c}{Cluster component means 1 and 2 combined: }\\
& GMM distributions  & Unnamed & 2   & 1 and 3 & Unnamed & 2   & 1 and 3 \\
\rotatebox{90}{\parbox{3cm}{PCA component 1}}  
& \includegraphics[scale=0.28,trim = 5mm 7mm 5mm 0mm, clip]{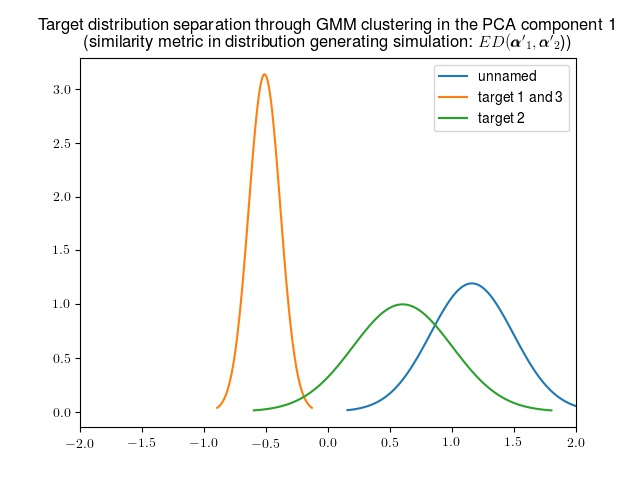}
& \includegraphics[scale=0.11]{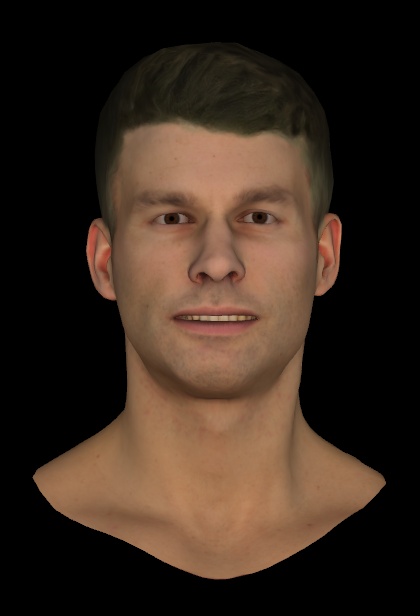}
& \includegraphics[scale=0.11]{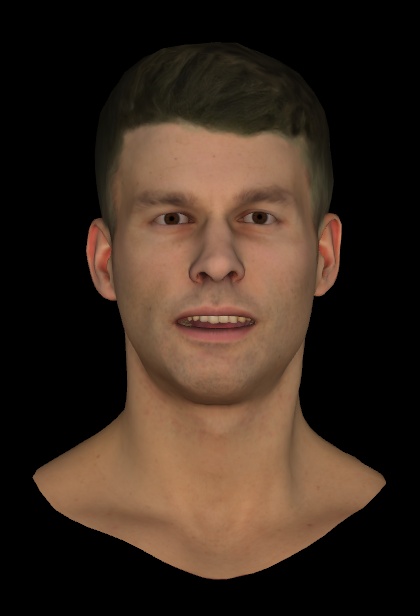}
&\includegraphics[scale=0.11]{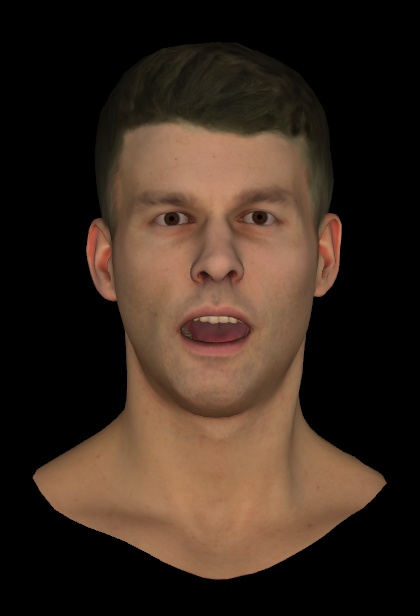} &  \\
\rotatebox{90}{\parbox{3cm}{PCA component 2}}  
& \includegraphics[scale=0.28,trim = 5mm 7mm 5mm 0mm, clip]{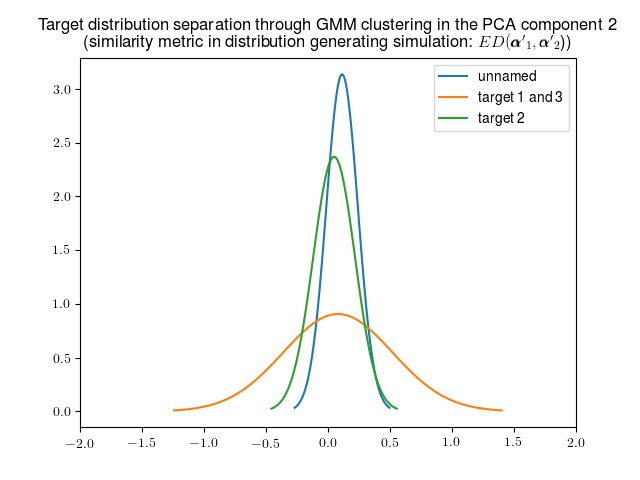}
& \includegraphics[scale=0.11]{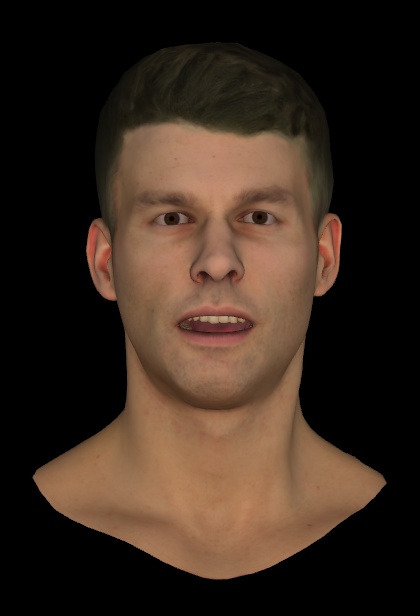}
& \includegraphics[scale=0.11]{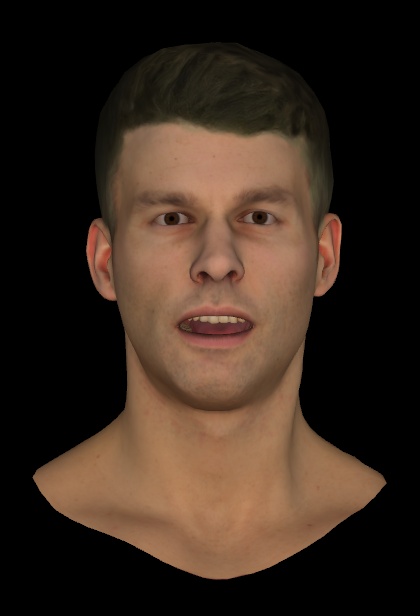}
& \includegraphics[scale=0.11]{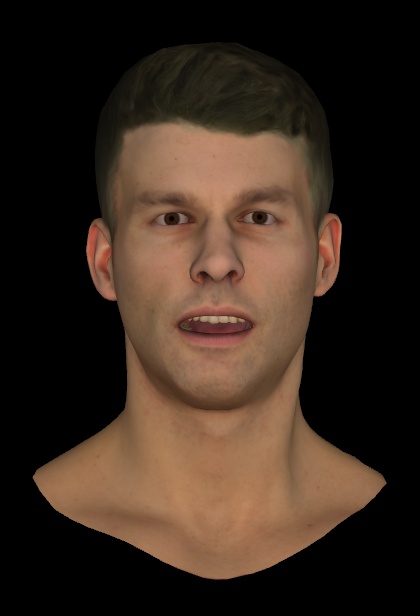}&
  \multirow{2}{*}[4cm]{\includegraphics[scale=0.12]{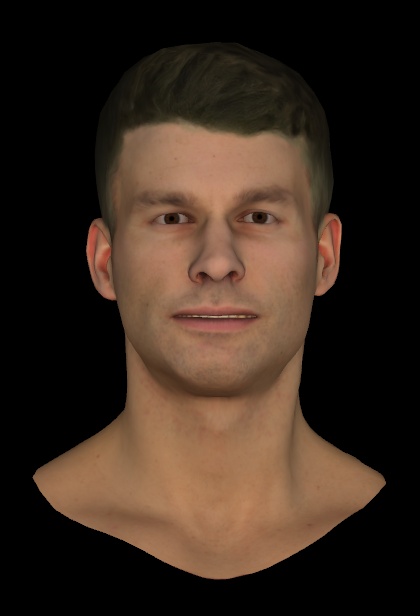} }
& \multirow{2}{*}[4cm]{\includegraphics[scale=0.12]{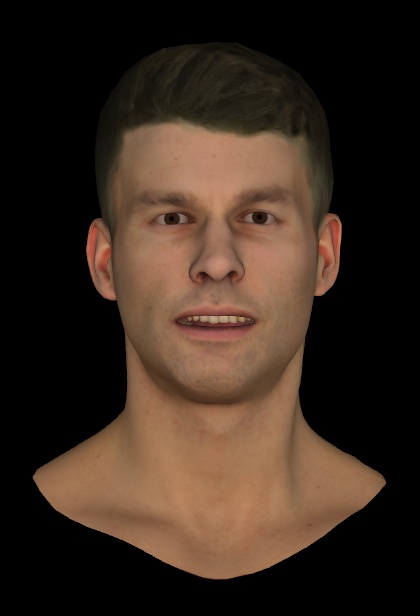}} 
& \multirow{2}{*}[4cm]{\includegraphics[scale=0.12]{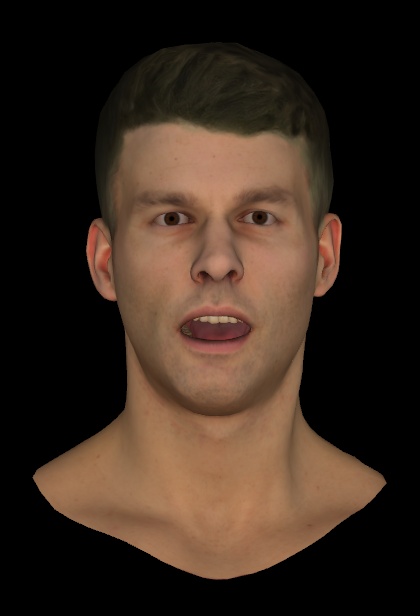}} \\
\hline
\multicolumn{8}{c}{$vrtx\_rms(\mathcal{F}_1,\mathcal{F}_2)$} \\
\hline
&& \multicolumn{3}{c}{Cluster component mean of target:} & \multicolumn{3}{c}{Cluster component means 1 and 2 combined: }\\
& GMM distributions  & 1 & 2   &  3 & 1 & 2 & 3 \\
\rotatebox{90}{\parbox{3cm}{PCA component 1}}  
& \includegraphics[scale=0.26,trim = 5mm 7mm 5mm 0mm, clip]{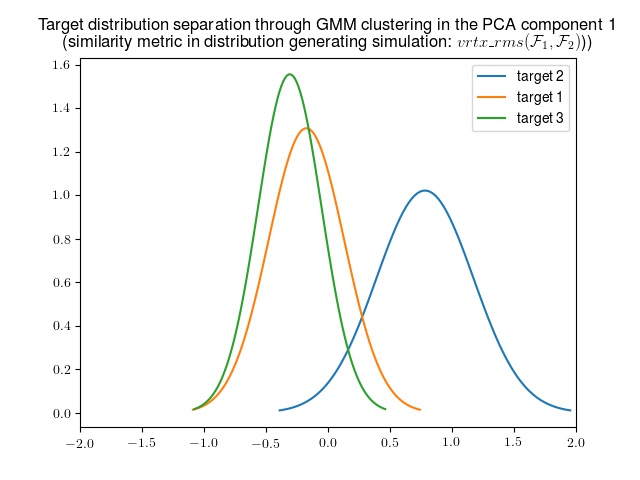}
& \includegraphics[scale=0.11]{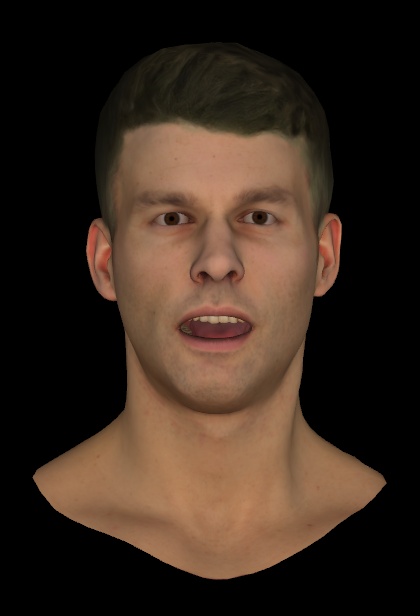}
& \includegraphics[scale=0.11]{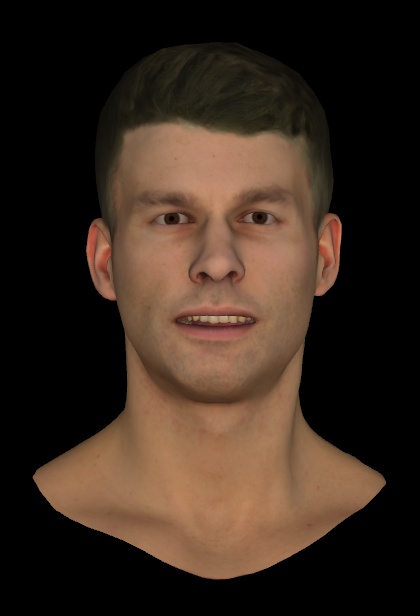}
&\includegraphics[scale=0.11]{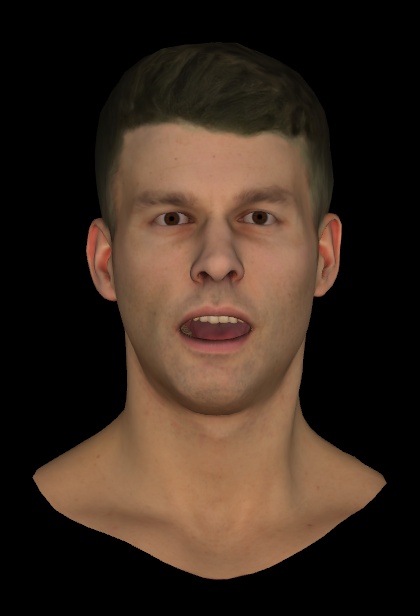} &  \\
\rotatebox{90}{\parbox{3cm}{PCA component 2}}  
& \includegraphics[scale=0.26,trim = 5mm 7mm 5mm 0mm, clip]{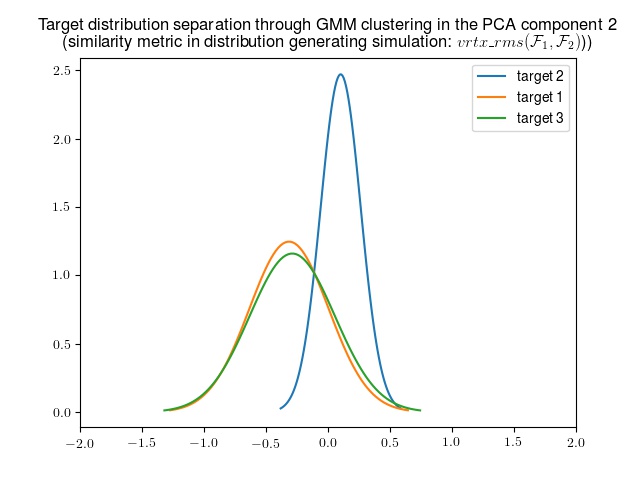}
& \includegraphics[scale=0.11]{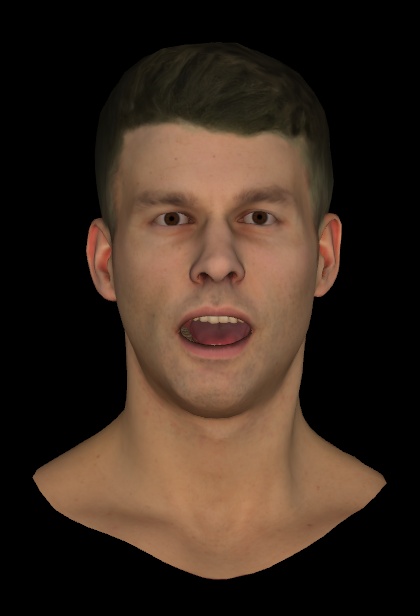}
& \includegraphics[scale=0.11]{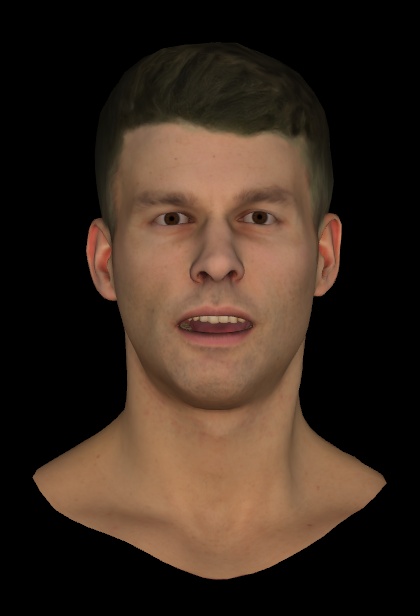}
& \fcolorbox{red}{white}{\includegraphics[scale=0.12]{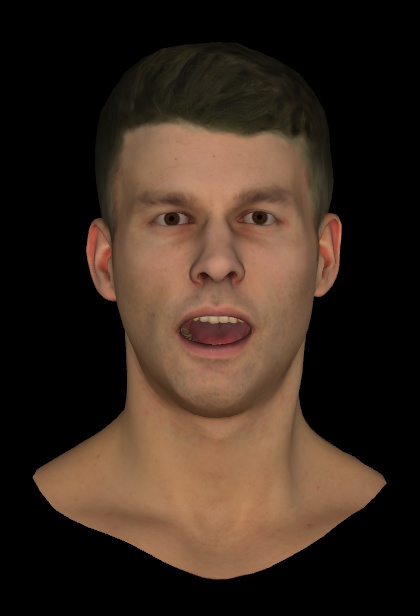}}&
  \multirow{2}{*}[4cm]{\includegraphics[scale=0.12]{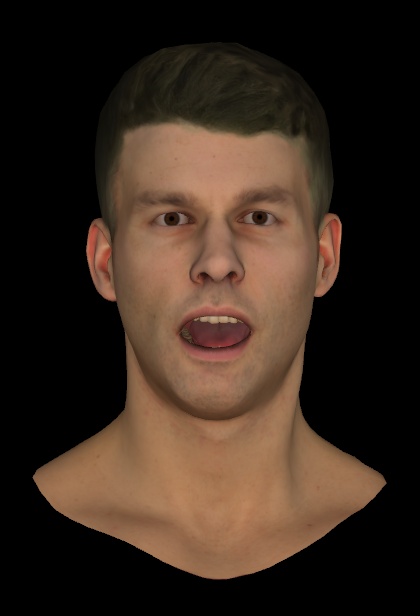} }
& \multirow{2}{*}[4cm]{\includegraphics[scale=0.12]{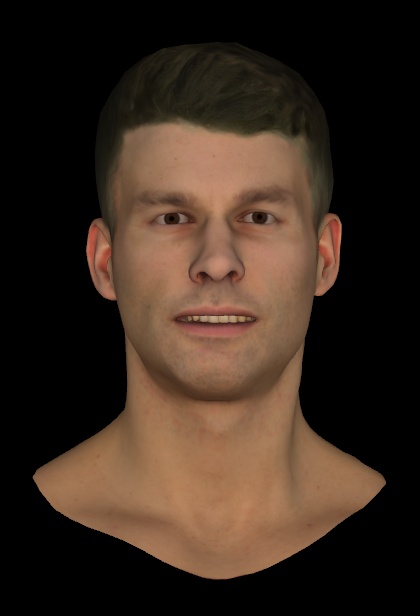}} 
& \multirow{2}{*}[4cm]{\fcolorbox{red}{white}{\includegraphics[scale=0.12]{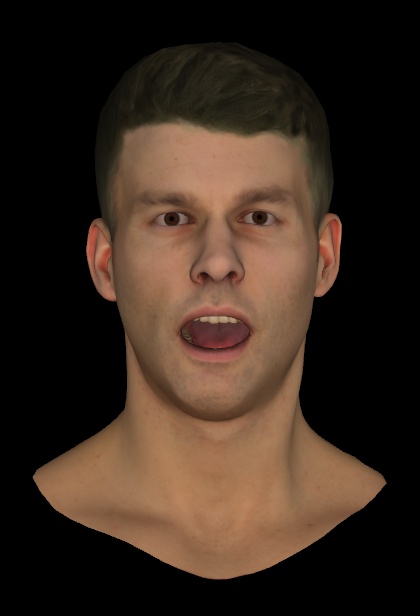}}} \\
\hline
\multicolumn{8}{c}{$CD(\boldsymbol{\alpha}_{1}, \boldsymbol{\alpha}_{2})$} \\
 \hline
 && \multicolumn{3}{c}{Cluster component means of target:} & \multicolumn{3}{c}{Cluster component means 1 and 2 combined: }\\
& GMM distributions  & 1 &  2   & 3 & 1 & 2 &3 \\
\rotatebox{90}{\parbox{3cm}{PCA component 1}}  
& \includegraphics[scale=0.26,trim = 5mm 7mm 5mm 0mm, clip]{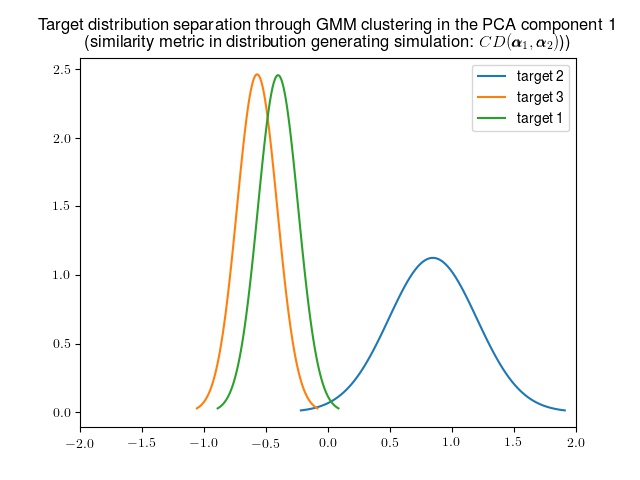}
& \includegraphics[scale=0.11]{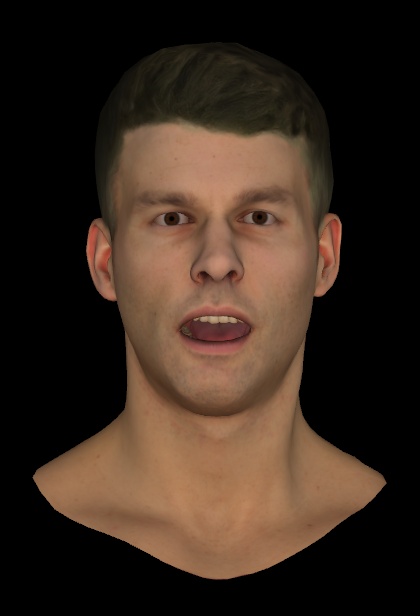}
& \includegraphics[scale=0.11]{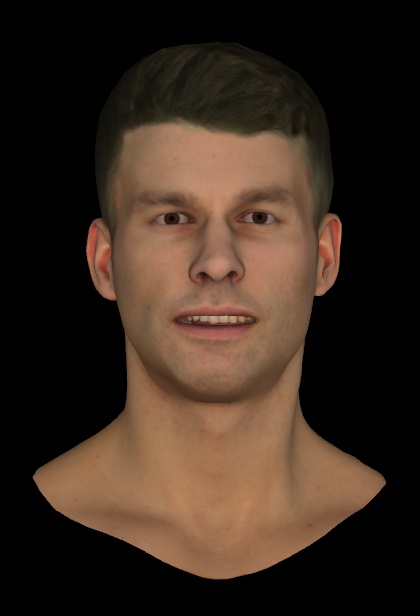}
&\includegraphics[scale=0.11]{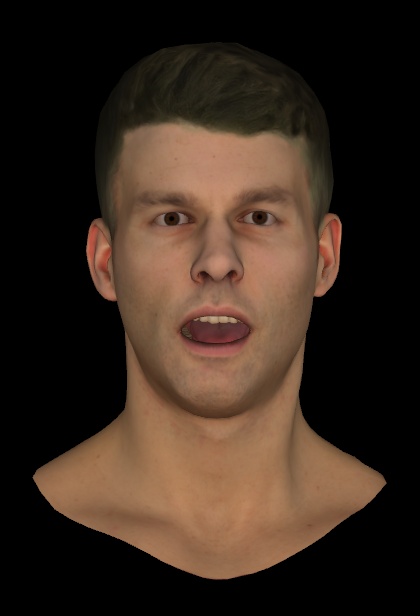} &  \\
\rotatebox{90}{\parbox{3cm}{PCA component 2}}  
& \includegraphics[scale=0.26,trim = 5mm 7mm 5mm 0mm, clip]{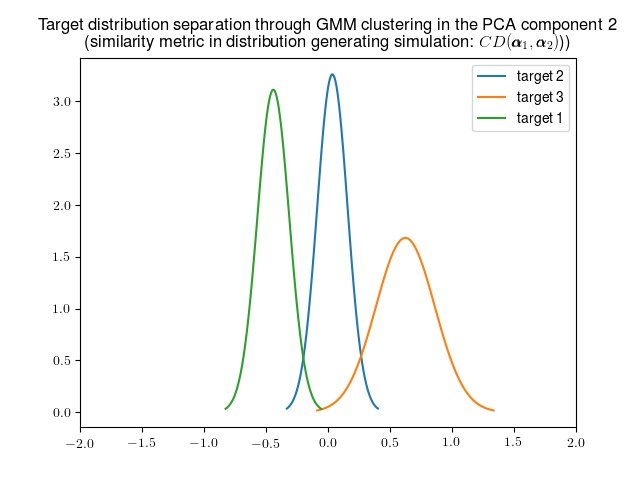}
& \includegraphics[scale=0.11]{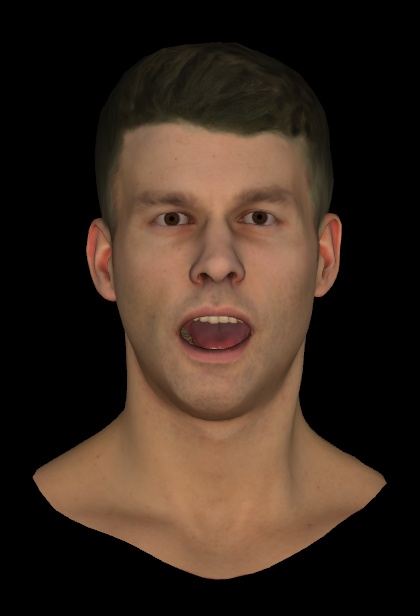}
& \includegraphics[scale=0.11]{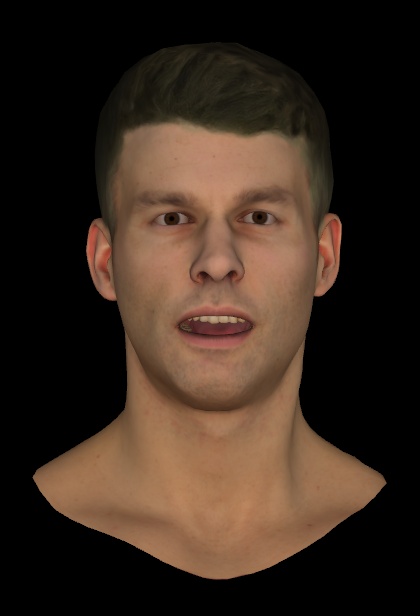}
& \fcolorbox{green}{white}{\includegraphics[scale=0.12]{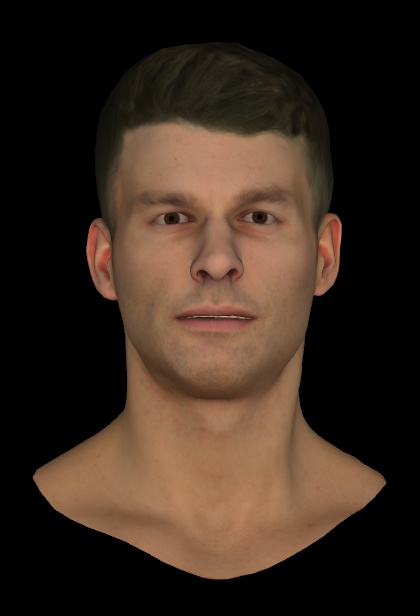}}&
  \multirow{2}{*}[4cm]{\includegraphics[scale=0.12]{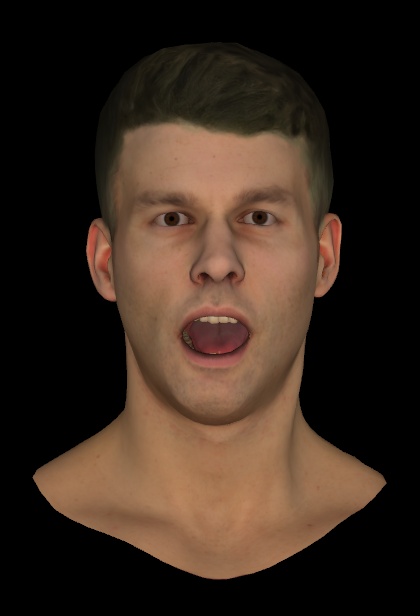} }
& \multirow{2}{*}[4cm]{\includegraphics[scale=0.12]{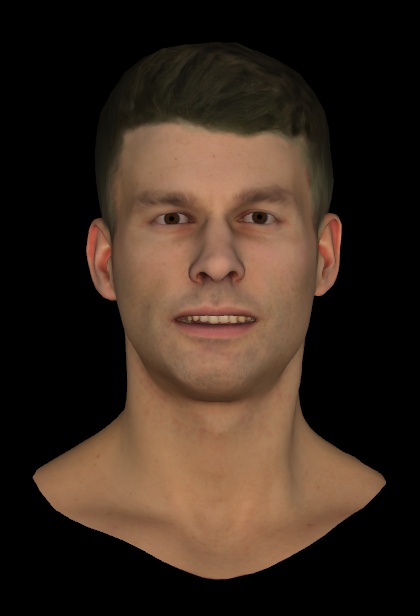}} 
&\multirow{2}{*}[4cm]{ \fcolorbox{green}{white}{\includegraphics[scale=0.12]{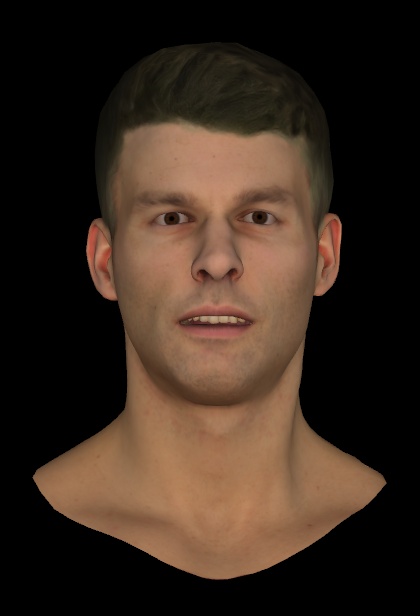}} }
\end{tabular}
\caption{Target distribution separability using Gaussian Mixture Model (GMM) fitting in the PCA space of the blendshape-model-based face representation. Data is collected by repeated EmoGen simulation towards three targets using $ED(\boldsymbol{\alpha}{'}_{1}, \boldsymbol{\alpha}{'}_{2})$, $vrtx\_rms(\mathcal{F}_1,\mathcal{F}_2)$ and $CD(\boldsymbol{\alpha}_{1}, \boldsymbol{\alpha}_{2})$ to automate similarity assessment for selection. A triple-cluster GMM is fit to combined target distributions for each similarity metric separately. Presented for each metric are the GMM representations (correspondences to targets inferred) in the first and second PCA component, visualisation of component cluster means and the total of combining the two components. From this figure, $CD(\boldsymbol{\alpha}_{1}, \boldsymbol{\alpha}_{2})$ facilitates the most accurate and consistent simulated convergence to target as its three cluster distributions are best separated in the first two PCA components and some of the cluster component means are clearly more representative e.g. the defining nose wrinkle of target 3 is only correctly attributed to its cluster in the combined distribution of $CD(\boldsymbol{\alpha}_{1}, \boldsymbol{\alpha}_{2})$. Green and red boxes highlight visualisations of respectively better and worse second PCA component cluster means and their effect on the combined total approximating targets.}
\label{gmm_viz}
\end{figure*}
\par \textbf{Cosine distance ($CD$)} is the theoretical metric comparing composition similarity of two blendshape weight vectors encoding facial expressions. The metric is defined as:
\begin{equation}
CD(\boldsymbol{\alpha}_{1}, \boldsymbol{\alpha}_{2}) = 1.0 - \frac{\boldsymbol{\alpha}_{1} \cdot \boldsymbol{\alpha}_{2}}{\|\boldsymbol{\alpha}_{1}\| \|\boldsymbol{\alpha}_{2}\|},
\label{cd}
\end{equation}
where $\boldsymbol{\alpha}_{1}$ and $\boldsymbol{\alpha}_{2}$ are the compared blendshape vectors. If the vectors align perfectly when normalised, the $CD(\boldsymbol{\alpha}_{1}, \boldsymbol{\alpha}_{2})$ error score will be 0 indicating that the expressions are a perfect compositional match in the blendshape space. Minimising the error in the blendshape vector composition means matching the identities of active (non-zero weight) blendshapes and the relative ratios of their activation levels (weights). Note that the blendshape composition error minimised by $CD(\boldsymbol{\alpha}_{1}, \boldsymbol{\alpha}_{2})$ is agnostic to the absolute magnitudes of the expression blendshape vectors as long the relative weight ratios are preserved. In other words, expressions variable in intensity along the same blendshape vector as in Figure~\ref{intensity}, obtained by scaling weight vector $\boldsymbol{\alpha}_{tgt_2}$ of target 2 by a constant, are indistinguishable by $CD(\boldsymbol{\alpha}_{1}, \boldsymbol{\alpha}_{2})$. The limitation is not prohibitive for use of $CD(\boldsymbol{\alpha}_{1}, \boldsymbol{\alpha}_{2})$ in convergence testing as the measure of blendshape composition already  provides a strict enough target definition, albeit as a line rather than as a point in the multi-dimensional blendshape space. In other contexts, expression intensity magnitude disambiguation can be done independently of blendshape composition.
\par \textbf{Root-mean-square (RMS) vertex error ($vrtx\_rms$) } is a theoretical metric defined in the geometric domain. Compared facial meshes are of the same topology meaning that the vertices are in correspondence and their total number $N$ is constant in all comparisons.  We define the metric as:
\begin{equation}
vrtx\_rms(\mathcal{F}_1,\mathcal{F}_2) = \sqrt{\frac{1}{N}\sum_{n=1}^{N}{\|\bm{v}_{1,i} - \bm{v}_{2,i} \|^2}},
\end{equation}
where $\bm{v}\in \mathcal{V}$ is a 3D vertex in the set of vertices $\mathcal{V}$ belonging to the topology of our face model. Subscripts $1$ and $2$ refer to the two face mesh instantiations $\mathcal{F}_1$ and $\mathcal{F}_2$ being compared.
\par \textbf{Evaluation using simulated data.} We assess performance of the three defined theoretical metrics by simulation. For each target in Figure~\ref{targets} we generate a distribution of evolved elites by EmoGen simulation with $ED(\boldsymbol{\alpha}{'}_{1}, \boldsymbol{\alpha}{'}_{2})$, $vrtx\_rms(\mathcal{F}_1,\mathcal{F}_2)$ and $CD(\boldsymbol{\alpha}_{1}, \boldsymbol{\alpha}_{2})$ as the similarity metric to automate selection. Thus three combined target distributions, one for each metric, result. The similarity metric performance is assessed using two indicators. Firstly, \textit{convergence accuracy} evaluates the closeness of the average face in the generated distributions of each metric to the target. Secondly, \textit{convergence consistency} measures how readily separable combined target distributions of each metric are into target clusters minimising confusion possibility.
\par For consistency assessment we utilise Gaussian Mixture Model (GMM) fitting to the combined distributions. Data, collected in the blendshape-model representation, is converted to the PCA space. The PCA space is built from the three generated combined target distributions and the targets themselves. We fit a triple-cluster GMM to each combined distribution in the PCA space. In the analysis, for each combined distribution we assess: 1.~whether the three identified clusters correspond to the target distributions and 2.~the degree of overlap between estimated labelled Gaussians and resulting misclassification of target distribution samples. The cluster-to-target correspondence is inferred by looking at what cluster label the GMM assigns to each original target or simulated target distribution mean.
\begin{table}
\center
\begin{tabular}{c|c|c|c}
\hline \hline
\multicolumn{4}{c}{Generation with $ED(\boldsymbol{\alpha}{'}_{1}, \boldsymbol{\alpha}{'}_{2})$ as similarity metric} \\
\hline
cluster identity$\rightarrow$   & unnamed &  target 2 &  target 1 and 3  \\
\hline
target 1 samples classified as      &   1   &     13            &  486             \\
target 2 samples classified as      &   191    &    309            & 0            \\
target 3 samples classified as      &  0  &      19           &  481             \\
\hline
\hline \hline
\multicolumn{4}{c}{Generation with $vrtx\_rms(\mathcal{F}_1,\mathcal{F}_2)$ as similarity metric} \\
\hline
cluster identity $\rightarrow$ & target 1 &  target 2 & target 3  \\
\hline
target 1 samples classified as  &   197  &    0      &  303 \\
target 2 samples classified as  &   2     &   497     &  1                 \\
target 3 samples classified as  &   151   &    0      &  349               \\
\hline
\multicolumn{4}{c}{Generation with $CD(\boldsymbol{\alpha}_{1}, \boldsymbol{\alpha}_{2})$ as similarity metric} \\
\hline
cluster identity $\rightarrow$  & target 1 &  target 2 & target 3 \\
\hline
target 1 samples classified as  &  472     &    8      &   20     \\
target 2 samples classified as  &    0     &    499    &    1    \\
target 3 samples classified as  &    3     &     1     &   496   \\
\hline \hline

\end{tabular}
\vspace{0.1cm}
\caption{Target distribution separability using Gaussian Mixture Models~(GMM) fitting quantitatively (also see Figure~\ref{gmm_viz}). $ED(\boldsymbol{\alpha}{'}_{1}, \boldsymbol{\alpha}{'}_{2})$, $vrtx\_rms(\mathcal{F}_1,\mathcal{F}_2)$ and $CD(\boldsymbol{\alpha}_{1}, \boldsymbol{\alpha}_{2})$ are the similarity metrics driving EmoGen simulation to generate per-target (Figure~\ref{targets}) distributions of evolved samples. A triple-cluster GMM is fit to the combined distribution of three targets for each metric separately, inferring cluster correspondence to target where possible. GMM predictions for target distribution samples are presented. From this table, $CD(\boldsymbol{\alpha}_{1}, \boldsymbol{\alpha}_{2})$ facilitates the most consistent simulated convergence of the GA as its GMM clusters clearly correspond to the targets with hardly any misclassification. In the combined distribution of $vrtx\_rms(\mathcal{F}_1,\mathcal{F}_2)$, there is misclassification between target 1 and 3 distribution samples. GMM fitting is not able to separate target 1 and 3 in the combined distribution of $ED(\boldsymbol{\alpha}{'}_{1}, \boldsymbol{\alpha}{'}_{2})$ at all and creates a cluster (``unnamed'') not corresponding to any target directly.}
\label{gmm}
\end{table}
\par Figure~\ref{gmm_viz} illustrates GMM cluster separation for the first two PCA components. In all three cases a distinct target 2 distribution is identified correctly, but there are marked differences in separability of more similar targets 1 and 3. Firstly, GMM completely fails to separate similar targets 1 and 3 in the combined distribution generated using $ED(\boldsymbol{\alpha}{'}_{1}, \boldsymbol{\alpha}{'}_{2})$. Instead cluster optimisation appears to split target 2 distribution into two clusters. Secondly, although the combined distribution of $vrtx\_rms(\mathcal{F}_1,\mathcal{F}_2)$ is separable into target clusters, the component means of the challengingly similar targets 1 and 3 are not sufficiently distinct resulting in a very close combined total. The component 2 mean in this case lacks the characteristic nose wrinkle of target 3. Thirdly, in contrast a clear separation of problematic targets 1 and 3, especially in the second PCA component, is found in the distribution of $CD(\boldsymbol{\alpha}_{1}, \boldsymbol{\alpha}_{2})$. Visually, this boils down to the correct identification of the nose wrinkle as a characteristic of the target 3 cluster. 
\par Confirming the observations, Table~\ref{gmm} presents classification of simulated target distribution samples by the GMM. Near perfect classification is achieved using the GMM built on the combined distribution of $CD(\boldsymbol{\alpha}_{1}, \boldsymbol{\alpha}_{2})$, while with the distributions of $ED(\boldsymbol{\alpha}{'}_{1}, \boldsymbol{\alpha}{'}_{2})$  and $vrtx\_rms(\mathcal{F}_1,\mathcal{F}_2)$ there is, respectively, lumping of target 1 and 3 into a single cluster and a substantial misclassification overlap. From this analysis we can conclude that $CD(\boldsymbol{\alpha}_{1}, \boldsymbol{\alpha}_{2})$ leads to the \textit{most consistent} convergence to target in simulation as the found clusters are in clear correspondence to targets and the Gaussian cluster variances are evidently small enough to practically avoid distribution sample misclassification.
\par
\begin{figure*}
\begin{tabular}{cccccccc}
& Target & \multicolumn{6}{c}{Average of target distributions (with distance error heatmaps)} \\
&        & \multicolumn{2}{c}{$ED(\boldsymbol{\alpha}{'}_{1},\boldsymbol{\alpha}{'}_{2})$} &
           \multicolumn{2}{c}{$vrtx\_rms(\mathcal{F}_1,\mathcal{F}_2)$  }     &
           \multicolumn{2}{c}{$CD(\boldsymbol{\alpha}_{1},\boldsymbol{\alpha}_{2})$ }       \\
1 & \includegraphics[scale=0.11]{target1.jpg} 
& \includegraphics[scale=0.11]{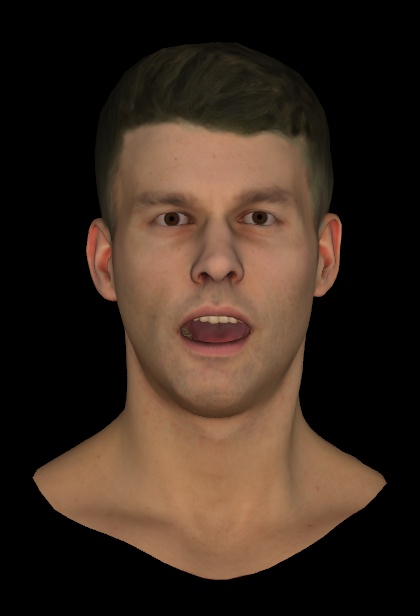}  & \includegraphics[scale=0.1]{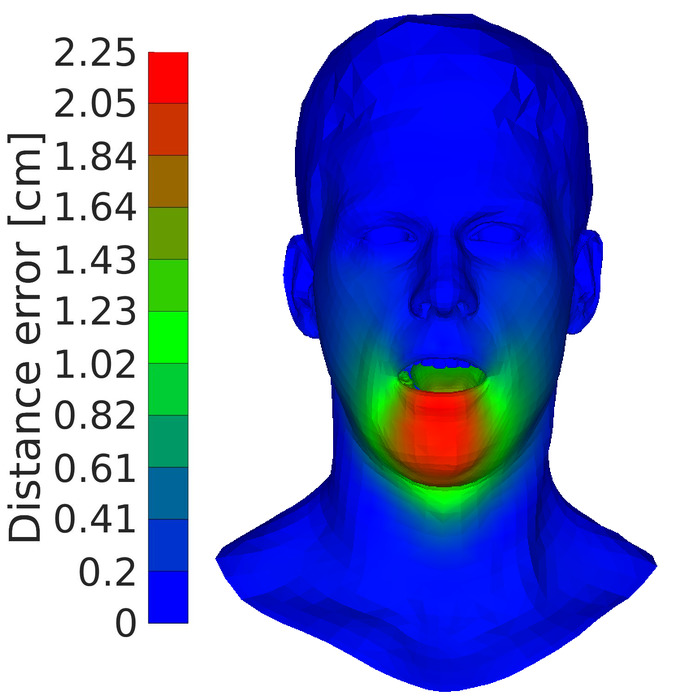}                                            
& \includegraphics[scale=0.11]{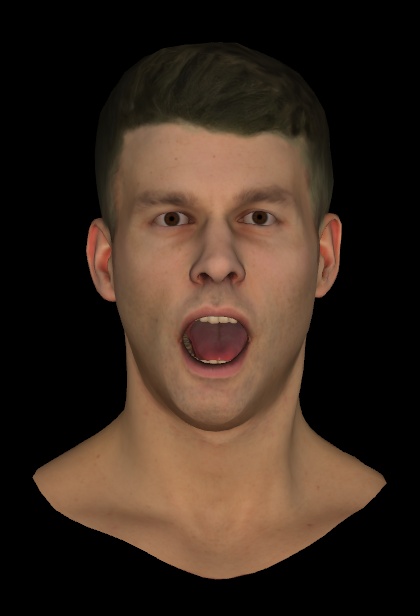} & \includegraphics[scale=0.1]{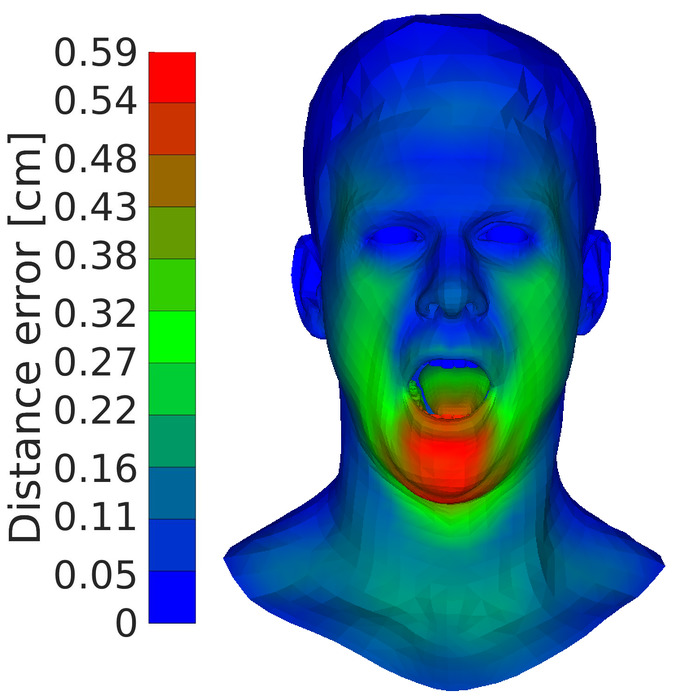}
& \includegraphics[scale=0.11]{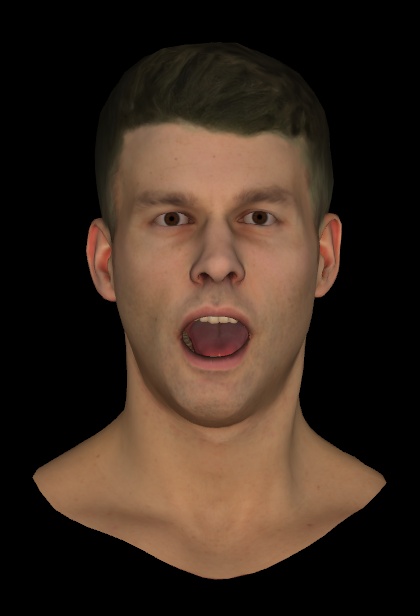}  & \includegraphics[scale=0.1]{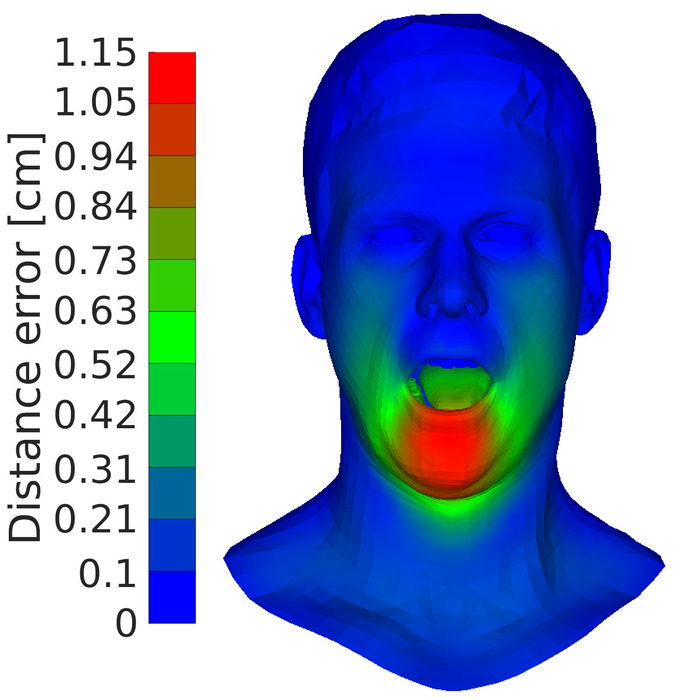}  \\
2 &
\includegraphics[scale=0.11]{target2.jpg} 
&\includegraphics[scale=0.11]{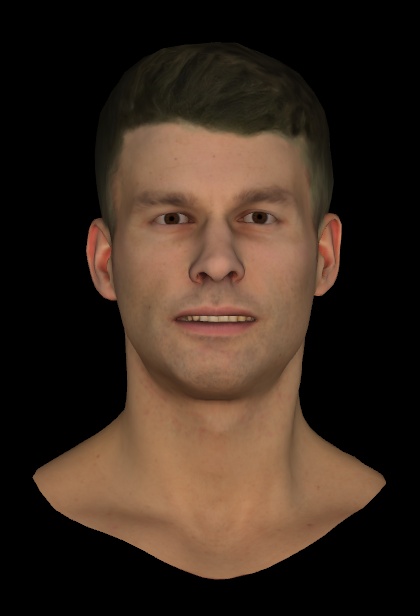}& \includegraphics[scale=0.1]{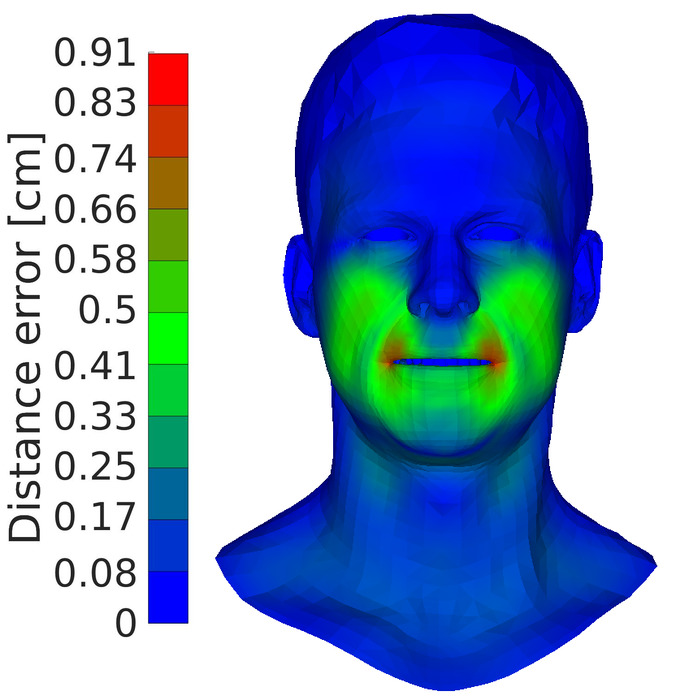}                                              &\includegraphics[scale=0.11]{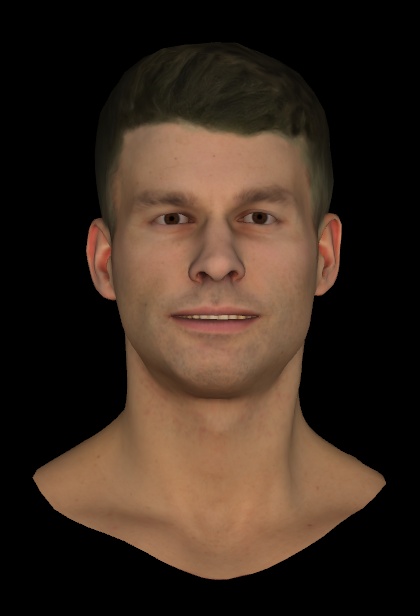}& \includegraphics[scale=0.1]{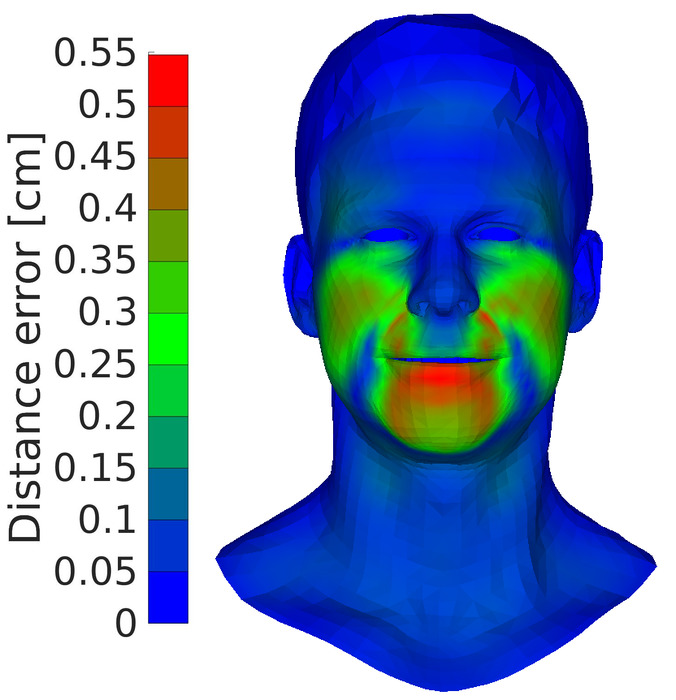}                                              &\includegraphics[scale=0.11]{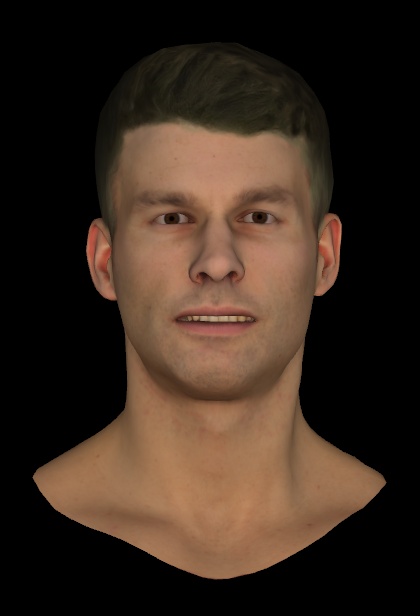} & \includegraphics[scale=0.1]{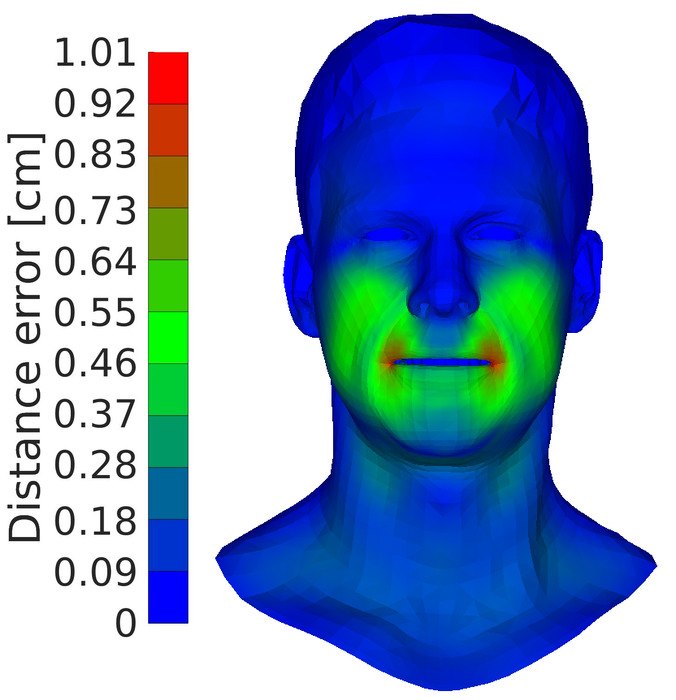}                                             \\
3 &
\includegraphics[scale=0.11]{target3.jpg} 
& \includegraphics[scale=0.11]{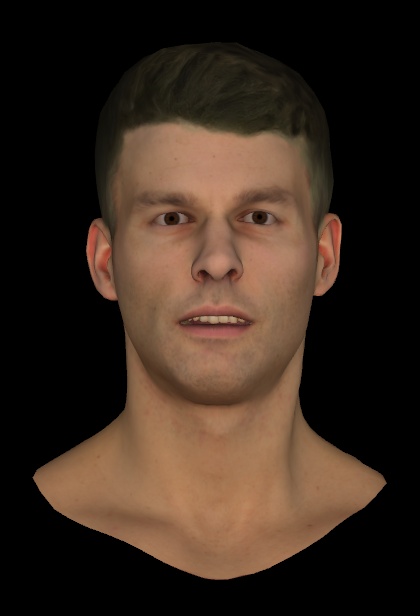}&\includegraphics[scale=0.1]{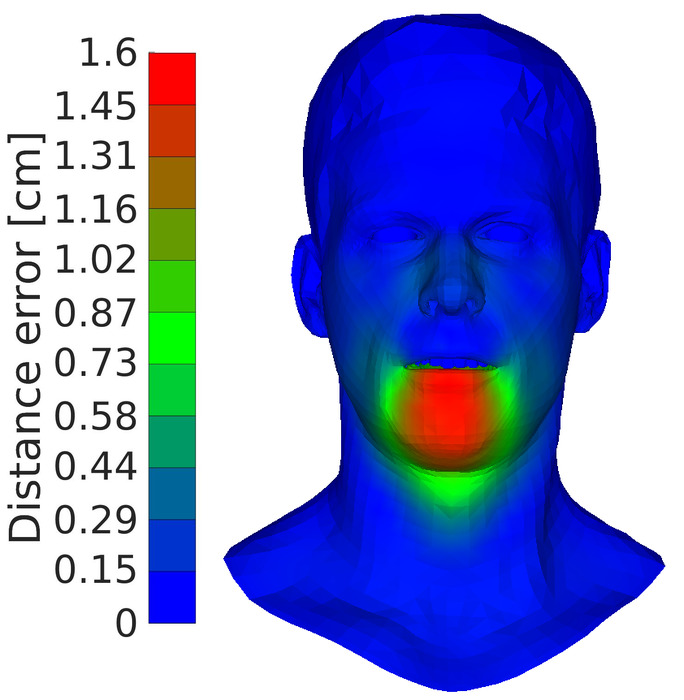}                                            &\includegraphics[scale=0.11]{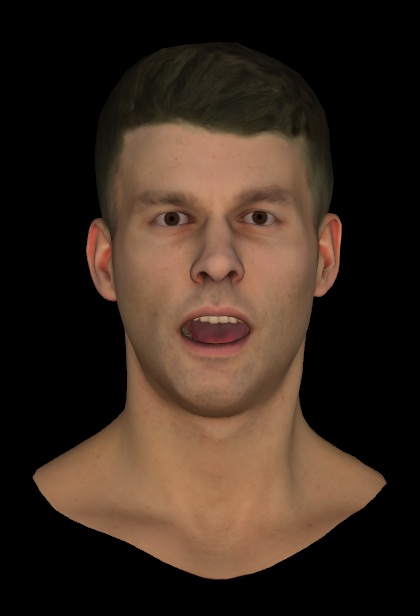} & \includegraphics[scale=0.1]{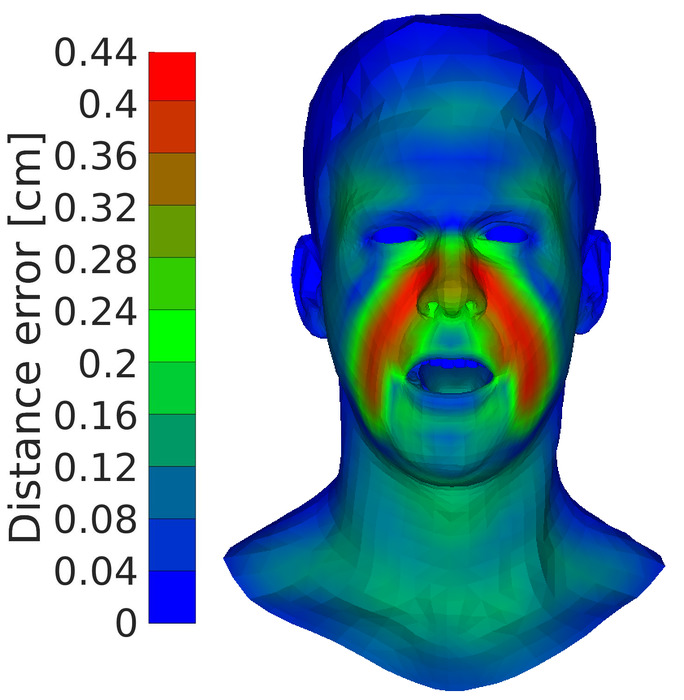}                                              & \includegraphics[scale=0.11]{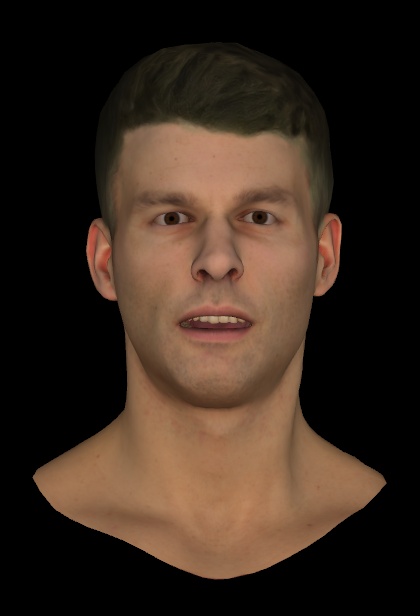}  & \includegraphics[scale=0.1]{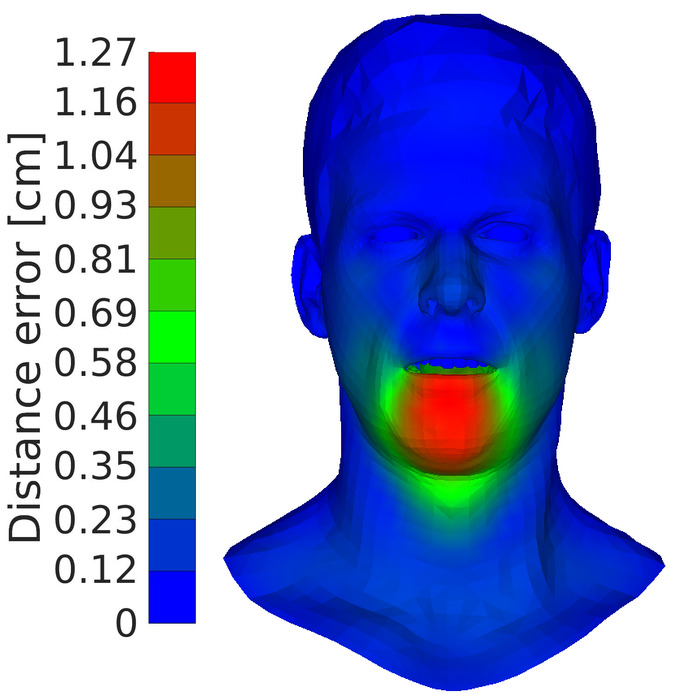}                                              \\
\end{tabular}
\caption{Convergence accuracy using $ED(\boldsymbol{\alpha}{'}_{1}, \boldsymbol{\alpha}{'}_{2})$, $vrtx\_rms(\mathcal{F}_1,\mathcal{F}_2)$ and $CD(\boldsymbol{\alpha}_{1}, \boldsymbol{\alpha}_{2})$ as similarity metrics to automate selection in EmoGen simulation. Presented for each metric is the average expression in its target distributions, rendered for assessment qualitatively and quantitatively i.e. as a heatmap of distance error relative to the target (note variable error ranges).}
\label{distribution_means}
\end{figure*}

 Cluster means do not reflect simulated target distribution means when separability by fitting a GMM is poor as is the case  with the data generated using $ED(\boldsymbol{\alpha}{'}_{1}, \boldsymbol{\alpha}{'}_{2})$ and $vrtx\_rms(\mathcal{F}_1,\mathcal{F}_2)$. Hence \textit{convergence accuracy} is assessed by comparing the simulated target distribution mean, generated using each similarity metric, to the targets in Figure~\ref{targets}. As the visual differences can be subtle, in Figure~\ref{distribution_means} we also show heatmaps  to draw the reader's attention to the key features. 
\par As can be expected, $vrtx\_rms(\mathcal{F}_1,\mathcal{F}_2)$ minimises the overall geometric error, which is below $<0.6\,\text{cm}$ for all targets. However, this comes at the expense of omitting subtler features such the the nose wrinkle of target 3.
In summary, $vrtx\_rms(\mathcal{F}_1,\mathcal{F}_2)$ will always optimise for the facial feature of the largest magnitude. 
\par $ED(\boldsymbol{\alpha}{'}_{1}, \boldsymbol{\alpha}{'}_{2})$ and  $CD(\boldsymbol{\alpha}_{1}, \boldsymbol{\alpha}_{2})$ perform more similarly in terms of the mean accuracy. Two noticeable shortcomings of $ED(\boldsymbol{\alpha}{'}_{1}, \boldsymbol{\alpha}{'}_{2})$ relative to $CD(\boldsymbol{\alpha}_{1}, \boldsymbol{\alpha}_{2})$ are the extent of jaw opening (local error $\approx2\,\text{cm}$ and $\approx1\,\text{cm}$ respectively) and the less pronounced nose wrinkle (local error $\approx0.5\,\text{cm}$ and $\approx0\,\text{cm}$ respectively). In conclusion, the distribution means achieved with $CD(\boldsymbol{\alpha}_{1}, \boldsymbol{\alpha}_{2})$ strike the best balance between the overall geometric accuracy and retaining subtle features. 
\par \textbf{Overall assessment of theoretical metrics.} In summary, we conclude that $CD(\boldsymbol{\alpha}_{1}, \boldsymbol{\alpha}_{2})$ exhibits the most desirable behaviour in terms of both convergence \textit{accuracy} and \textit{consistency} of all the theoretical metrics considered. The advantages outweigh its one major limitation of being agnostic to the absolute expression intensity as illustrated in Figure~\ref{intensity}. The measure of convergence in terms of blendshape composition alone already provides a strong indication of our GA's performance in expression sample generation. Henceforth we assess convergence to target through cosine distance error distributions. If necessary, further disambiguation by intensity can be performed using $vrtx\_rms(\mathcal{F}_1,\mathcal{F}_2)$ afterwards.
\subsubsection{Empirical metrics}
These metrics are classed as empirical because they are derived from expression distributions generated by human participants. The aim is to investigate whether any additional accuracy or robustness of facial similarity assessment can be won by exploring observed covariances in data. 
\begin{table}
\begin{tabular}{c|c|c}
notation target   &PCA mode variance  & Target relative to neutral \\
\hline
$\boldsymbol{\beta}_{l}$     & low:   component 41   & $\beta_{l,41} = \beta_{n,41} + 1.0$ \\ \hline
$\boldsymbol{\beta}_{h}$     & high:  component 1    & $\beta_{l,1} = \beta_{n,1} + 1.0$ \\ \hline
$\boldsymbol{\beta}_{m_{1}}$ & mixed: component 1 \& 41,  & $\beta_{m_{1},1} = \beta_{n,1} + 1.0$ \\ 
                      &    equally                    & $\beta_{m_{1},41} = \beta_{n,41} + 1.0$ \\
\hline
\end{tabular}
\vspace{0.2cm}
\caption{Target face definition for control experiment on relative performance of $ED(\boldsymbol{\beta}_{1},\boldsymbol{\beta}_{2})$ and $std\_ED(\boldsymbol{\beta}_{1},\boldsymbol{\beta}_{2})$ in the PCA space. $\boldsymbol{\beta}_{n}$ is the PCA representation of the neutral of the blendshape face model $\|\boldsymbol{\alpha}\|= 0$.}
\label{defs}
\end{table}

\begin{figure*}
\begin{tabular}{cccccccc}
\multicolumn{2}{c}{Target} & Metric & \multicolumn{5}{c}{Five most similar faces according to metric }\\
 &&& CD=0.970 & CD=0.970 & CD=1.030 & CD=0.674 & CD=0.990 \\
\multirow{2}{*}{\rotatebox[origin=c]{90}{\parbox{1.5cm}{\textbf{Target $\boldsymbol{\beta}_{l}$}}}}&
\multirow{2}{*}[1.5cm]{\includegraphics[scale=0.20]{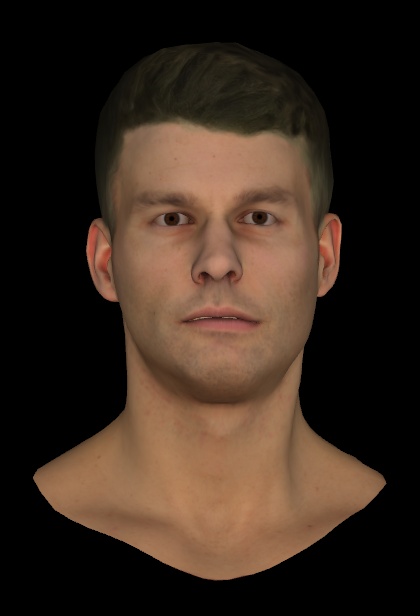}}&
\rotatebox[origin=c]{90}{\parbox{0.2cm}{$ED(\boldsymbol{\beta}_{1},\boldsymbol{\beta}_{2})$}} &
\includegraphics[scale=0.15]{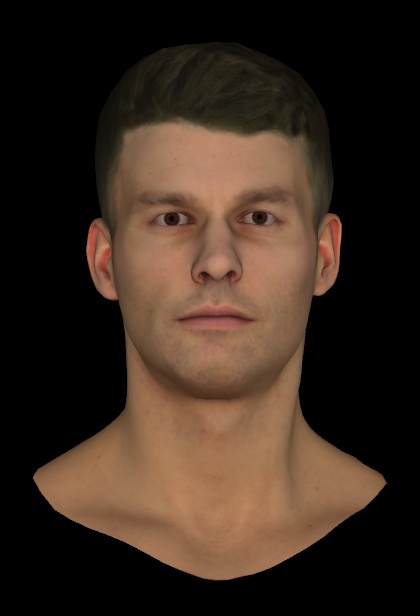}&
\includegraphics[scale=0.15]{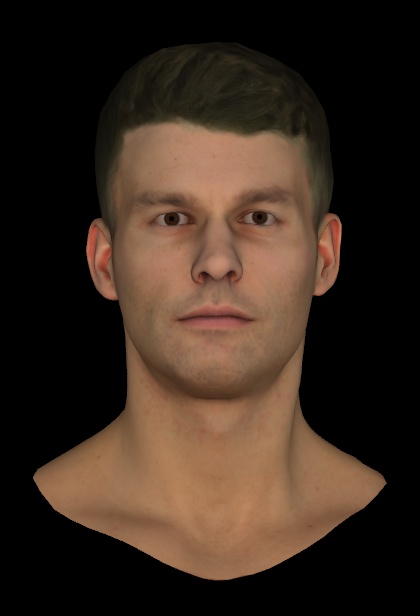}&
\includegraphics[scale=0.15]{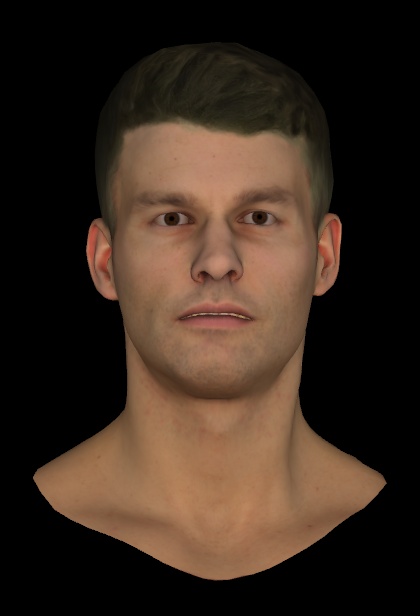}&
\includegraphics[scale=0.15]{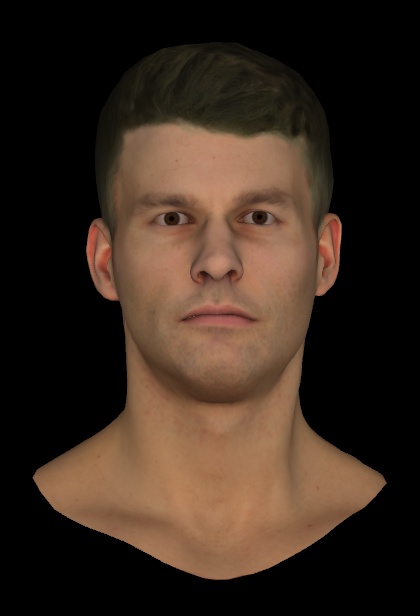}&
\includegraphics[scale=0.15]{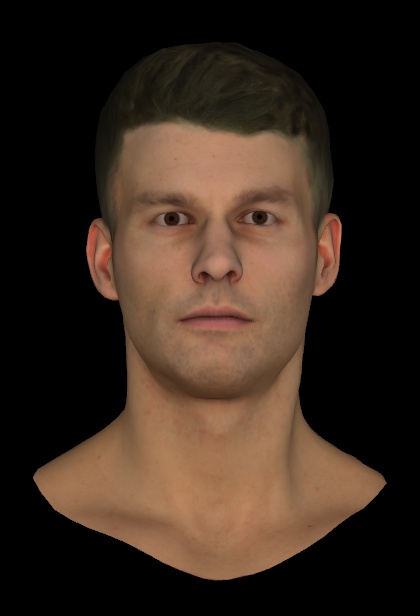}  \\
 &&& CD=0.572 & CD=0.494 & CD=0.469 & CD=0.567 &  CD=0.642 \\
                                            &
                                            &
\rotatebox[origin=c]{90}{\parbox{0.2cm}{$std\_ED(\boldsymbol{\beta}_{1},\boldsymbol{\beta}_{2})$}} &                                            
\includegraphics[scale=0.15]{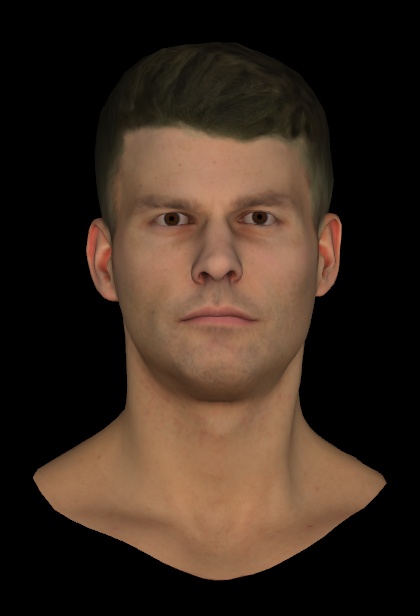}&
\includegraphics[scale=0.15]{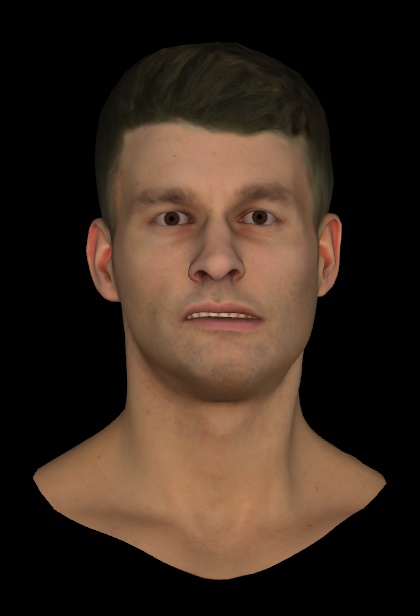}&
\includegraphics[scale=0.15]{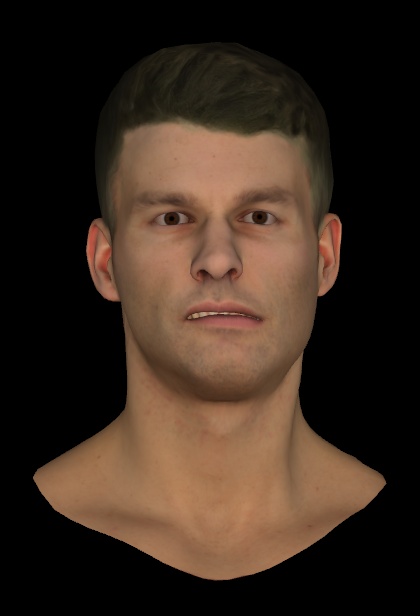}&
\includegraphics[scale=0.15]{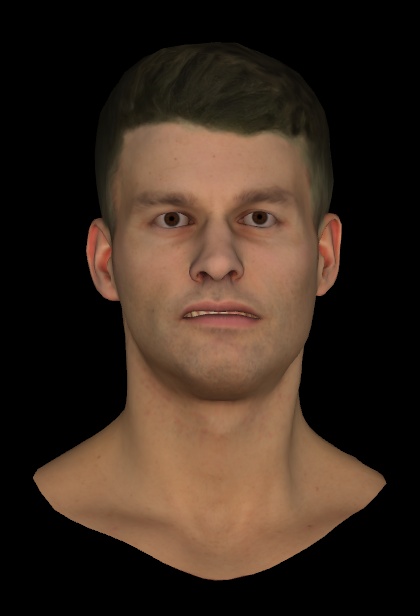}&
\includegraphics[scale=0.15]{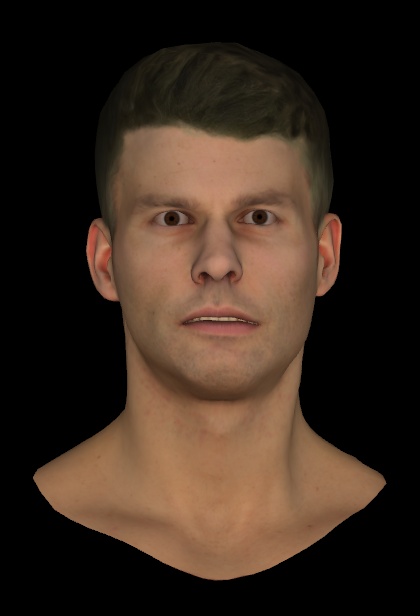} \\
\hline
 &&& CD=0.351 & CD=0.523 & CD= 0.978 & CD=0.978 & CD=0.491 \\
\multirow{2}{*}{\rotatebox[origin=c]{90}{\parbox{1.5cm}{\textbf{Target $\boldsymbol{\beta}_{h}$}}}}&
\multirow{2}{*}[1.5cm]{\includegraphics[scale=0.20]{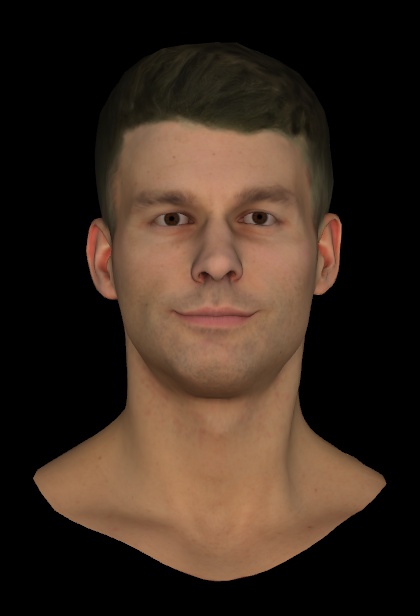}}&
\rotatebox[origin=c]{90}{\parbox{0.2cm}{$ED(\boldsymbol{\beta}_{1},\boldsymbol{\beta}_{2})$}} &
\includegraphics[scale=0.15]{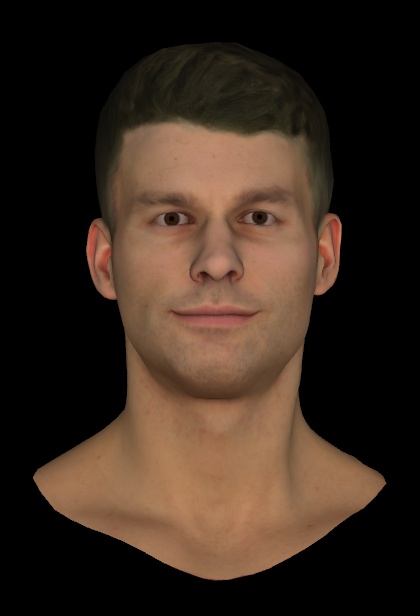}&
\includegraphics[scale=0.15]{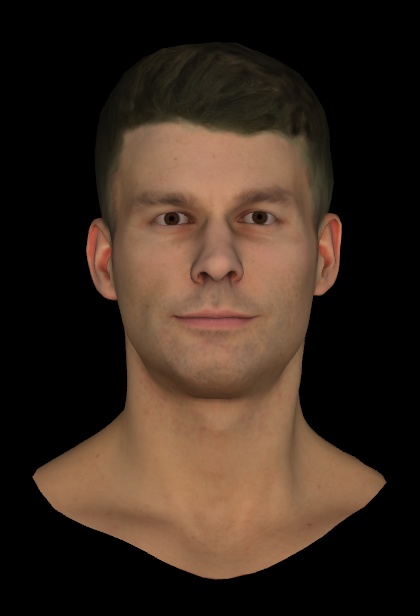}&
\includegraphics[scale=0.15]{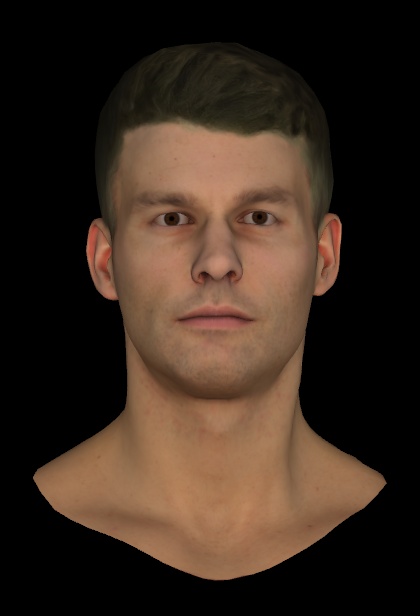}&
\includegraphics[scale=0.15]{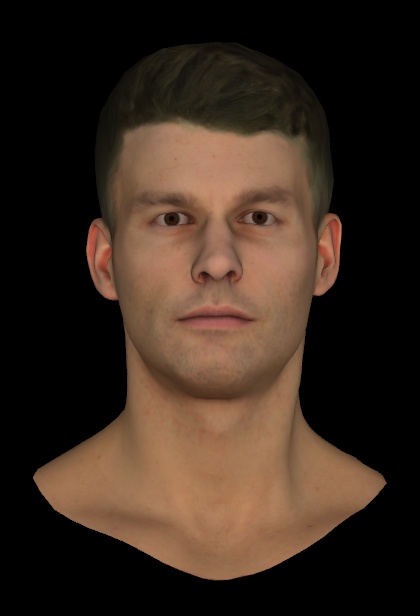}&
\includegraphics[scale=0.15]{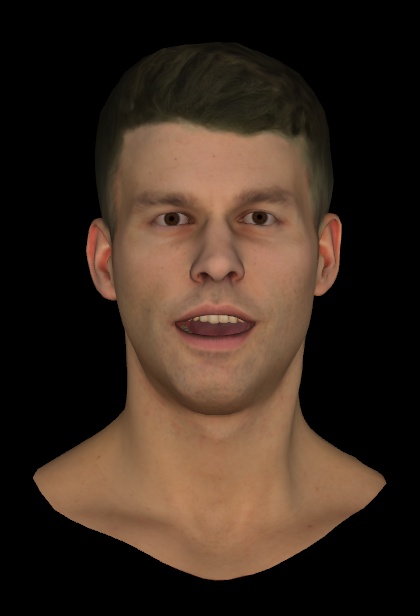}  \\
 &&& CD=0.978  & CD=0.978 & CD=1.141 & CD=0.968 &  CD= 1.032 \\
                                            &
                                            &
\rotatebox[origin=c]{90}{\parbox{0.2cm}{$std\_ED(\boldsymbol{\beta}_{1},\boldsymbol{\beta}_{2})$}} &                                            
\includegraphics[scale=0.15]{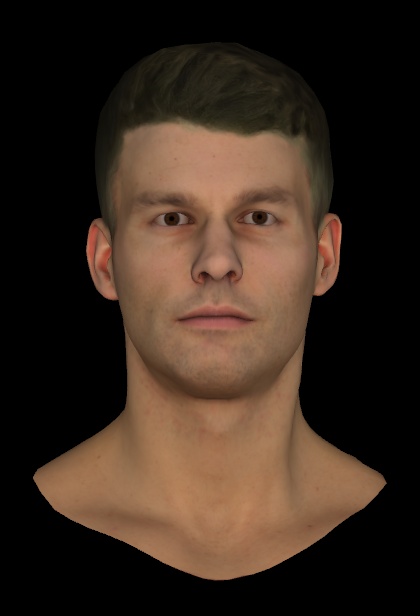}&
\includegraphics[scale=0.15]{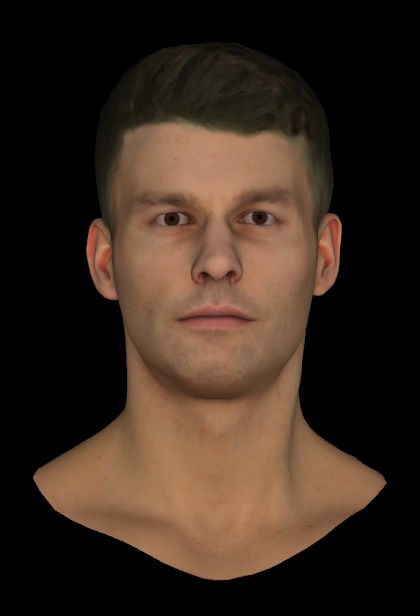}&
\includegraphics[scale=0.15]{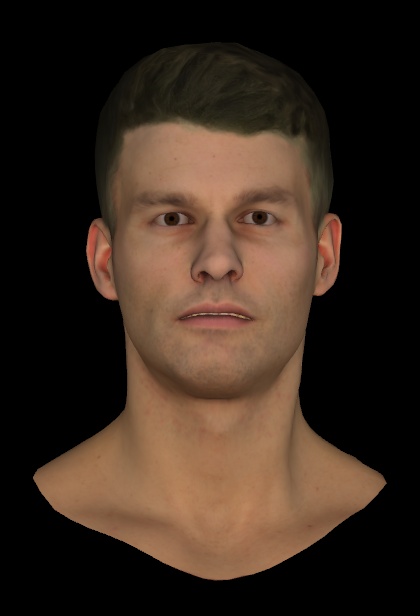}&
\includegraphics[scale=0.15]{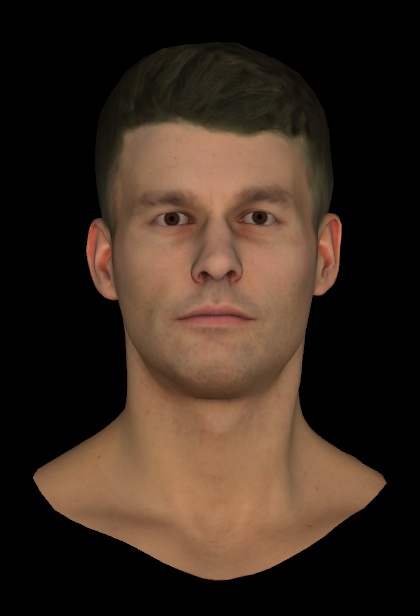}&
\includegraphics[scale=0.15]{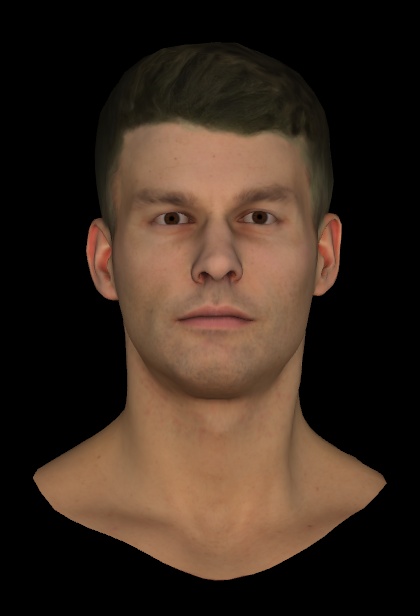}\\
\hline
 &&& CD=0.534 & CD= 0.653 & CD=0.621  & CD=0.621 & CD=0.605 \\
\multirow{2}{*}{\rotatebox[origin=c]{90}{\parbox{2cm}{\textbf{Target $\boldsymbol{\beta}_{m_{1}}$}}}}&
\multirow{2}{*}[1.5cm]{\includegraphics[scale=0.20]{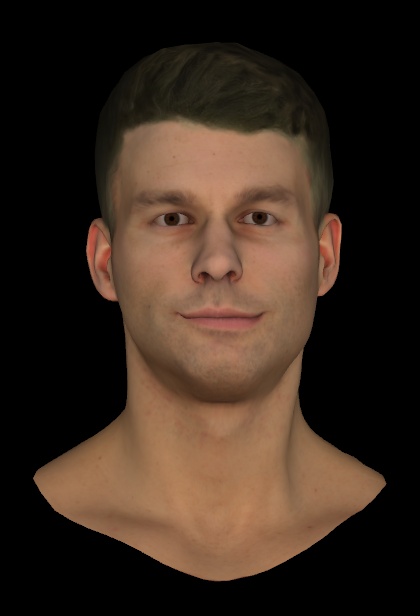}}&
\rotatebox[origin=c]{90}{\parbox{0.2cm}{$ED(\boldsymbol{\beta}_{1},\boldsymbol{\beta}_{2})$}} &
\includegraphics[scale=0.15]{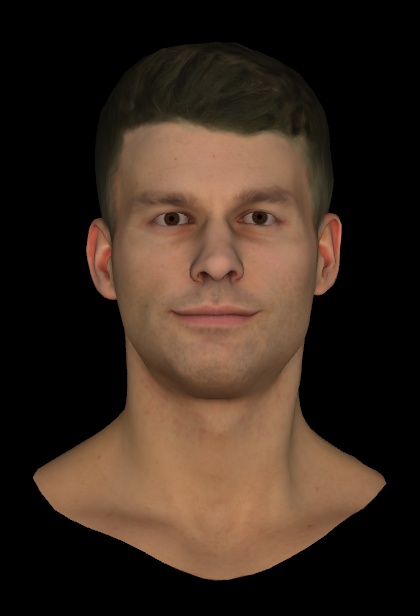}&
\includegraphics[scale=0.15]{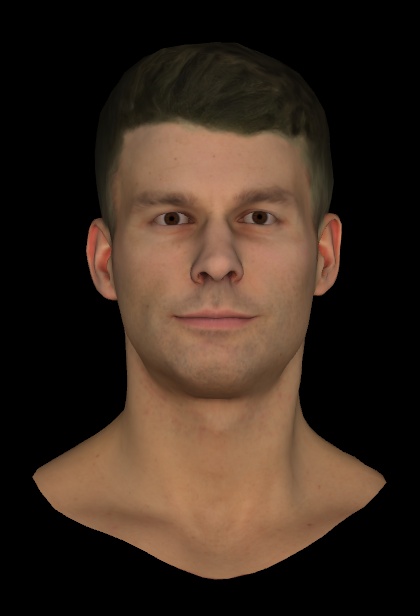}&
\includegraphics[scale=0.15]{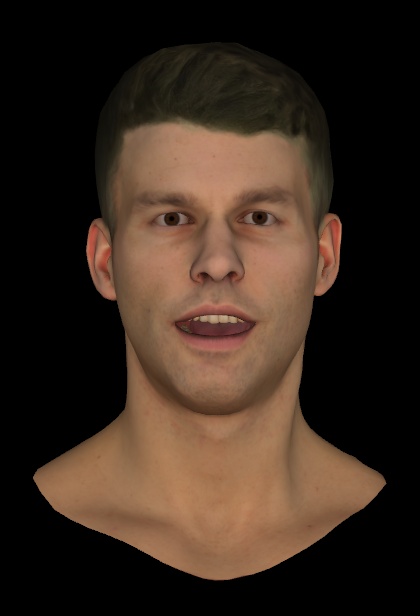}&
\includegraphics[scale=0.15]{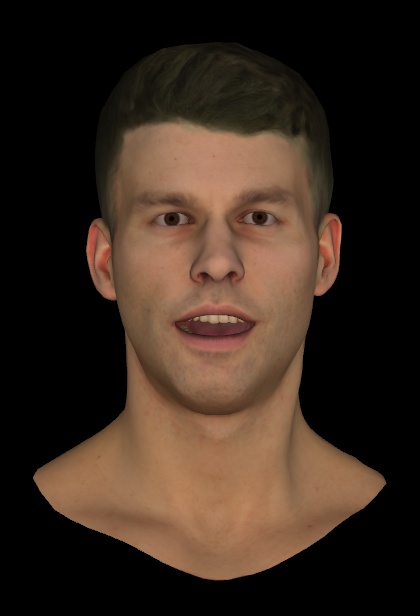}&
\includegraphics[scale=0.15]{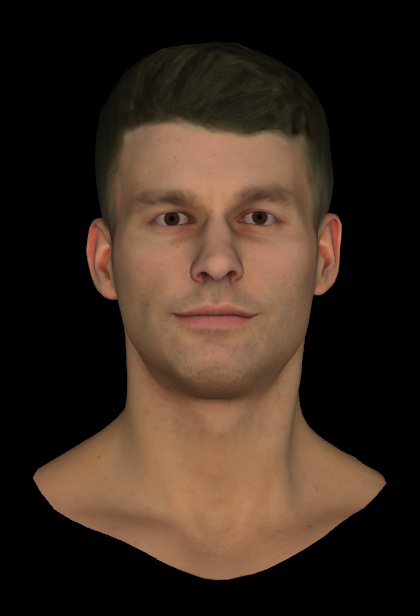}  \\
 &&& CD=0.767  & CD=0.907 & CD=0.676 & CD=0.794 &  CD= 1.004  \\
                                            &
                                            &
\rotatebox[origin=c]{90}{\parbox{0.2cm}{$std\_ED(\boldsymbol{\beta}_{1},\boldsymbol{\beta}_{2})$}} &                                            
\includegraphics[scale=0.15]{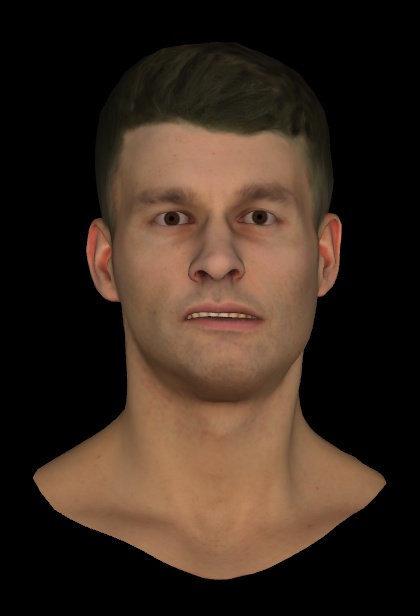}&
\includegraphics[scale=0.15]{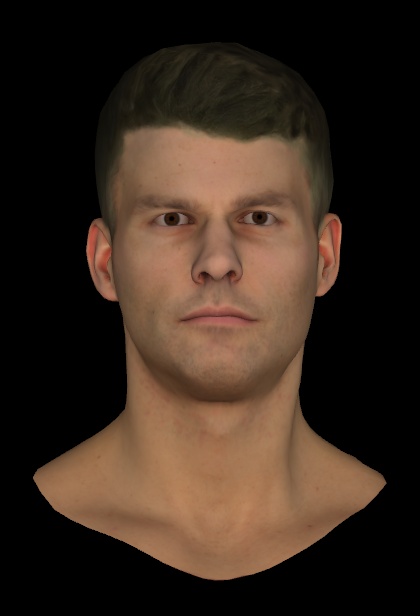}&
\includegraphics[scale=0.15]{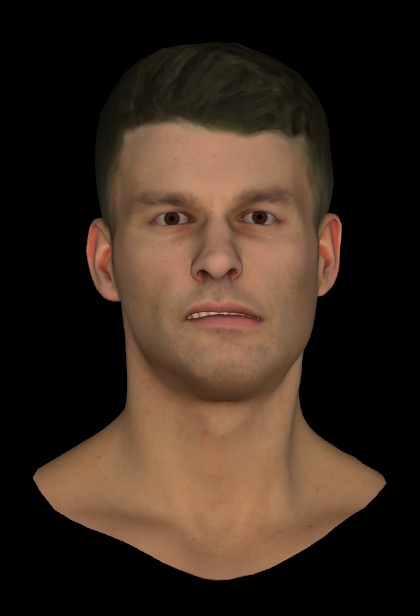}&
\includegraphics[scale=0.15]{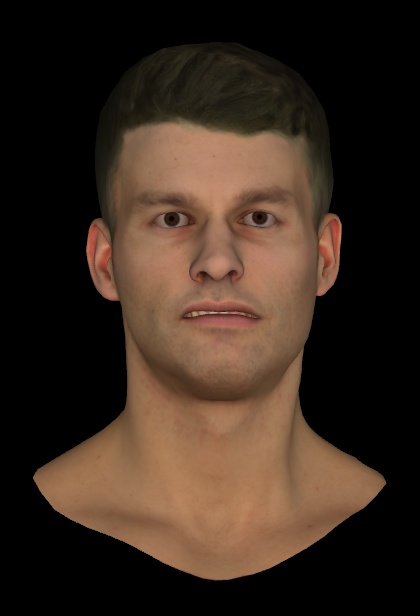}&
\includegraphics[scale=0.15]{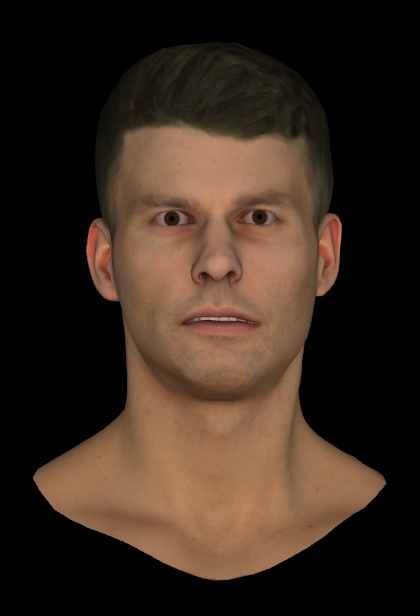}\\
\hline
\end{tabular}
\caption{Similarity assessing performance of $ED(\boldsymbol{\beta}_{1},\boldsymbol{\beta}_{2})$ and $std\_ED(\boldsymbol{\beta}_{1},\boldsymbol{\beta}_{2})$ metrics in the PCA space for different targets. For each target and metric five most similar faces in the dataset of 9592 user generated faces are sought. Corresponding \textbf{cosine distances ($CD$) to target in the blendshape space} are also indicated for each match found.}
\label{found_faces}
\end{figure*}

\begin{figure*}
\begin{tabular}{c|cc|cc}
\multirow{4}{*}[-2.0cm]{\rotatebox[origin=c]{90}{target 1}}&\multicolumn{2}{c}{Selection metric $ED(\boldsymbol{\beta}_{1},\boldsymbol{\beta}_{2})$} 
& \multicolumn{2}{c}{Selection metric $std\_ED(\boldsymbol{\beta}_{1},\boldsymbol{\beta}_{2})$} \\
\hline
&\scriptsize{average gen.0} &    &\scriptsize{average gen.0} &\\
&\includegraphics[scale=0.09]{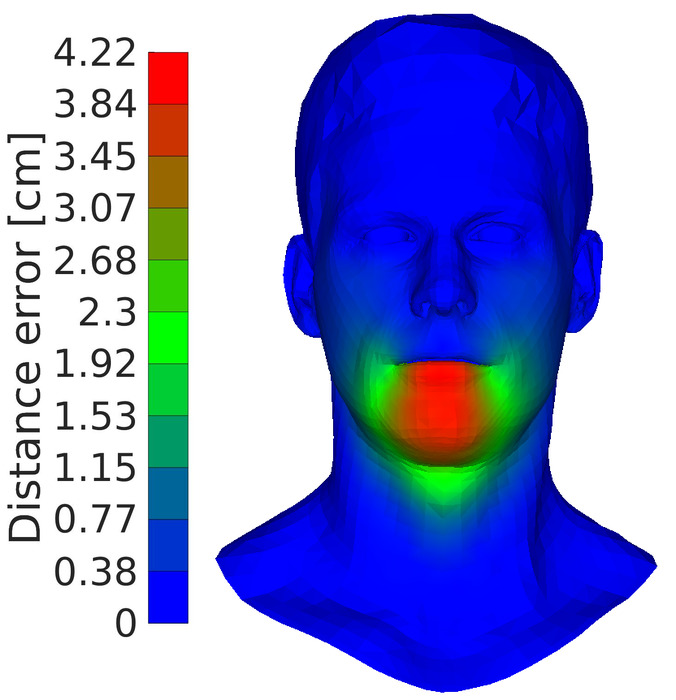} & 
\multirow{4}{*}[2.0cm]{\includegraphics[scale=0.41, trim = 5mm 0mm 8mm 0mm, clip]{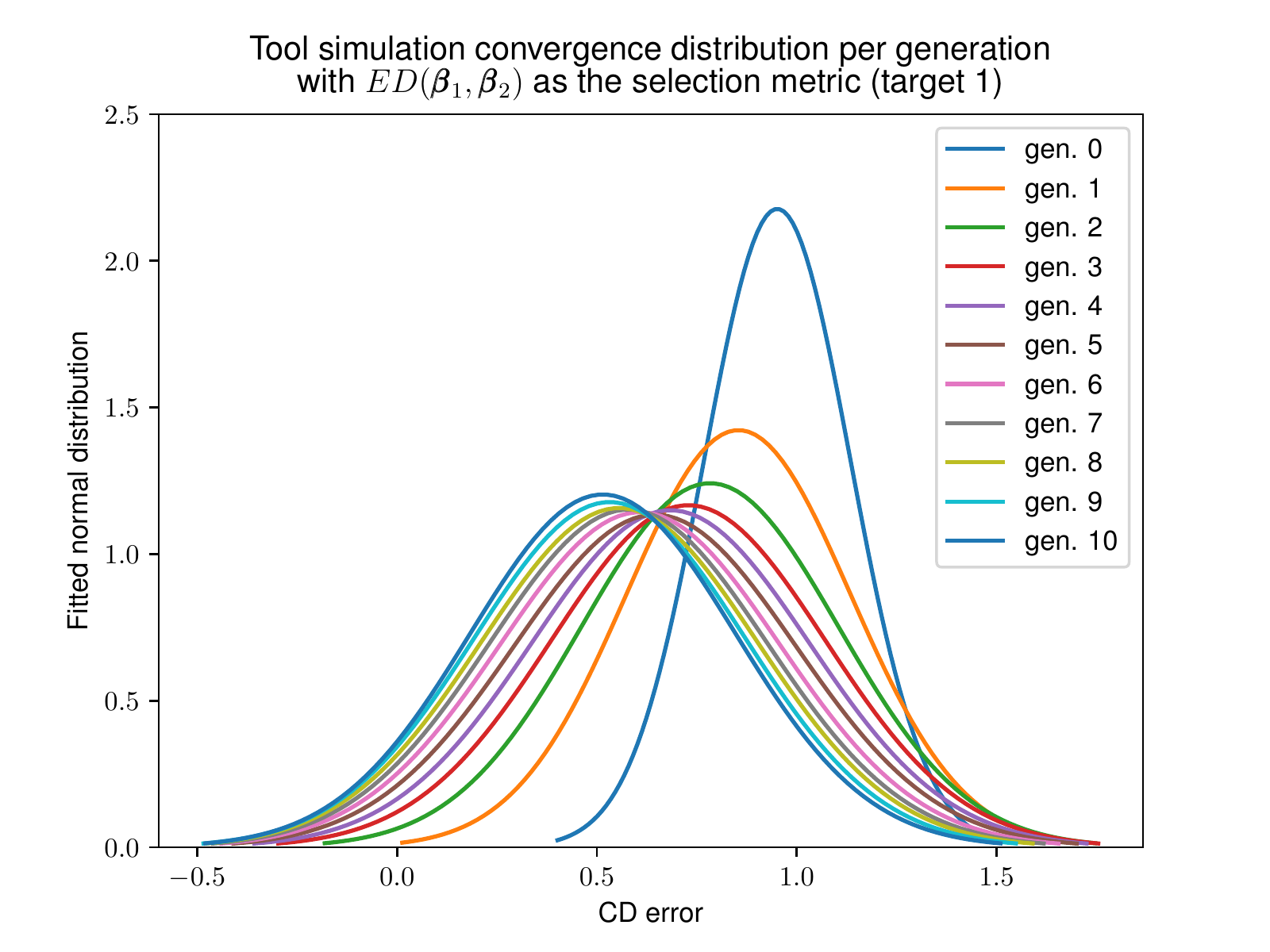}}  & \includegraphics[scale=0.09]{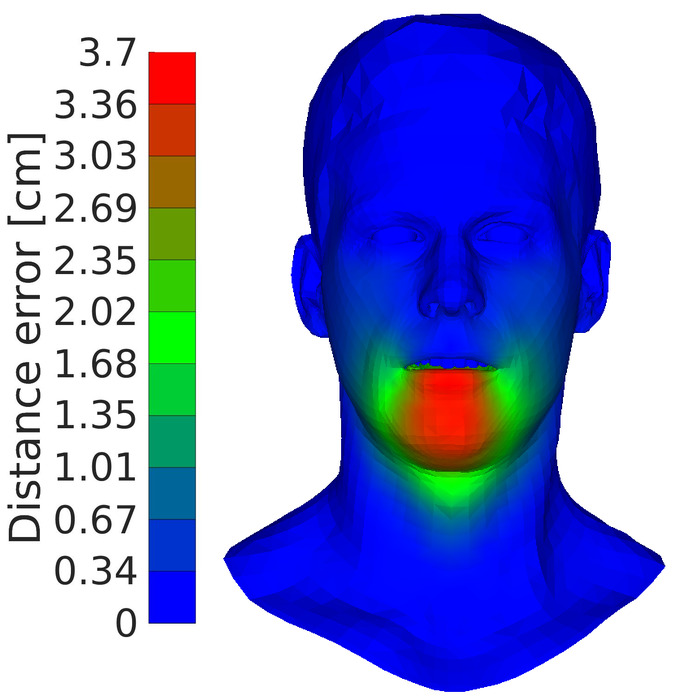} &
\multirow{4}{*}[2.0cm]{\includegraphics[scale=0.41, trim = 5mm 0mm 8mm 0mm, clip]{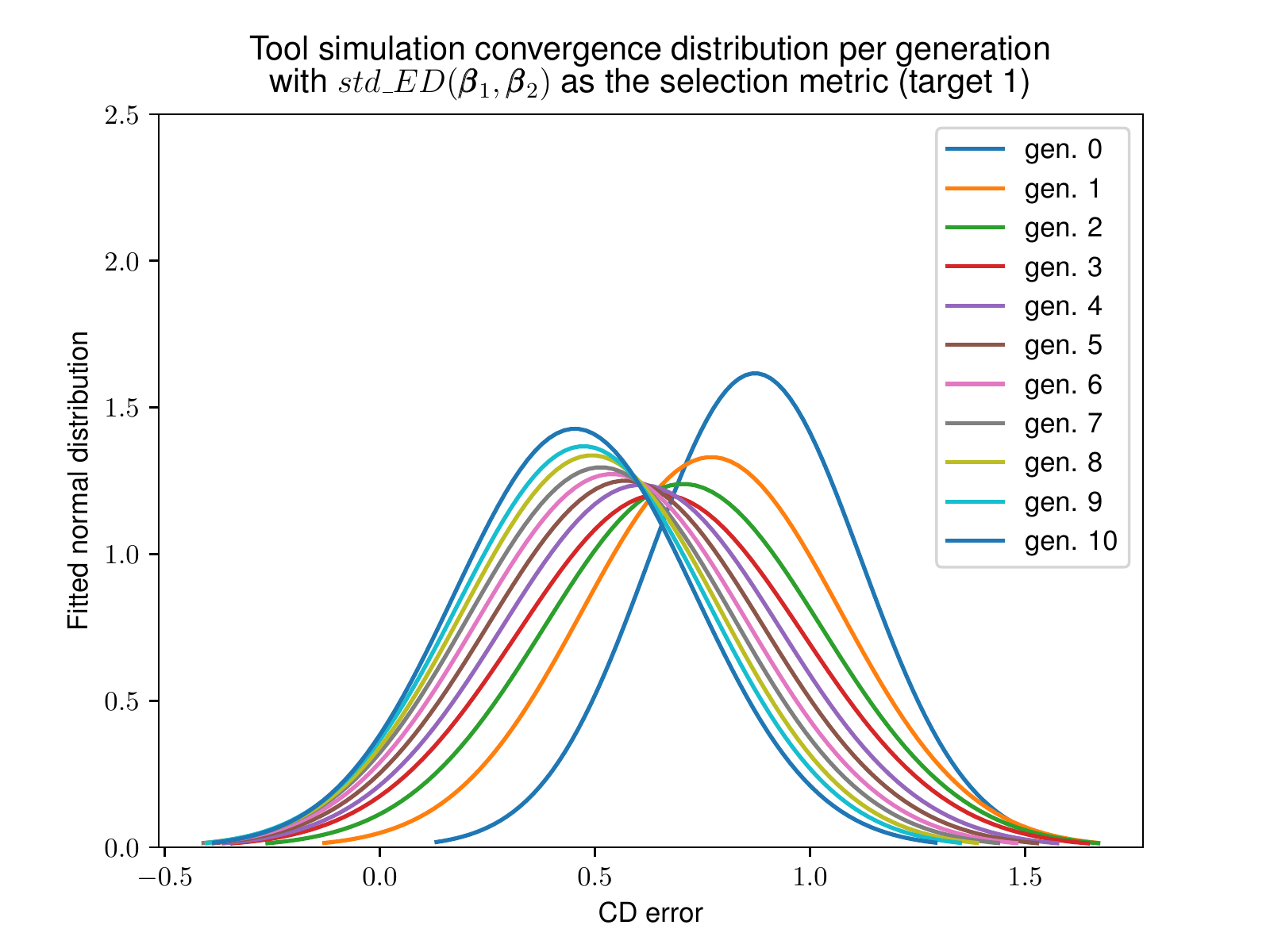}} \\
&\scriptsize{average gen.10} &  & \scriptsize{average gen.10} &\\
&\includegraphics[scale=0.09]{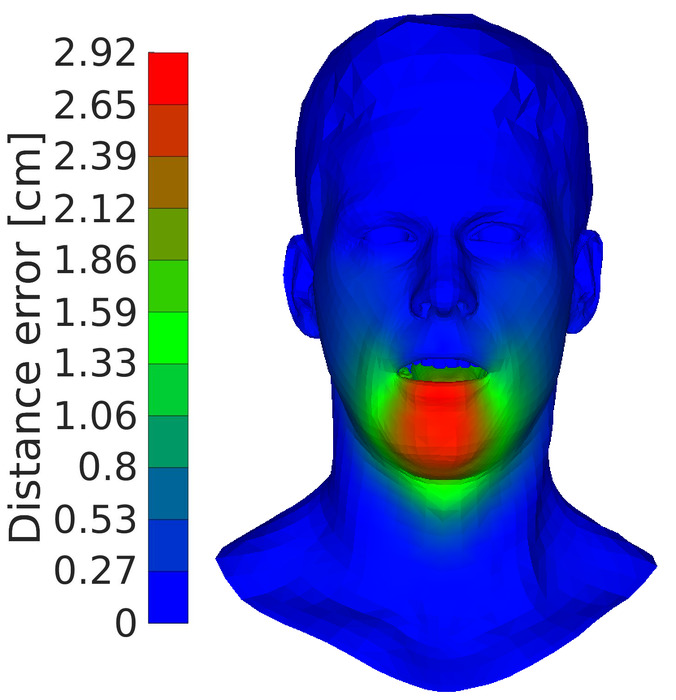}& & \includegraphics[scale=0.09]{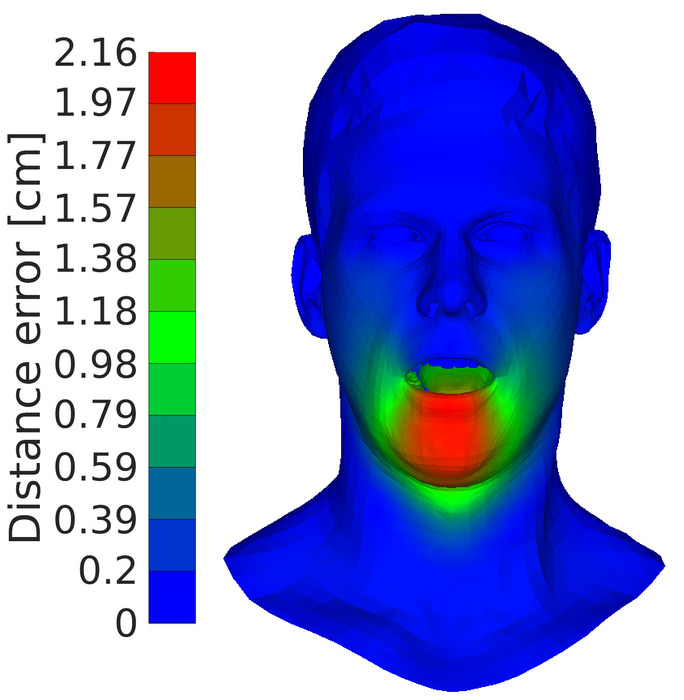} &\\
\hline \hline
\multirow{4}{*}[-2.0cm]{\rotatebox[origin=c]{90}{target 2}}& \scriptsize{average gen.0} &    &\scriptsize{average gen.0} &\\
&\includegraphics[scale=0.09]{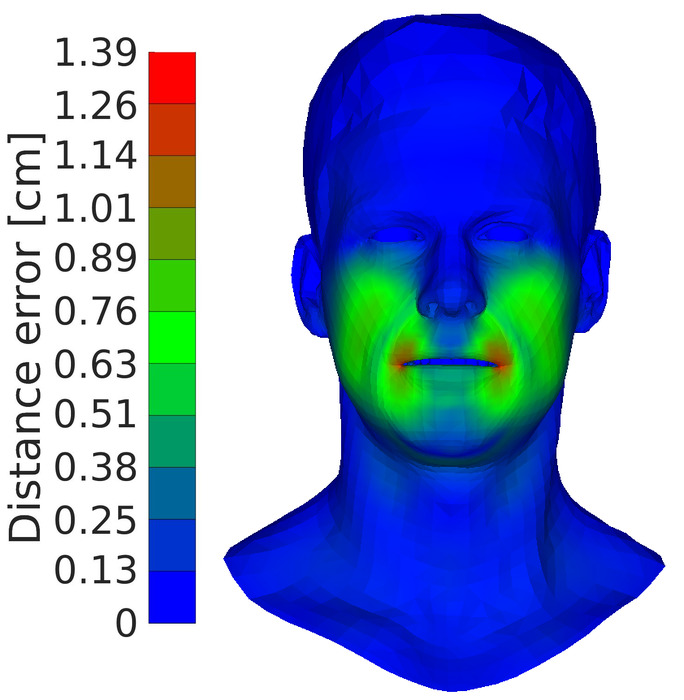} & 
\multirow{4}{*}[2.0cm]{\includegraphics[scale=0.41, trim = 5mm 0mm 8mm 0mm, clip]{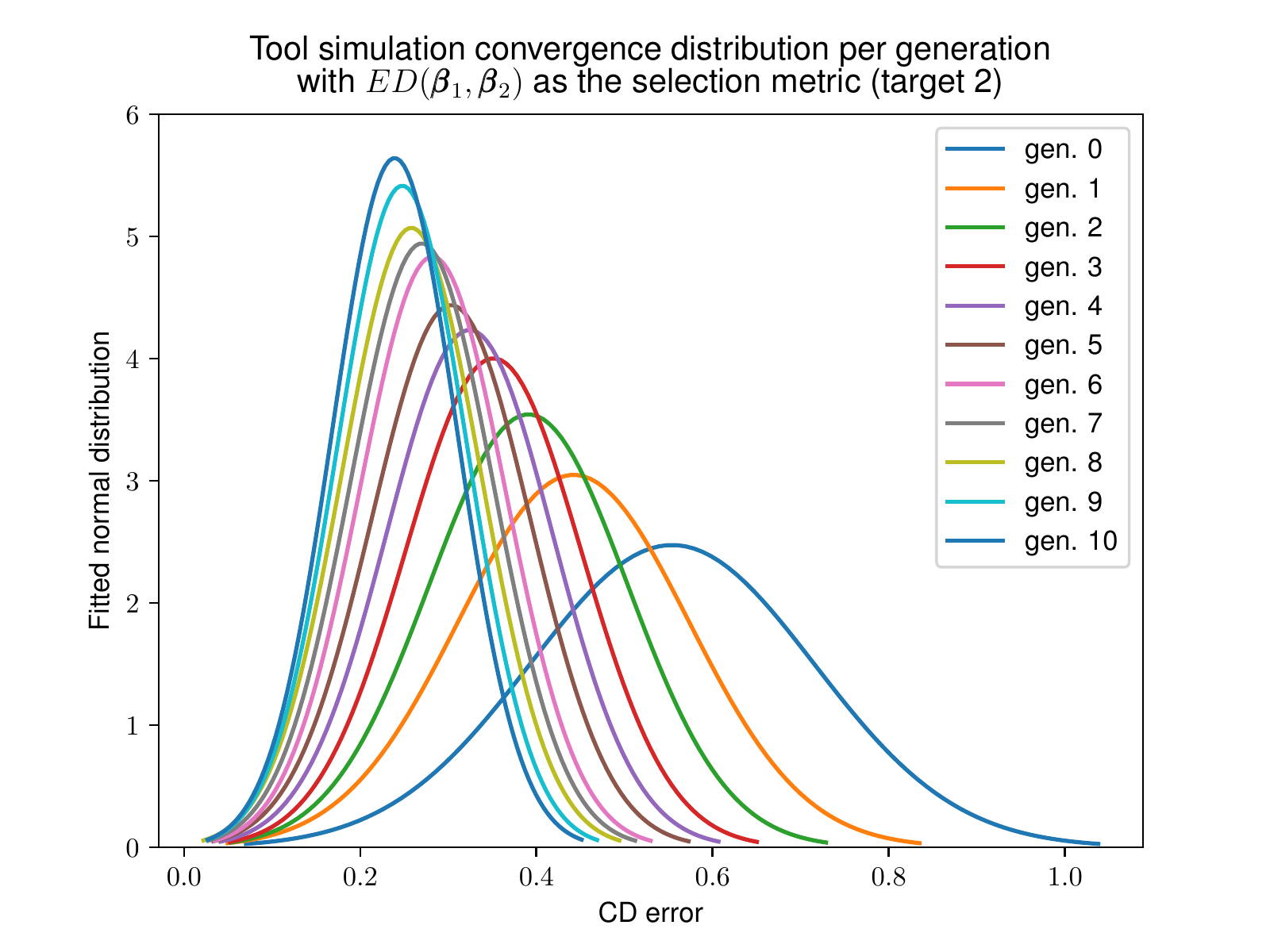}}  & \includegraphics[scale=0.09]{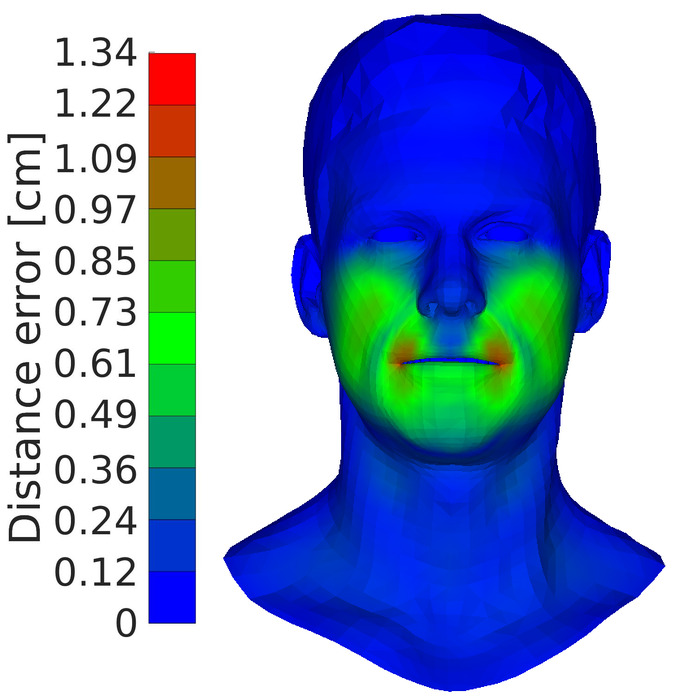} &
\multirow{4}{*}[2.0cm]{\includegraphics[scale=0.41, trim = 5mm 0mm 8mm 0mm, clip]{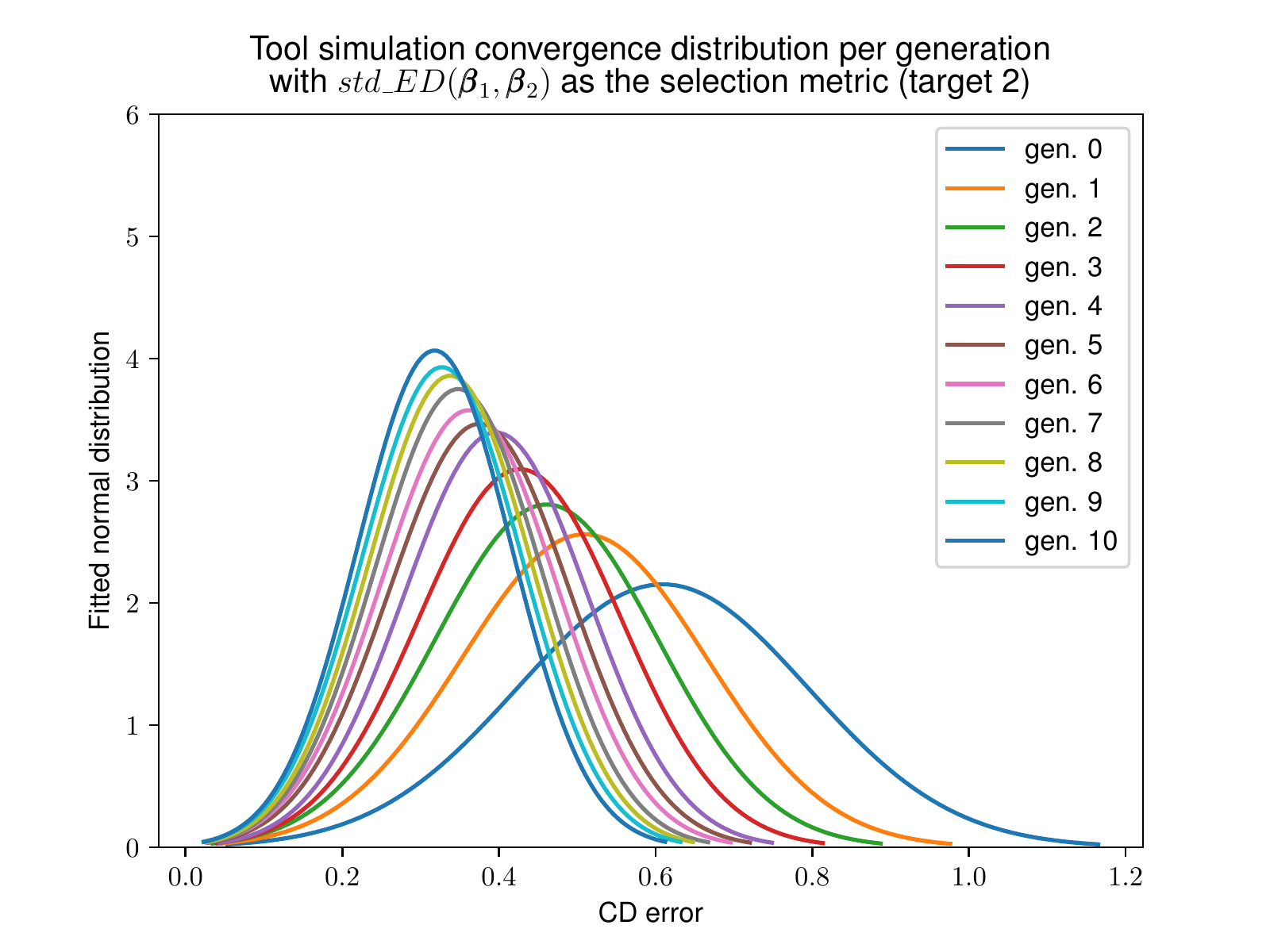}} \\
&\scriptsize{average gen.10} &  & \scriptsize{average gen.10} &\\
&\includegraphics[scale=0.09]{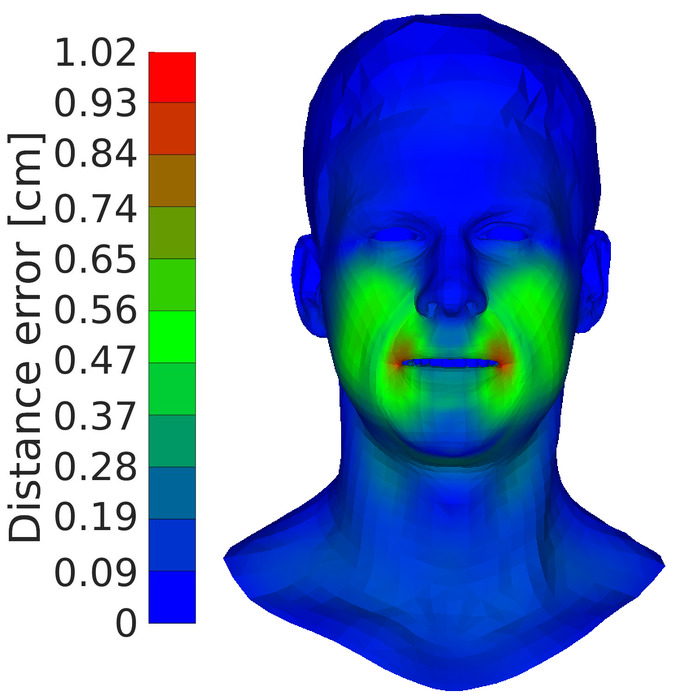}& & \includegraphics[scale=0.09]{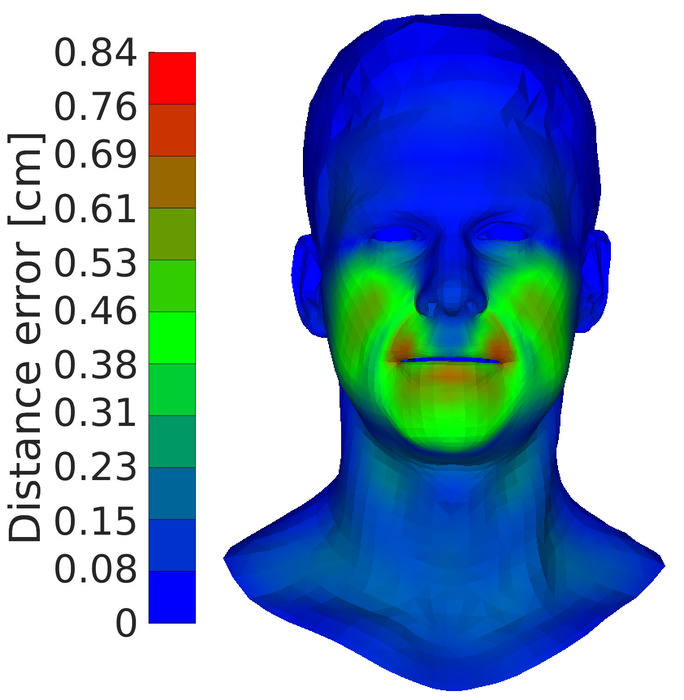} &\\
\end{tabular}
\caption{Convergence to target in EmoGen simulation using $ED(\boldsymbol{\beta}_{1},\boldsymbol{\beta}_{2})$ and $std\_ED(\boldsymbol{\beta}_{1},\boldsymbol{\beta}_{2})$ as the automatic selection metric. The targets are 1 and 2 from Figure~\ref{targets}. Shown for each selection metric are: 1. spatial distribution of distance error (``heatmap'') of the average selected elite at initialisation (gen.0) and at the end of the process (gen.10) relative to the target; 2. a set of per-generation cosine distance error distributions illustrating convergence behaviour.}
\label{pca_sim}
\end{figure*}
\begin{figure}[htb!]
\begin{tabular}{c@{\hspace{0mm}}|c@{\hspace{0mm}}c}
\multirow{4}{*}[-2.0cm]{\rotatebox[origin=t]{90}{target 1}}
&\scriptsize{average gen.0} & \multirow{4}{*}{\includegraphics[scale=0.41, trim = 5mm 0mm 8mm 0mm, clip]{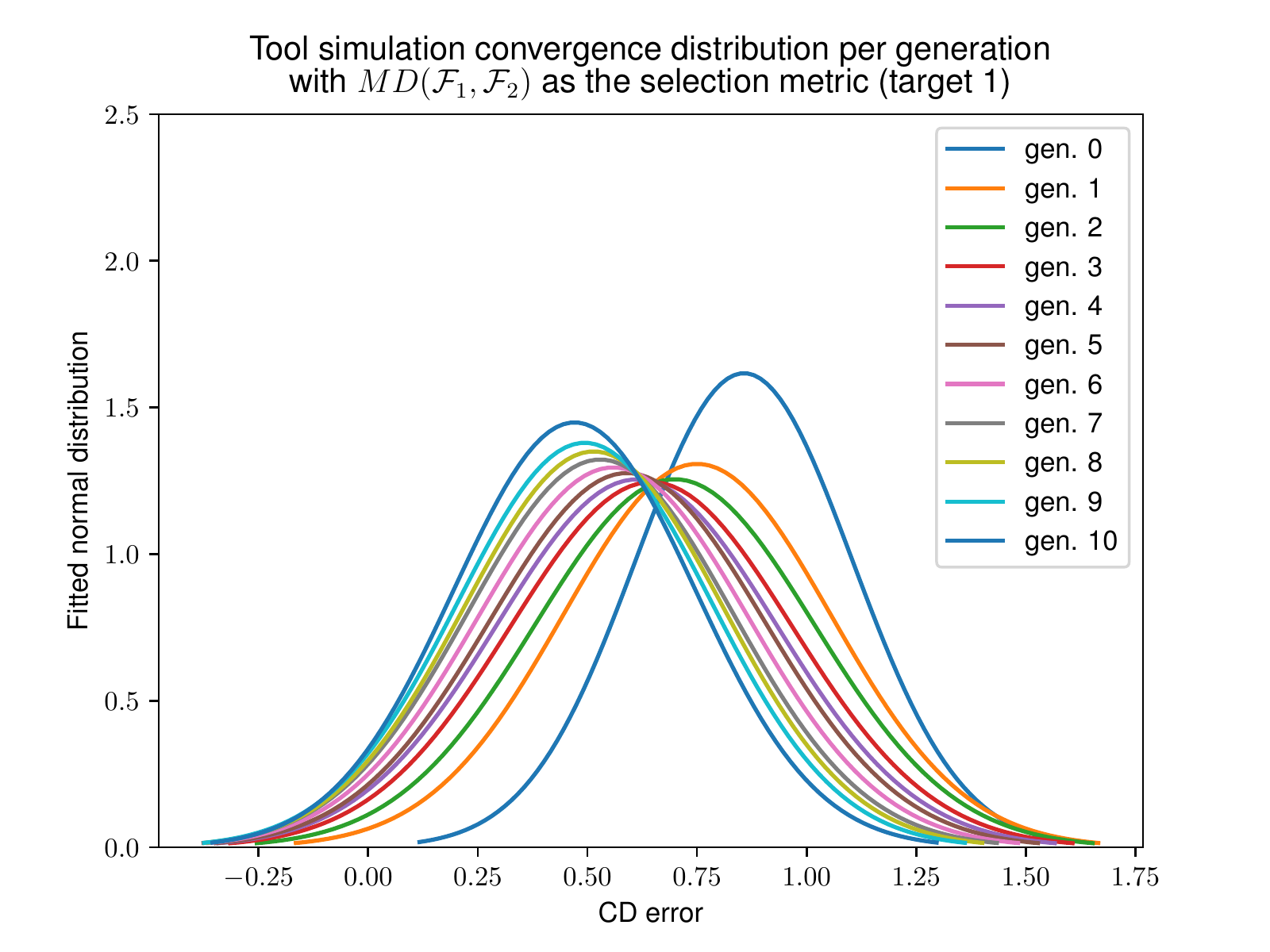}}  \\
&\includegraphics[scale=0.09]{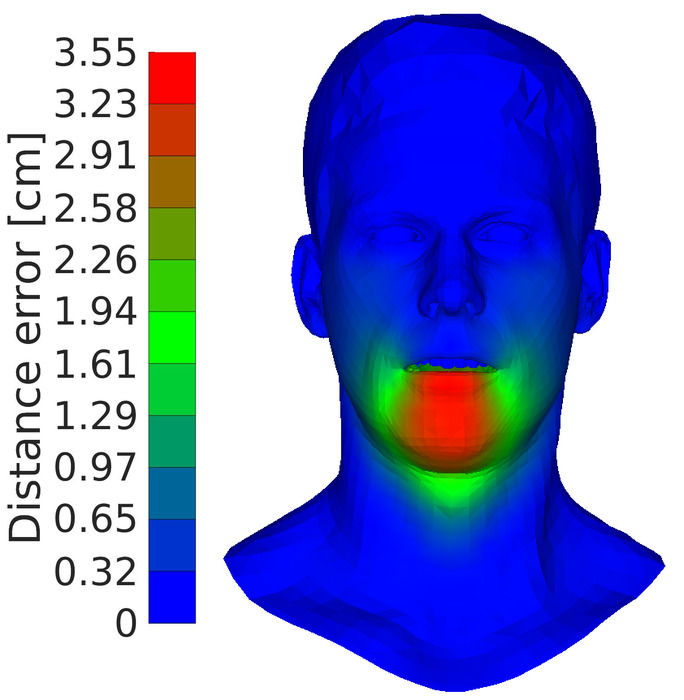} &\\
&\scriptsize{average gen.10} &\\
&\includegraphics[scale=0.09]{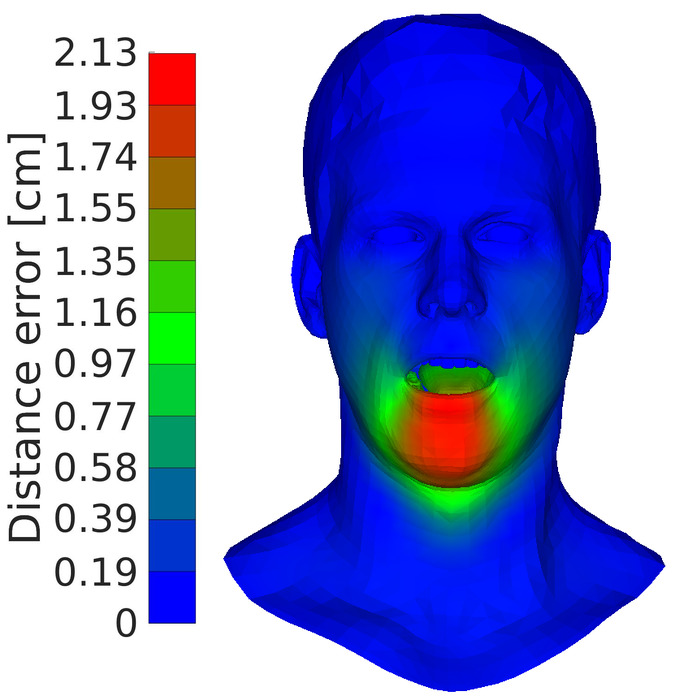} & \\ \hline \hline
\multirow{4}{*}[-2.0cm]{\rotatebox[origin=t]{90}{target 2}}
&\scriptsize{average gen.0} & \multirow{4}{*}{\includegraphics[scale=0.41, trim = 5mm 0mm 8mm 0mm, clip]{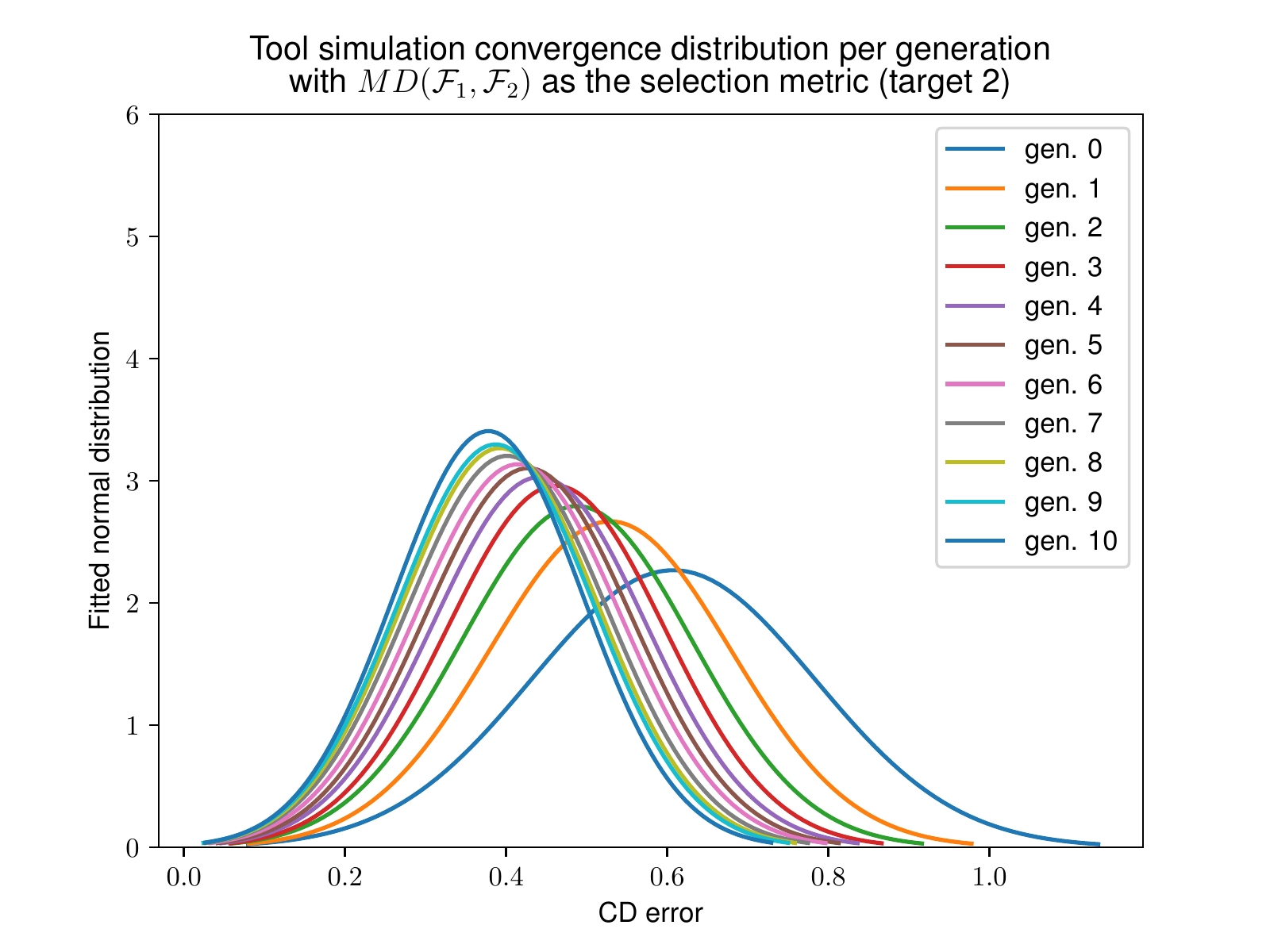}}  \\
&\includegraphics[scale=0.09]{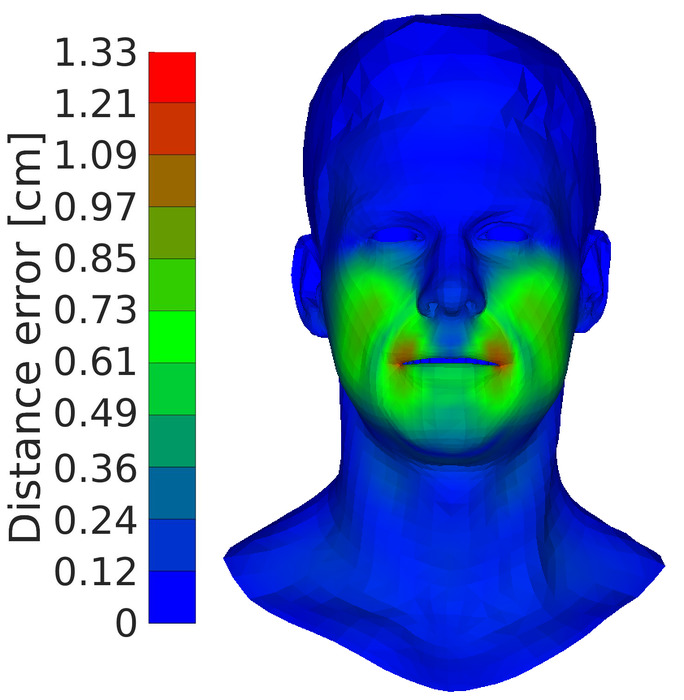} &\\
&\scriptsize{average gen.10} &\\
&\includegraphics[scale=0.09]{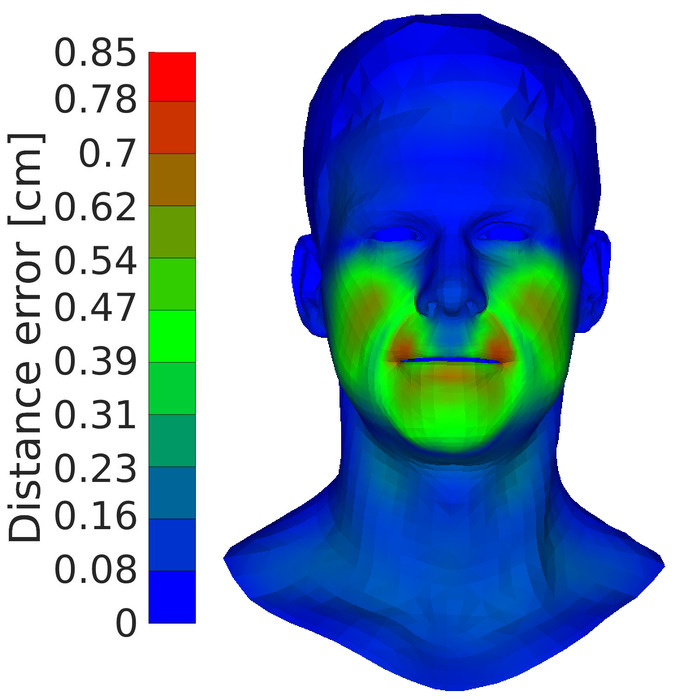} & \\
\end{tabular}
\caption{Convergence to target in EmoGen simulation using $MD(\mathcal{F}_{1},\mathcal{F}_{2})$  as the automatic selection metric. The targets are 1 and 2 from Figure~\ref{targets}. Shown for each selection metric are: 1. spatial distribution of distance error (``heatmap'') of the average facial expression at initialisation (gen.0) and at the end of the process (gen.10) relative to the target; 2. a set of per-generation cosine distance error distributions illustrating convergence behaviour. Compare to the performance of the empirical blendshape-domain selection metrics in Figure~\ref{pca_sim}.}
\label{MD_vertex_sim}
\end{figure}
\begin{figure*}[ht!]
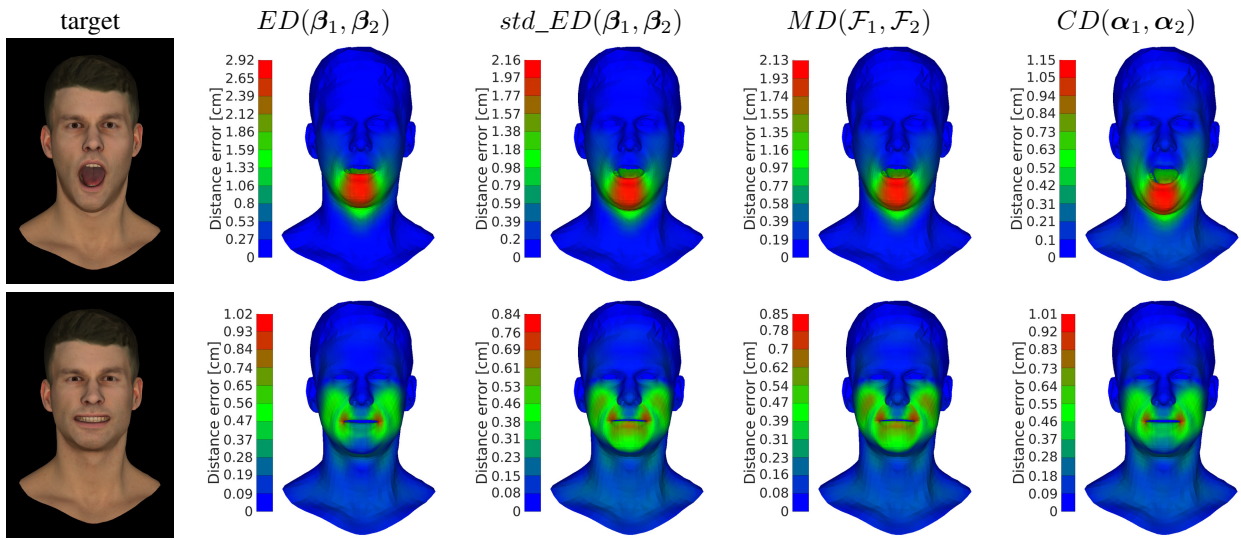

\begin{tabular}{ccccc}
 target & $ED(\boldsymbol{\beta}_{1},\boldsymbol{\beta}_{2})$ & $std\_ED(\boldsymbol{\beta}_{1},\boldsymbol{\beta}_{2})$ & $MD(\mathcal{F}_{1}, \mathcal{F}_{2})$ & $CD(\boldsymbol{\alpha}_{1}, \boldsymbol{\alpha}_{2})$ \\
\includegraphics[scale=0.15]{target1.jpg}&
\includegraphics[scale=0.13]{sim18_10_heatmap.jpg} &
\includegraphics[scale=0.13]{sim19_10_heatmap.jpg} &
\includegraphics[scale=0.13]{sim22_gen10_heatmap.jpg} &
\includegraphics[scale=0.13]{sim2_heatmap.jpg} \\
\includegraphics[scale=0.15]{target2.jpg}&
\includegraphics[scale=0.13]{sim20_10_heatmap.jpg} &
\includegraphics[scale=0.13]{sim21_10_heatmap.jpg} &
\includegraphics[scale=0.13]{sim23_gen10_heatmap.jpg} &
\includegraphics[scale=0.13]{sim10b_heatmap.jpg} \\
\end{tabular}
\caption{Spatial error distributions of the average face converged to using the empirical similarity metrics
$ED(\boldsymbol{\beta}_{1},\boldsymbol{\beta}_{2})$, $std\_ED(\boldsymbol{\beta}_{1},\boldsymbol{\beta}_{2})$ and $MD(\mathcal{F}_{1}, \mathcal{F}_{2})$ compared to the best performing theoretical metric $CD(\boldsymbol{\alpha}_{1}, \boldsymbol{\alpha}_{2})$.}
\label{theoretical_emperical}
\end{figure*}

\par \textbf{Principle component analysis (PCA) decomposition in the blendshape space} reduces dimensionality of expression descriptors mapping onto orthogonal axes. Apart from removing correlations between the core shapes and correctives, as well as symmetrical shape redundancy, the transformation will also find any correlations between blendshapes introduced by the users tasked with generating expressions of emotion classes. Let us define PCA component vector $\boldsymbol{\beta} = PCA(\boldsymbol{\alpha})$, which is blendshape vector $\boldsymbol{\alpha}$ mapped onto the PCA space.
\par Two vectors in the PCA space can be compared using \textit{simple Euclidean distance} $ED(\boldsymbol{\beta}_{1},\boldsymbol{\beta}_{2})$ defined analogously to Equation~\ref{ed_blnd} or \textit{Mahalanobis distance} 
($MD$) that takes data variances into account. We define Mahalanobis distance as:
\begin{equation}
MD(\boldsymbol{\beta}_{1},\boldsymbol{\beta}_{2}) = \sqrt{(\boldsymbol{\beta}_{1} - \boldsymbol{\beta}_{2})^{T}\mathbf{S}^{-1}(\boldsymbol{\beta}_{1} - \boldsymbol{\beta}_{2})},
\end{equation}
where $\mathbf{S}$ is the covariance matrix derived from the training distribution. Since PCA by definition diagonalises $\mathbf{S}$, $MD$ reduces to \textit{standardised Euclidean distance}:
\begin{equation}
std\_ED(\boldsymbol{\beta}_{1},\boldsymbol{\beta}_{2}) = \sqrt{\sum_{i=1}^K{ 
\frac{ ( \beta_{1,i} - \beta_{2,i} )^2 } { \boldsymbol{\sigma}^2_{i} } } },
\end{equation}
where $\beta_{i}$ is the $i^{th}$ PCA component in $\boldsymbol{\beta}$ of cardinality $K$ and $\boldsymbol{\sigma}^2_{i}$ is the component's variance observed in the training distribution.
\par To assess the defined similarity metrics, we build a model of the $\boldsymbol{\beta}$ facial representation space by performing PCA decomposition on a training set of 9592 blendshape vectors. These vectors correspond to faces selected by 240 human participants while evolving their representations of happy, sad, angry or fearful emotions through EmoGen's GA. Not all samples are independent as elites from all generations of each evolution process are included in the data. The training set approximates the span of the blendshape model space. The resultant PCA model, explaining $99\%$ of variance, consists of 46 components. We compare 
$ED(\boldsymbol{\beta}_{1},\boldsymbol{\beta}_{2})$ and $std\_ED(\boldsymbol{\beta}_{1},\boldsymbol{\beta}_{2})$ as competing similarity metrics in the PCA space.
\par The relative performance of $ED(\boldsymbol{\beta}_{1},\boldsymbol{\beta}_{2})$ and $std\_ED(\boldsymbol{\beta}_{1},\boldsymbol{\beta}_{2})$ for tasks related to similarity assessment has been found to depend on the mode of the PCA model where the difference resides. Specifically, $std\_ED$ amplifies the importance of the PCA components characterised by lower variance in the training data. It will be more accurate in finding similarity to targets mostly defined in lower variance PCA modes. $ED$ on the other hand will be comparatively more accurate for targets defined by higher variance PCA modes because, unlike $std\_ED$, the metric is unbiased and does not overfit lower variance components. For targets with both low and high variance components, the relative metric performance is hard to predict exactly but can be said to loosely correlate to the dominance of each type of component. The selective acute sensitivity to errors in some components makes $std\_ED$ unsuitable for efficient data clustering within a distribution. 
\par We can validate the observations in the following control experiment. We define three target faces relative to the neutral expression in Table~\ref{defs}. The targets are visualised in Figure~\ref{found_faces}. For clarity of illustration, the targets represent specific modes in the PCA model i.e. activation in the low and high variance components only and activation in both equally. Next we use $ED(\boldsymbol{\beta}_{1},\boldsymbol{\beta}_{2})$ and $std\_ED(\boldsymbol{\beta}_{1},\boldsymbol{\beta}_{2})$ to find five most similar faces to each target in the entire training dataset. The accuracy in each case is compared quantitatively using cosine distance to targets in the blendshape space and qualitatively by visual inspection. 
\par Figure~\ref{found_faces} confirms the previously stated expectations on the relative performance of $ED(\boldsymbol{\beta}_{1},\boldsymbol{\beta}_{2})$ and $std\_ED(\boldsymbol{\beta}_{1},\boldsymbol{\beta}_{2})$. $ED(\boldsymbol{\beta}_{1},\boldsymbol{\beta}_{2})$ finds just one plausible match for $\boldsymbol{\beta}_{l}$, characterised by a lower jaw shift, compared to five excellent ones  found with $std\_ED(\boldsymbol{\beta}_{1},\boldsymbol{\beta}_{2})$. Equally true is the inability of $std\_ED(\boldsymbol{\beta}_{1},\boldsymbol{\beta}_{2})$ to find matches for the high variance PCA component target $\boldsymbol{\beta}_{h}$ because of the biasing towards low variance components. Especially interesting is the observation that with a mixed target $\boldsymbol{\beta}_{m_{1}}$,  defined by an equal activation of the high and low component, each metric optimises one axis of the PCA space. While the preference of $std\_ED(\boldsymbol{\beta}_{1},\boldsymbol{\beta}_{2})$ towards the low variance components is predictable, it is  important to note that $ED(\boldsymbol{\beta}_{1},\boldsymbol{\beta}_{2})$ does not have an inherent bias towards higher variance component faces. There are just likely more options of that category to choose from in the pool. With a mixed target, cosine distance indicates the superiority of  
$ED(\boldsymbol{\beta}_{1},\boldsymbol{\beta}_{2})$. Visually, this superiority is not apparent and is presumably due to the high variance PCA component representing more blendshapes. 
\par Also in EmoGen simulation, when aiming for targets representing variation in the low variance PCA modes, better convergence is achieved with $std\_ED(\boldsymbol{\beta}_{1},\boldsymbol{\beta}_{2})$ compared to $ED(\boldsymbol{\beta}_{1},\boldsymbol{\beta}_{2})$. Consider using the similarity metrics to evolve target 1 in Figure~\ref{targets} whose key PCA component is 31 (out of 46) is one of the lower variance ones. Quantitatively, with $std\_ED(\boldsymbol{\beta}_{1},\boldsymbol{\beta}_{2})$ the final generation (gen.10) distribution (Figure~\ref{pca_sim}, row 1, right) shows a lower mean and standard deviation ($0.453\pm0.279$) of cosine distance error than the corresponding distribution with $ED(\boldsymbol{\beta}_{1},\boldsymbol{\beta}_{2})$ ($0.515\pm0.334$) in Figure~\ref{pca_sim}, row 1, left. The heatmap error visualisations of the average converged to face at generation 10 in  Figure~\ref{pca_sim} confirm the significance of the difference in the CD scores: while the spatial error distribution pattern is similar in both case (i.e. around the jaw), its magnitude is lower with $std\_ED(\boldsymbol{\beta}_{1},\boldsymbol{\beta}_{2})$. Also revealing is the much lower variance with $ED(\boldsymbol{\beta}_{1},\boldsymbol{\beta}_{2})$ at the initial generation: the metric, due to its averaging tendency, is more prone than $std\_ED(\boldsymbol{\beta}_{1},\boldsymbol{\beta}_{2})$ to choosing the neutral expression amongst the initialisation options resulting in an average with a still closed jaw at generation 0. 
\par In contrast, for target 2 (Figure~\ref{pca_sim}, row~2), which is  mainly described by the higher variance PCA components 4 and 11, $ED(\boldsymbol{\beta}_{1},\boldsymbol{\beta}_{2})$ significantly outperforms $std\_ED(\boldsymbol{\beta}_{1},\boldsymbol{\beta}_{2})$ in the mean and standard deviation of the final convergence error distribution. As a visual confirmation, the mid-range distance error (green) is more widespread in the heatmaps of $std\_ED(\boldsymbol{\beta}_{1},\boldsymbol{\beta}_{2})$ for this target.
\par \textbf{Mahalanobis distance in the vertex space} empirically models spatial correlations between vertices. Given a fixed face model topology, mesh $\mathcal{F}_{i}$ is fully defined by the set $\mathcal{V}$ of N 3D vertices $\boldsymbol{v}_{i} =[x, y,z]^\top$. For independence from the absolute position, we  model each vertex $n$ of mesh $\mathcal{F}_{i}$ as a 3D vector offset from the corresponding vertex in the neutral expression $\mathcal{F}_{0}$: $\delta \boldsymbol{v}_{n,i} = \boldsymbol{v}_{n,i} - \boldsymbol{v}_{n,0}$.  The data covariance matrix $\mathbf{S}_{3N\times3N}$ in the vertex representation is $3N\times3N$ i.e. N vertex offsets with 3 dimensions each. 
\par Next, comparing two face meshes $\mathcal{F}_{1}$ and $\mathcal{F}_{2}$, we represent each mesh as a column vector of offsets
$ \delta \boldsymbol{V}_{1} = [\delta \boldsymbol{v}_{1,1}^\top, \delta \boldsymbol{v}_{2,1}^\top .. \delta \boldsymbol{v}_{N,1}^\top]^\top$ and 
$ \delta \boldsymbol{V}_{2} = [\delta \boldsymbol{v}_{1,2}^\top, \delta \boldsymbol{v}_{2,2}^\top .. \delta \boldsymbol{v}_{N,2}^\top]^\top$ respectively. The vertex-domain Mahalanobis distance between $\mathcal{F}_{1}$ and $\mathcal{F}_{2}$  is defined as:
\begin{equation}
MD(\mathcal{F}_{1}, \mathcal{F}_{2}) = \sqrt{ (\delta \boldsymbol{V}_{1} -\delta \boldsymbol{V}_{2})^\top \mathbf{S}^{-1}_{3N\times3N}(\delta \boldsymbol{V}_{1} -\delta \boldsymbol{V}_{2})}
\end{equation}
\par Figure~\ref{MD_vertex_sim} presents convergence to target with $MD(\mathcal{F}_{1}, \mathcal{F}_{2})$ as the selection metric in simulation. Interestingly, the presented convergence behaviour with $MD(\mathcal{F}_{1}, \mathcal{F}_{2})$, both visually in the spatial error heatmaps and in terms of the cosine distance error distributions, is similar to that of $std\_ED(\boldsymbol{\beta}_{1},\boldsymbol{\beta}_{2})$ in Figure~\ref{pca_sim}. This similarity is particularly remarkable for target 1 with the respective final error means and standard deviations of ($0.471\pm0.275$) and ($0.453\pm0.279$).  Both empirical metrics are trained on the same user-generated data but in different representations, specifically the PCA-reduced blendshape-space representation of $std\_ED(\boldsymbol{\beta}_{1},\boldsymbol{\beta}_{2})$ and the 3D vertex-space one of $MD(\mathcal{F}_{1}, \mathcal{F}_{2})$. This observation of behavioural similarity indicates that the blendshape representation accurately reflects the key modes of geometric facial deformation. The question whether the use of blendshape representation might be misleading in assessing facial similarity arises because all blendshape weights are in the $[0,1]$ range regardless of the vertex offsets they drive. Hence geometric differences of substantially different amplitudes may be treated as equally significant. However, comparing the equivalent metrics $std\_ED(\boldsymbol{\beta}_{1},\boldsymbol{\beta}_{2})$ and $MD(\mathcal{F}_{1}, \mathcal{F}_{2})$ in the blendshape and vertex domains respectively, no strong evidence in favour of either representation can be observed based on convergence patterns. Hence the effect of unequal offsets from blendshape modelling is not prohibitive.
\par \textbf{Overall assessment of empirical metrics.} In summary, the empirical metrics in the blendshape or vertex domains do not outperform theoretical $CD(\boldsymbol{\alpha}_{1}, \boldsymbol{\alpha}_{2})$ presented in Section~\ref{def_theoretical} in expression similarity assessment. This is clear comparing the spatial error distributions visualised as heatmaps for target 1 and 2 in Figures~\ref{distribution_means},~\ref{pca_sim} and \ref{MD_vertex_sim}. For ease of comparison, the heatmaps are presented again next to each other in Figure~\ref{theoretical_emperical}. Note that the conclusion holds despite the presented empirical metrics having the advantage of not being agnostic to expression intensity unlike $CD(\boldsymbol{\alpha}_{1}, \boldsymbol{\alpha}_{2})$. 
\par The empirical metrics have their biases from the training data: e.g. a particular PCA decomposition or covariance matrix. Such biases are best exemplified by the previously presented comparison of $ED(\boldsymbol{\beta}_{1},\boldsymbol{\beta}_{2})$ and $std\_ED(\boldsymbol{\beta}_{1},\boldsymbol{\beta}_{2})$. Straightforward Euclidean distance $ED(\boldsymbol{\beta}_{1},\boldsymbol{\beta}_{2})$ does not always isolate the defining facial characteristic ($\boldsymbol{\beta}_{l}$ in Figure~\ref{found_faces}, target 1 in Figure~\ref{pca_sim}) given the full principle component vector representation, essentially over-averaging in the distance computation. Standardised Euclidean distance $std\_ED(\boldsymbol{\beta}_{1},\boldsymbol{\beta}_{2})$ acutely depends on the computed relative mode variances being truly representative of the space to justify biased weighting of the low variance components. Generation of a sufficiently expressive training set for this purpose is not trivial.
Due to these limitations, the empirical metrics will not be used to drive simulation for assessment of EmoGen's GA. They are also not recommended in data analysis of psychology experiments.
\begin{figure*}[h!]
\centering
\begin{tabular}{c@{\hspace{0mm}}c|c@{\hspace{0mm}}c}
\multicolumn{2}{c}{fixed} & \multicolumn{2}{c}{protocol-generated} \\ \hline
 \multirow{2}{*}[3.5cm]{\includegraphics[scale=0.12]{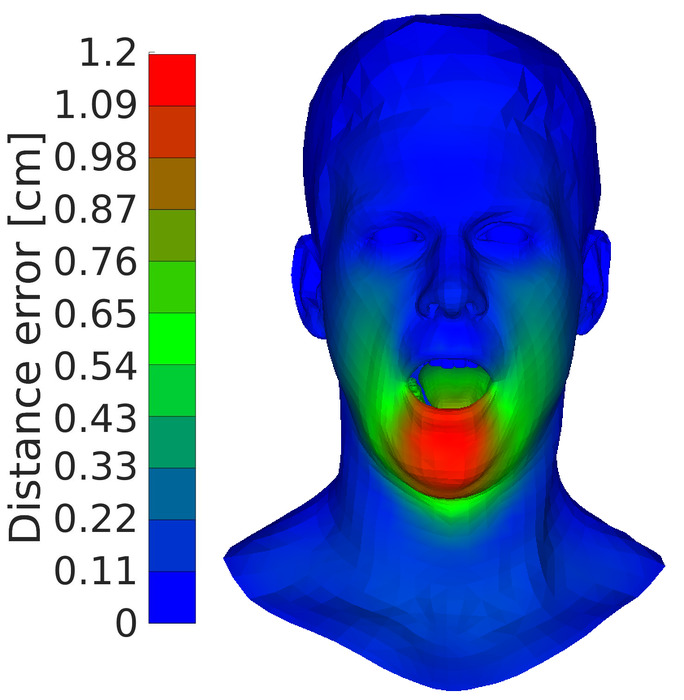} }& 
\includegraphics[scale=0.39, trim = 5mm 0mm 10mm 0mm, clip]{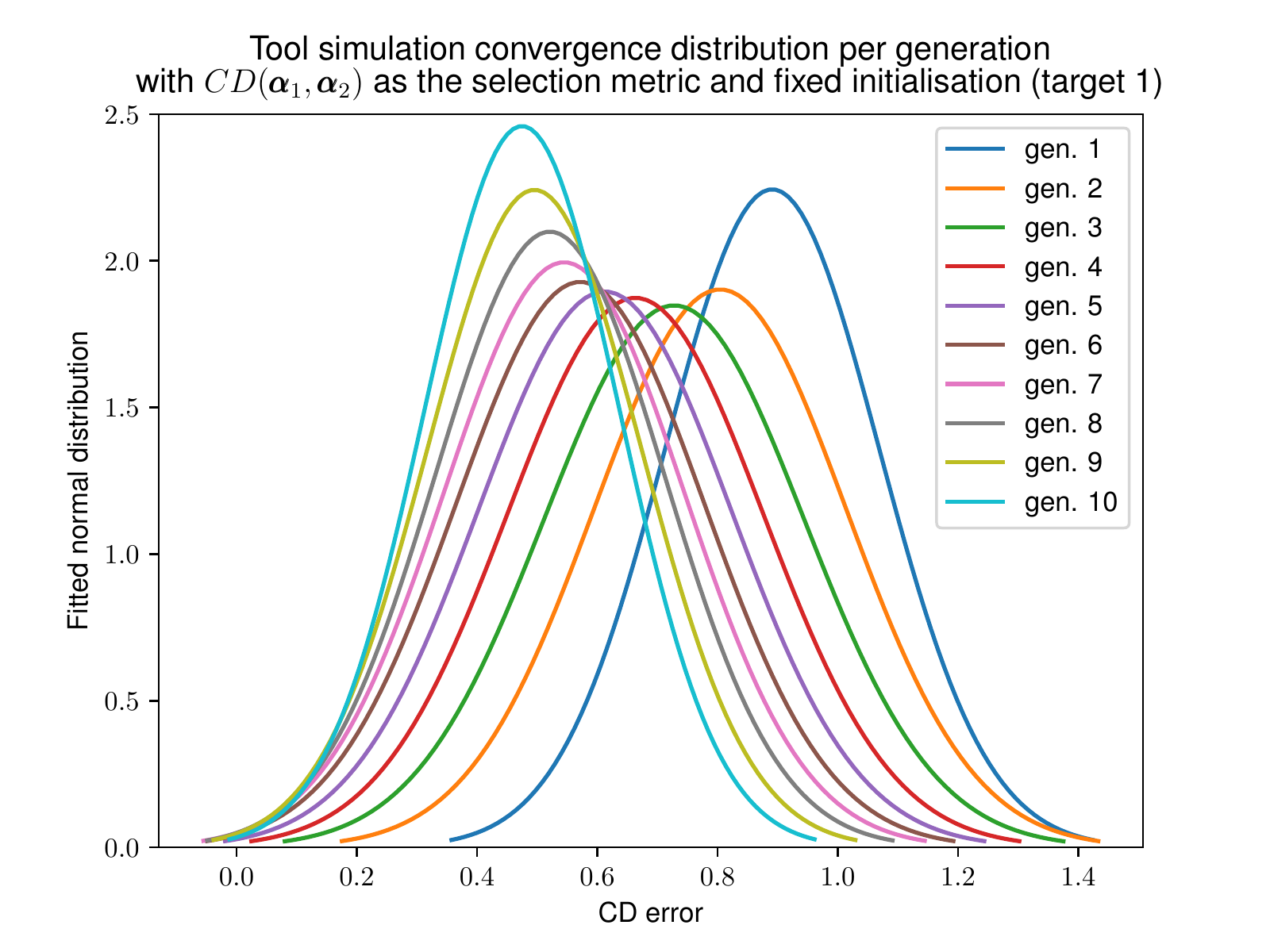} &
\includegraphics[scale=0.39, trim = 2mm 0mm 0mm 0mm, clip]{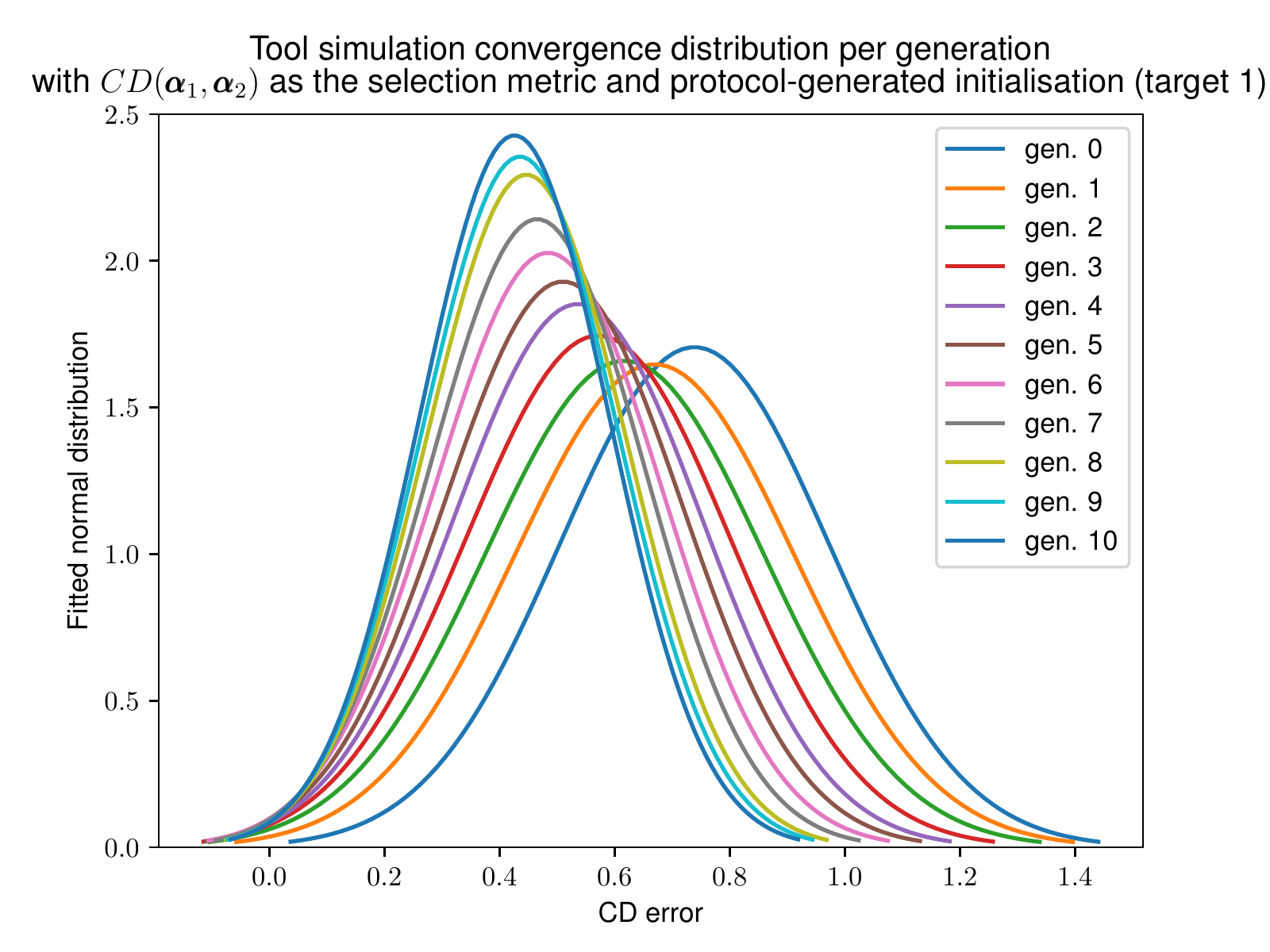} &
 \multirow{2}{*}[3.5cm]{\includegraphics[scale=0.12]{sim2_heatmap.jpg}}
\end{tabular}
\caption{Convergence to target 1 from fixed and protocol-generated initialisation using $CD(\boldsymbol{\alpha}_{1}, \boldsymbol{\alpha}_{2})$ as the selection metric in EmoGen simulation. Convergence behaviour is illustrated by the set of per-generation cosine distance error distributions. The heatmaps show spatial distance error distribution of the average face in the final generation relative to the target.}
\label{fixed_protocol}
\end{figure*}
\begin{figure}[h!]
\includegraphics[scale=0.5, trim = 0mm 0mm 0mm 0mm, clip]{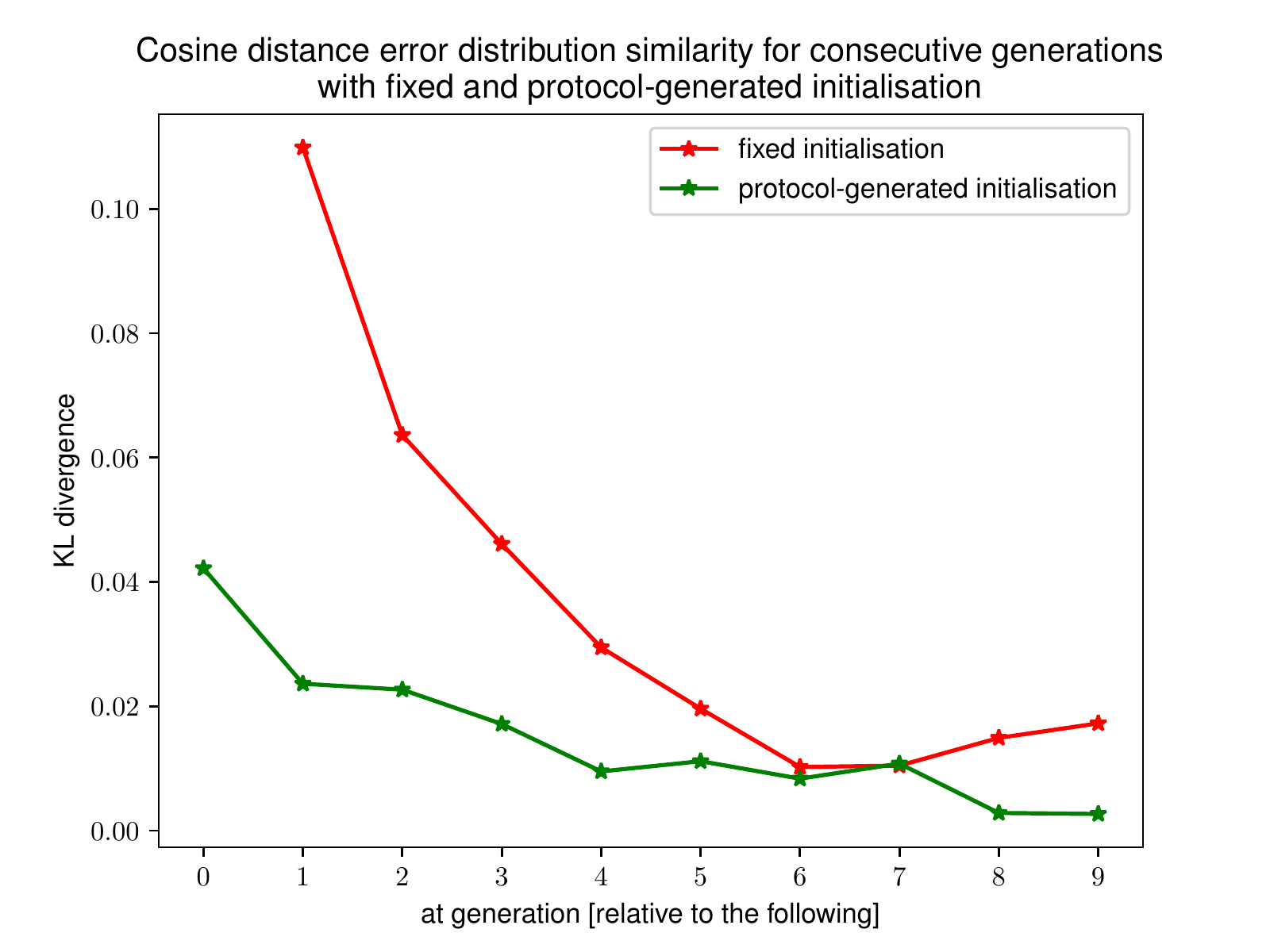}
\caption{Temporal convergence quality assessment: cosine distance error distribution similarity for consecutive generations measured as KL divergence for simulation data generated with fixed and protocol-generated genetic algorithm initialisation.}
\label{distribution_similarity}
\end{figure}
\begin{figure*}[h!]
\centering
\begin{tabular}{c@{\hspace{0mm}}c|c@{\hspace{0mm}}c}
\multicolumn{2}{c}{fixed} & \multicolumn{2}{c}{protocol-generated} \\ \hline
 \multirow{2}{*}[3.5cm]{\includegraphics[scale=0.12]{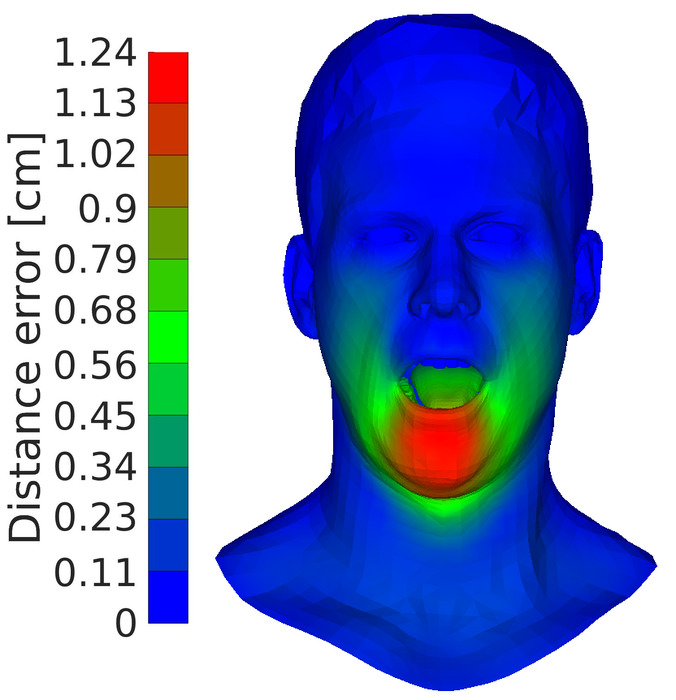} }& 
\includegraphics[scale=0.39, trim = 5mm 0mm 10mm 0mm, clip]{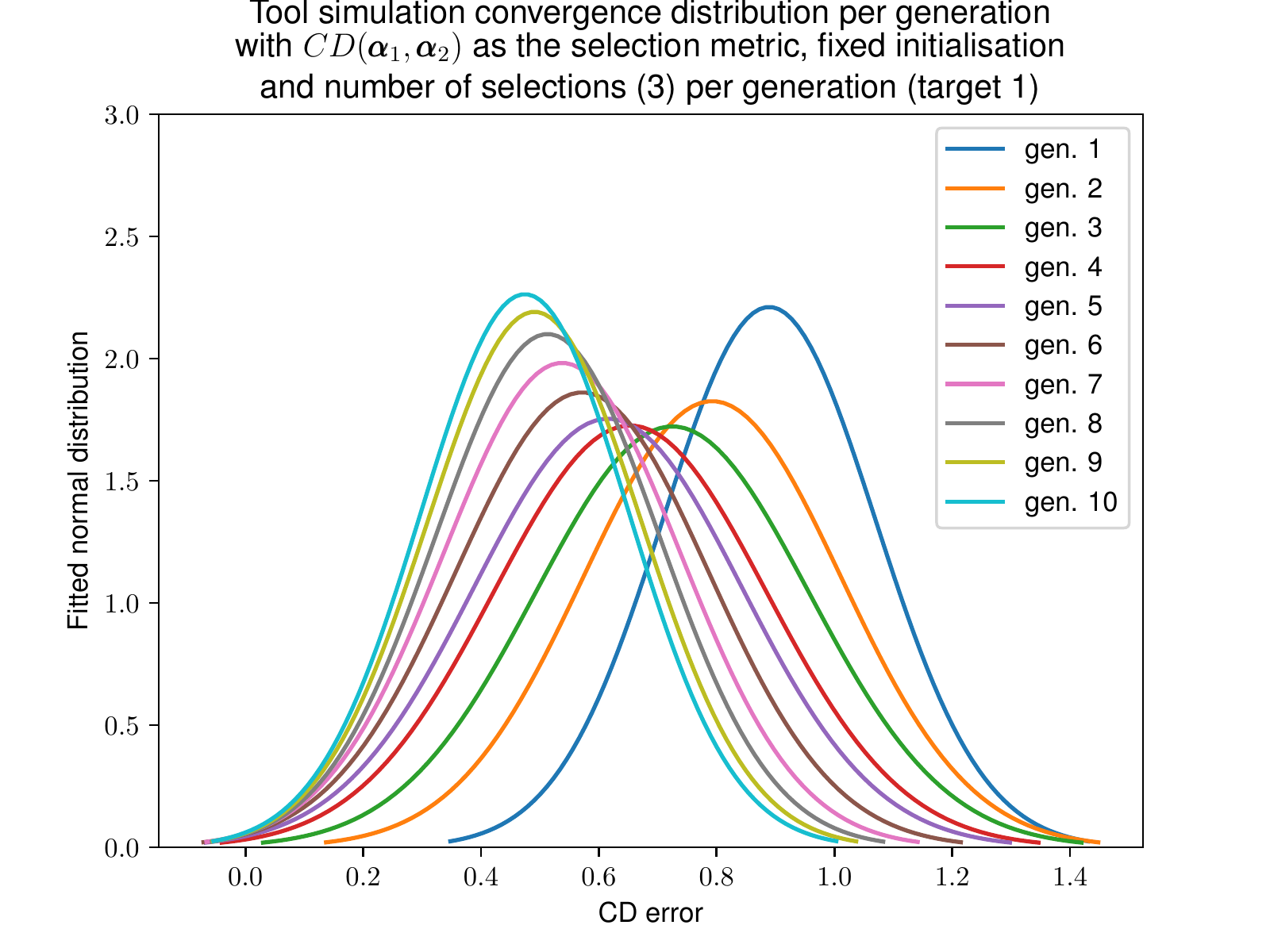} &
\includegraphics[scale=0.39, trim = 2mm 0mm 0mm 0mm, clip]{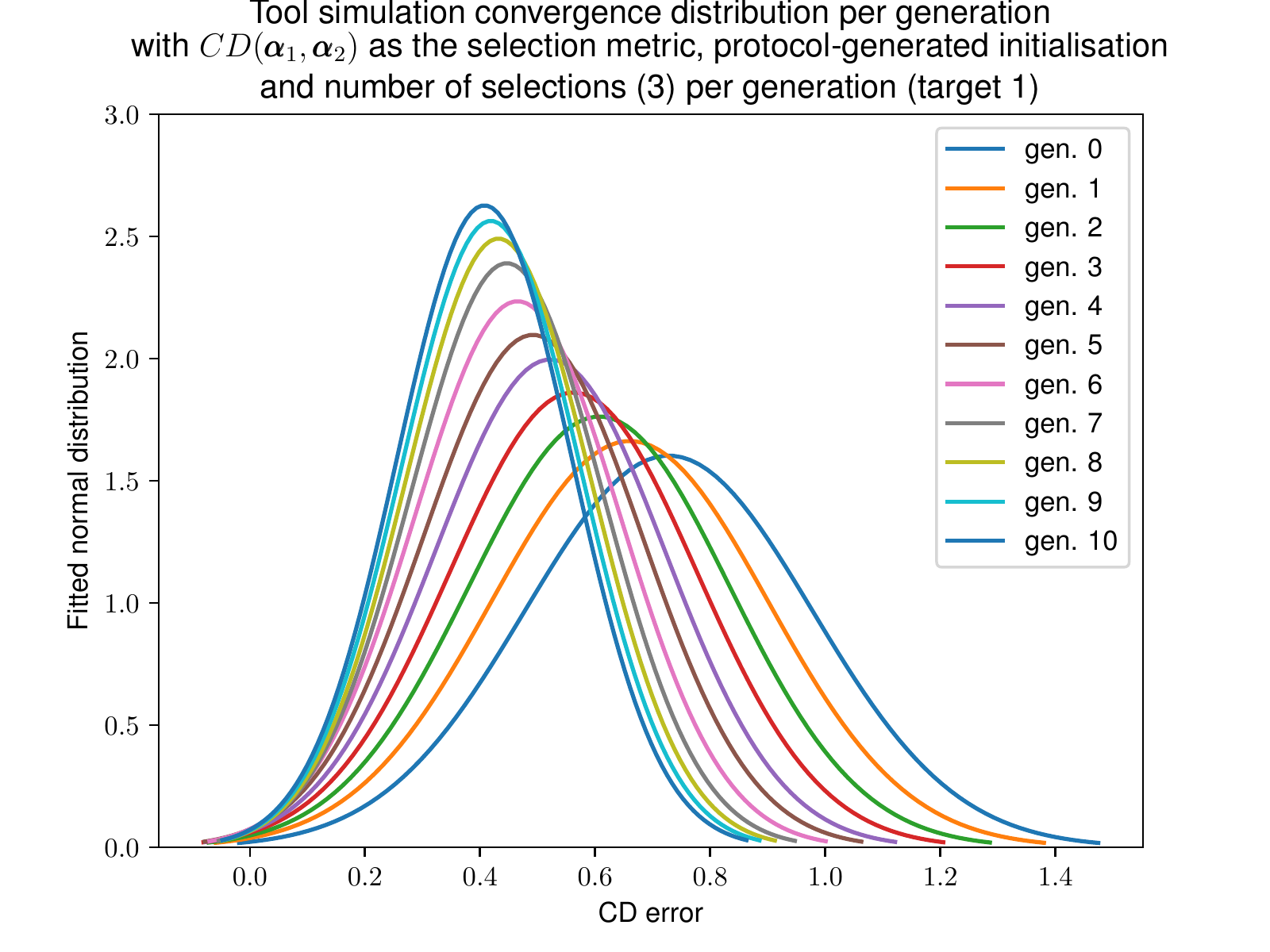} &
 \multirow{2}{*}[3.5cm]{\includegraphics[scale=0.12]{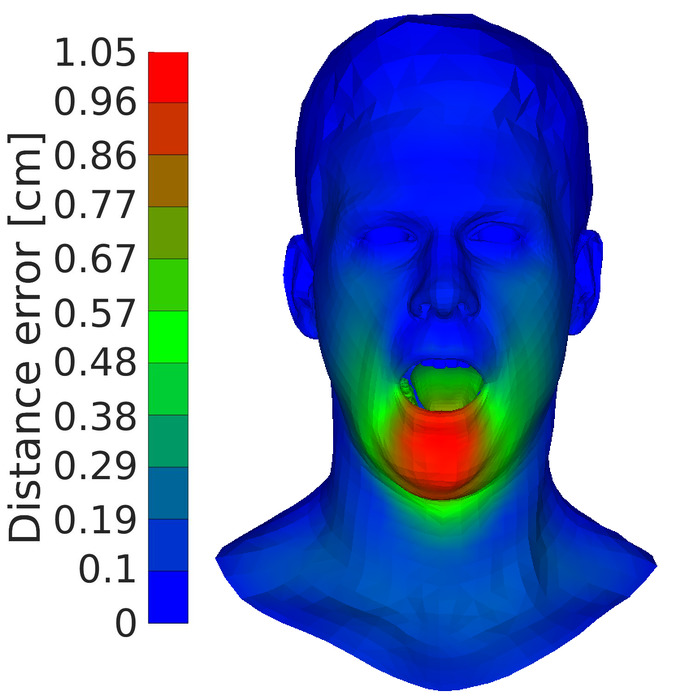}}
\end{tabular}
\caption{Convergence to target 1 from fixed and protocol-generated initialisation using $CD(\boldsymbol{\alpha}_{1}, \boldsymbol{\alpha}_{2})$ as the selection metric with a fixed number of selections per generation (3). Convergence behaviour is illustrated by the set of per-generation cosine distance error distributions. The heatmaps show spatial distance error distribution of the average face in the final generation relative to the target. Compare to Figure~\ref{fixed_protocol} that presents the corresponding result with the default setting on the number of selections.}
\label{fixed_protocol_3_selections}
\end{figure*}
\begin{figure}[h!]
\includegraphics[scale=0.5, trim = 0mm 0mm 0mm 0mm, clip]{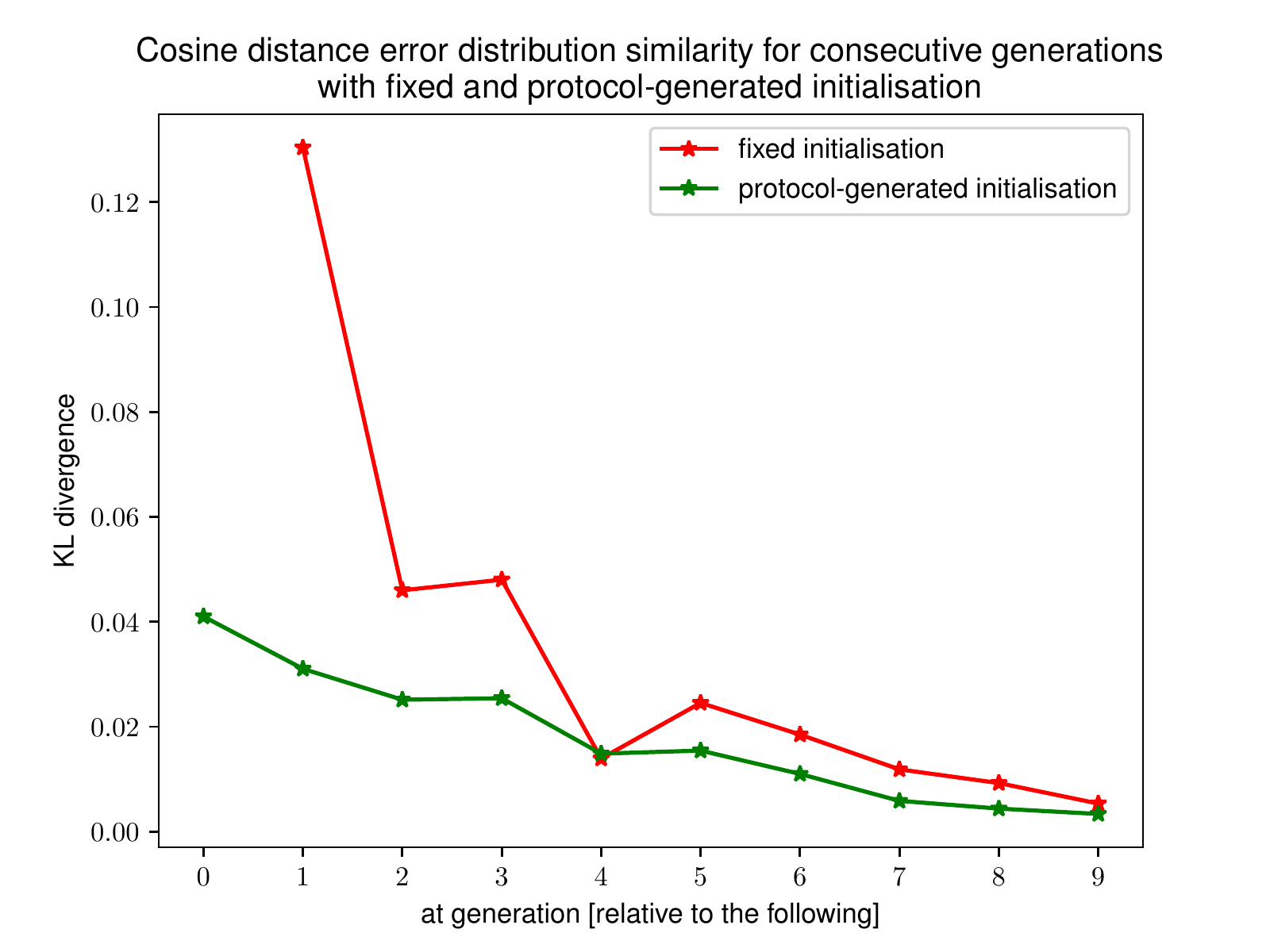}
\caption{Temporal convergence quality assessment with a fixed number (3) of selections per generation: cosine distance error distribution similarity for consecutive generations measured as KL divergence for simulation data generated with fixed and protocol-generated GA initialisation. Compare to Figure~\ref{distribution_similarity} that presents the corresponding result with the default setting on the number of selections.}
\label{distribution_similarity2}
\end{figure}
\subsection{EmoGen's GA assessment by simulation}
Through the preceding analysis the theoretical metric of cosine distance defined in Equation~\ref{cd} has emerged as having the best combination of desirable properties in expression similarity assessment. In this section, we use this metric in simulations to assess the performance of EmoGen's GA in expression sample generation under different parameter configurations. The metric is used both to automate selections in simulations and to quantify results as convergence error distributions.
\par \textbf{Initialisation effect.} Recall that EmoGen offers two options to initialise genetic algorithm for expression generation: 
\begin{itemize}
\item \textit{fixed:} a pre-defined expression set read from file;
\item \textit{protocol-generated:} as described in Section~\ref{config}.
\end{itemize}
We evaluate whether initialisation randomisation through protocol-based generation of the first face set leads to poorer GA convergence to the desired target. Specifically, this evaluation concerns the relative performance with a target \textit{not obviously biased} towards the pre-defined initialisation set. The effect of target bias on each type of initialisation is addressed further on. We choose target 1 from Figure~\ref{targets} because its single blendshape is not active in any of the options of the fixed initialisation used. 
\par Figure~\ref{fixed_protocol} shows cosine distance error distributions per generation for fixed and protocol-generated initialisations. The behaviour is predictably different at the early stage of the process: specifically fixed initialisation permits lower variance and starts further away from the target (see distribution means at the earlier generations). Also from Figure~\ref{distribution_similarity} we observe that with protocol-generated initialisation the generation-to-generation error distribution change measured as KL divergence is more gradual and consistent, with the eventual flattening out of the curve towards the end of the process. Fixed initialisation first shows a rapid decrease followed by a clear increase in KL divergence at the later generations. This increase is mildly indicative of divergent behaviour with fixed initialisation. Yet comparing only the final generation distributions, the convergence statistics (cosine distance error mean and standard deviation $\mu\pm\sigma$) are very similar for both initialisations: $0.47\pm 0.162$ and $0.43\pm0.164$ for respectively fixed and protocol-generated initialisations. Visually, this is confirmed in Figure~\ref{fixed_protocol} by the spatial distance error heatmaps of the elite distribution averages at the final generation that are nearly identical for the two initialisation types.
\par In summary, we can definitively conclude that protocol-generated initialisation does not impair convergence of EmoGen's GA. Furthermore, there is evidence in the error distribution evolution analysis of Figure~\ref{distribution_similarity} that protocol-based initialisation actually promotes more stable convergence. It will be discussed further on that protocol-generated initialisation helps to mitigate target bias in the initialisation set.
\par \textbf{Number of selections per generation} appears to marginally impact the cosine distance error behaviour. If the fixed number of three selections per generation is used, the standard deviation of the cosine distance error  increases to $0.176$ when using fixed initialisation. At the same time, for protocol-generated initialisation there is a general cosine distance error decrease in both mean and variance to $0.41\pm0.152$ in the final generation's distribution. Visually, the marginal improvement does not translate into a substantially different final average face (a fractionally more open jaw in Figure~\ref{fixed_protocol_3_selections} compared to Figure~\ref{fixed_protocol}). Also with the fixed number of selections, fixed initialisation shows a non-monotonic decrease in the generation-to-generation KL divergence of error distributions (Figure~\ref{distribution_similarity2}) unlike protocol-generated initialisation. Please for a detailed quantitative and qualitative comparison view Figures~\ref{fixed_protocol_3_selections} and \ref{distribution_similarity2} relative to Figures~\ref{fixed_protocol} and \ref{distribution_similarity}.
\par \textbf{Target bias and initialisation.} Bias occurs when the initialisation is likely to have options with a substantial similarity to the target in terms of blendshape composition. In simulation, bias is quantified by computing the mean cosine distance error of the best option at initialisation: the distance is constant with a fixed initialisation and variable given protocol generation of initialisation.
\par In Table~\ref{target_bias} we illustrate the relative susceptibility to target bias of each type of initialisation and its effect on result repeatability. Along with target 1 from Figure~\ref{targets}, we employ an average expression target generated by EmoGen users starting from either type of initialisation across 80 trials. Predictably, the expression average target has a greater bias for both initialisations, having been evolved from them. The repeatability behaviour for each bias level is illustrated in Table~\ref{target_bias} as percentages of inter-sample cosine distances in the final elite distributions falling into the four bins: $[0, 0.25)$, $[0.25. 0.5)$, $[0.5, 0.75)$ and $[0.75, 1.0]$. 
\begin{table*}[h!]
\centering
\begin{tabular}{|c|c|c|c|c|c|c|} 
\hline
\multirow{2}{*}{init. type} & \multirow{2}{*}{fixed init. no.} &  target bias  & \multicolumn{4}{c|}{inter-sample cosine distance distribution bins  }  \\ 
          &&$CD(\boldsymbol{\alpha}_{\text{target}}, \boldsymbol{\alpha}_{\text{elite}_{0}})$  &$[0, 0.25)$ [\%] &$[0.25. 0.5)$ [\%]  &$[0.5, 0.75)$ [\%] & $[0.75, 1.0]$ [\%] \\ \hline
fixed & 1 & 1.0  & 1.0 & 33.3 & 53.2 & 12.5 \\ \hline  
fixed & 3 & 0.87 & 1.4 & 37.6 & 53.2 & 7.8 \\\hline
fixed & 1 & 0.57 & 29.8 & 69.1 & 1.1 & 0 \\\hline
fixed & 3 & 0.38 & 99.0 & 1.0 & 0 & 0 \\\hline
protocol & n/a & 0.74 & 2.0 & 46.2 & 45.3 & 6.5 \\\hline
protocol & n/a & 0.55 & 6.6 &  78.9 & 14.4 & 0.1 \\  \hline  
\end{tabular}
\vspace{0.1cm}
\caption{Simulated elite repeatability for different types of initialisation and level of target bias. Note that $CD(\boldsymbol{\alpha}_{\text{target}}, \boldsymbol{\alpha}_{\text{elite}_{0}})=1.0$ indicates no target bias. Repeatability is assessed by presenting the distribution within the indicated bins of the cosine distances across all combinations of elites in the final generation.}
\label{target_bias} 
\end{table*}
\par The first observation from Table~\ref{target_bias} is that, due to the stochastic element in protocol generation, a particularly high target bias is difficult to achieve even with the average expression target (e.g. $0.55$ with protocol generation compared to $0.38$ with one of the fixed initialisation options). Secondly, the repeatability behaviour with protocol generation appears to be less sensitive to increased target bias: the increase of bias from $0.74$ to $0.55$ does not lead to mass accumulation of inter-sample distances in the first bin, unlike the trend with fixed initialisation.
In practical terms, the results make a case for the use of protocol generation in initialisation while participant testing to lower the chance for target bias and ensure independence of repeated trials. 
\par \textbf{Blendshape activation in the initialisation} measures the complexity of the population samples that serve as the starting point for the GA. Complexity is quantified in terms of the number of shapes with a non-zero weight $\alpha$. In Table~\ref{target_bias}, fixed initialisation no.~1 consists of more simplistic facial configurations than no.~3: quantitatively, the average count of non-zero activations per initial population member is 5.2 and 19.9 respectively in the two sets. However, the different initialisation complexity clearly does not affect \textit{repeatability} behaviour given similar target biases (e.g. $1.0$ and $0.87$, the first two rows of Table~\ref{target_bias}).
\par Next we present a study to investigate the effect of the number of blendshape activations in the initialisation on \textit{convergence accuracy}. To minimise the effect of target bias, we initialise by protocol generation. We test three blendshape activation ranges (Table~\ref{activation}), specifying the allowed number of unique core shapes in the initialisation, excluding the ones automatically added for symmetry and correctives. The simulation to study convergence is performed with targets 1 and 2 from Figure~\ref{targets} that are mutually different in complexity having one and twelve active blenshapes respectively. 
\par In Table~\ref{activation}, initial bias and its consistency are respectively the average elite-to-target cosine distance error $\mu_0$ and its standard deviation $\sigma_0$ at initialisation (generation 0). Convergence accuracy and consistency refer to the corresponding metrics $\mu_{10}$ and $\sigma_{10}$ at the final (10th) generation. 
\begin{table*}
\centering
\begin{tabular}{ccccc}
\hline
\multicolumn{5}{c}{target 1}\\ \hline
blendshape activation in initialisation & initial bias & bias consistency & convergence accuracy & convergence consistency \\
(number of core shapes) &  $\mu_0$ & $\sigma_0$ &  $\mu_{10}$ & $\sigma_{10}$ \\ \hline
1 - 3  & 0.752  & 0.370 & 0.362 & 0.206 \\
3 - 8  & 0.744  & 0.238 & 0.430 & 0.162 \\
8 - 16 & 0.737  & 0.150 & 0.434 & 0.130 \\ \hline
\multicolumn{5}{c}{target 2}\\ \hline
blendshape activation in initialisation & initial bias & bias consistency & convergence accuracy & convergence consistency \\
(number of core shapes) &  $\mu_0$ & $\sigma_0$ &  $\mu_{10}$ & $\sigma_{10}$ \\ \hline
1 - 3  & 0.557 & 0.132 & 0.222  & 0.075 \\
3 - 8  & 0.530  & 0.129 & 0.234 & 0.082 \\
8 - 16 & 0.501  & 0.102 & 0.265 & 0.072 \\ \hline
\end{tabular}
\vspace{0.1cm}
\caption{Initial bias and convergence behaviour with varying levels of blenshape activation in protocol-generated initialisation for two targets of varying complexity. Bias and convergence are measured as the average and standard deviation of cosine distance (elite-to-target) error at initialisation ($\mu_0$, $\sigma_0$) and final generation ($\mu_{10}$, $\sigma_{10}$) respectively.}
\label{activation}
\end{table*} 
The results presented in Table~\ref{activation} testify that, although having more active blendshapes at initialisation increases the average target similarity initially, it also makes it more difficult to converge at the final generation. For both targets the best accuracy $\mu_{10}$ is achieved when initialising with samples consisting of the fewest shapes. However, as the simplest initialisation also seems to lower consistency  $\sigma_{10}$  for simpler targets (like target 1), we compromise by setting initial blendshape activation to the default range of 3 to 8 core shapes in EmoGen's GA.
\begin{figure*}[h!]
\begin{tabular}{c@{\hspace{0mm}}c@{\hspace{0mm}}c c@{\hspace{0mm}} c@{\hspace{0mm}}c@{\hspace{0mm}}}
\rotatebox{90}{\parbox{3cm}{1-blendshape target}}  
&\includegraphics[scale=0.4, trim = 2mm 0mm 0mm 0mm, clip]{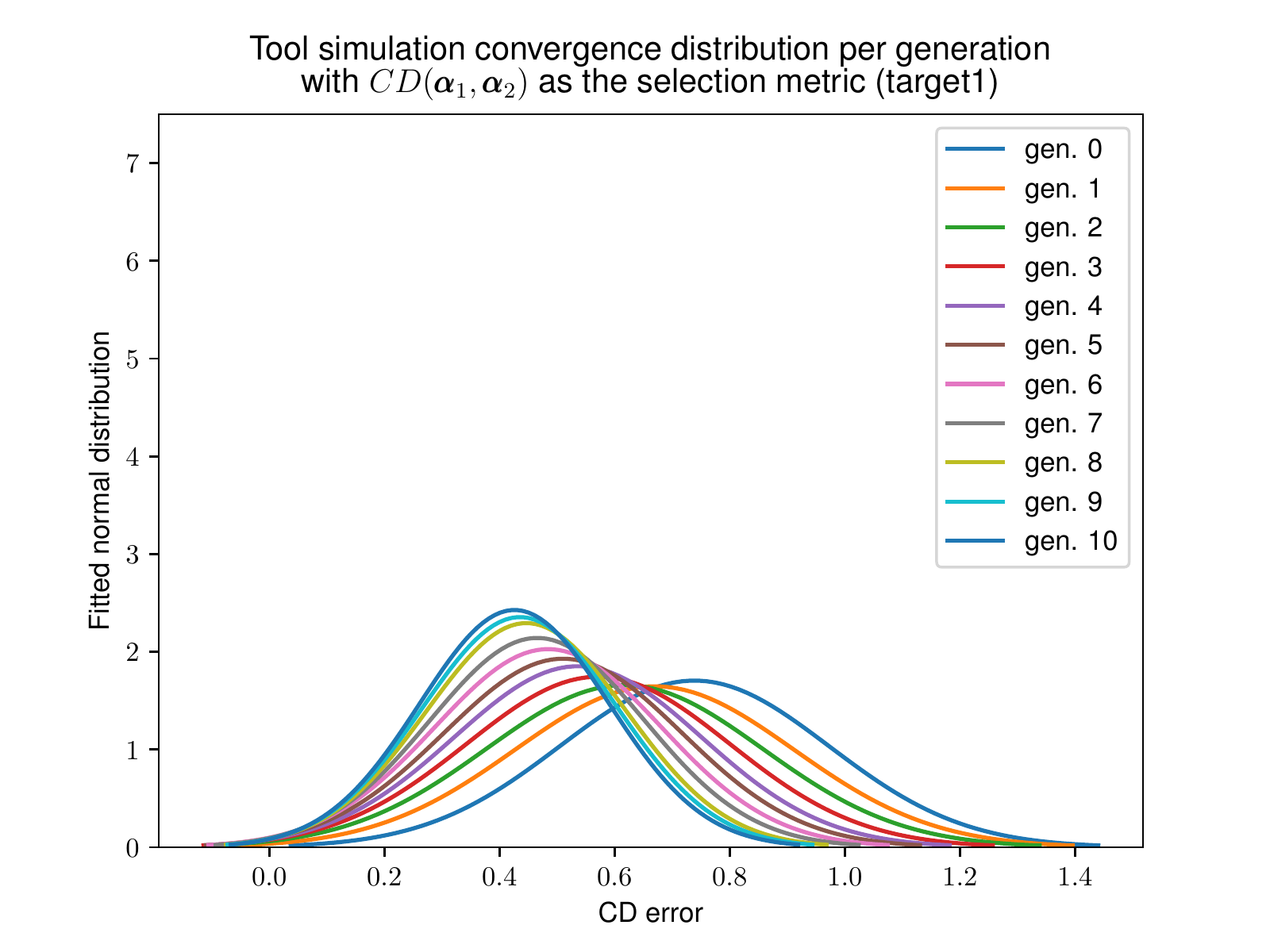} 
&\includegraphics[scale=0.11]{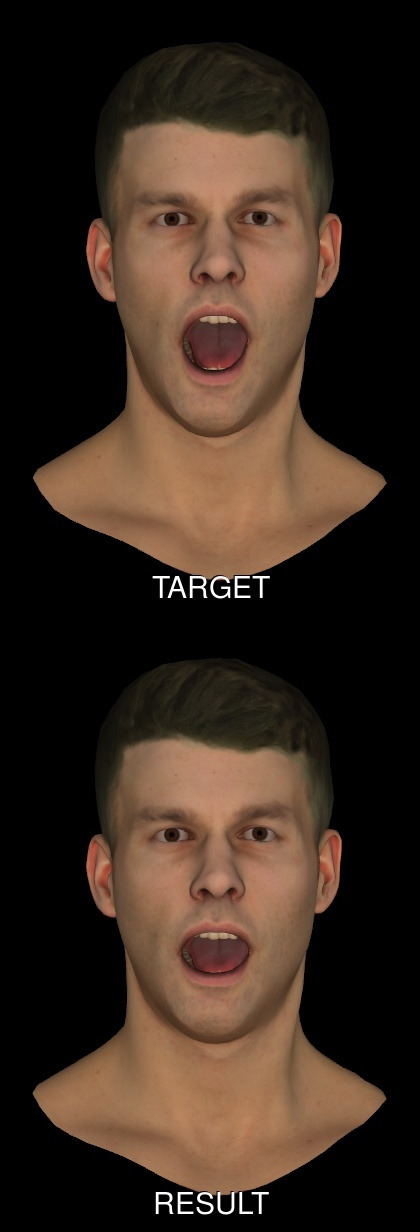}
&\rotatebox{90}{\parbox{3cm}{3-blendshape target}} 
& \includegraphics[scale=0.4, trim = 2mm 0mm 0mm 0mm, clip]{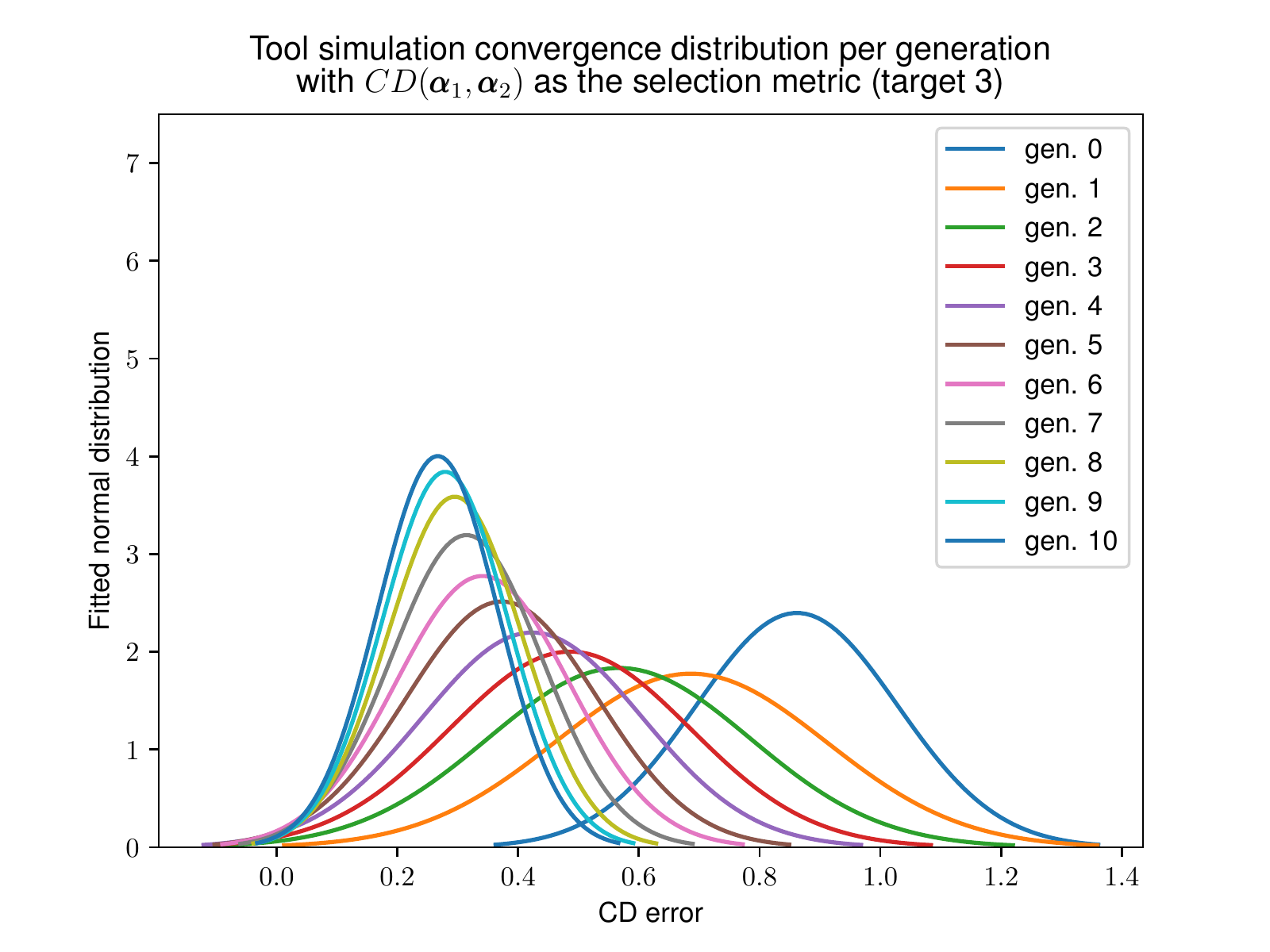}
& \includegraphics[scale=0.11]{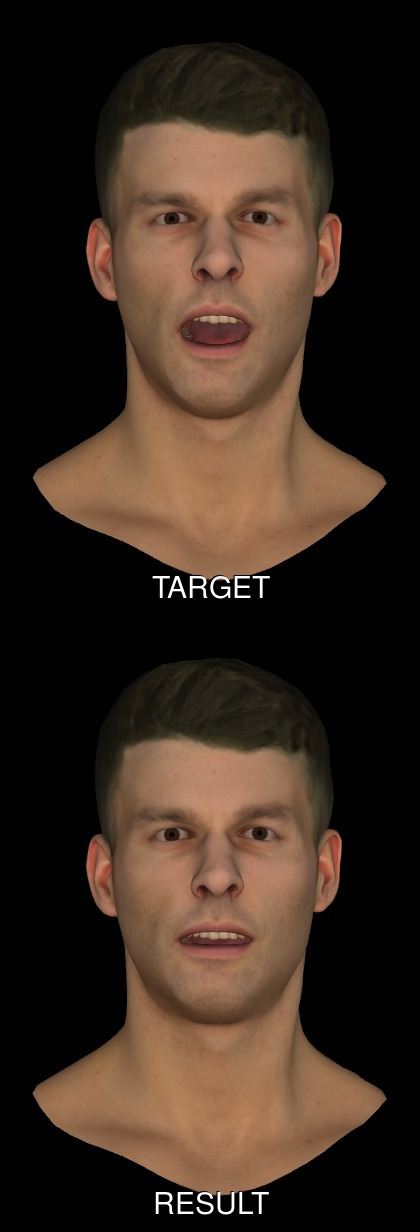}\\
\rotatebox{90}{\parbox{3cm}{12-blendshape target}}  
&\includegraphics[scale=0.4, trim = 2mm 0mm 0mm 0mm, clip]{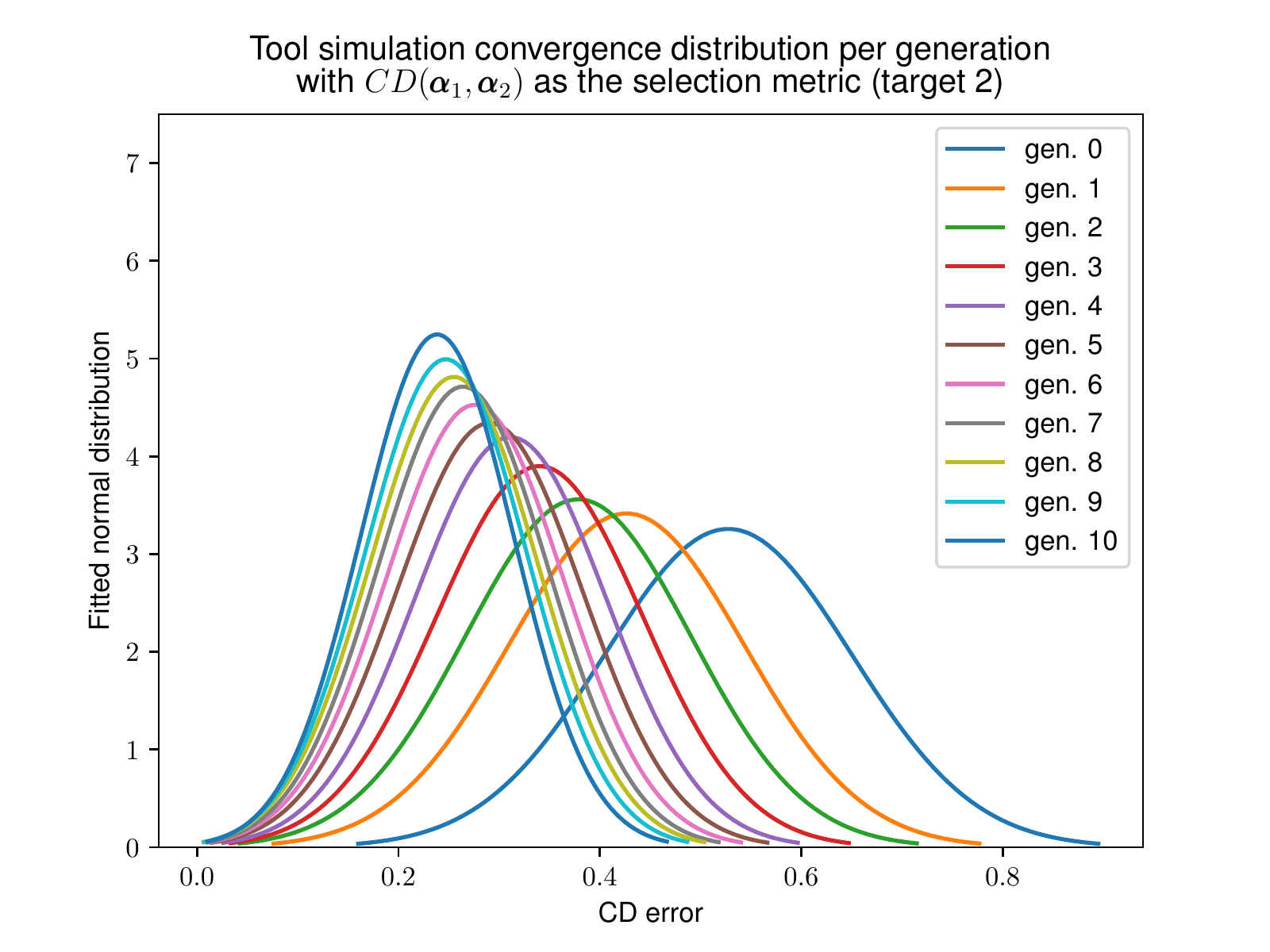}
& \includegraphics[scale=0.11]{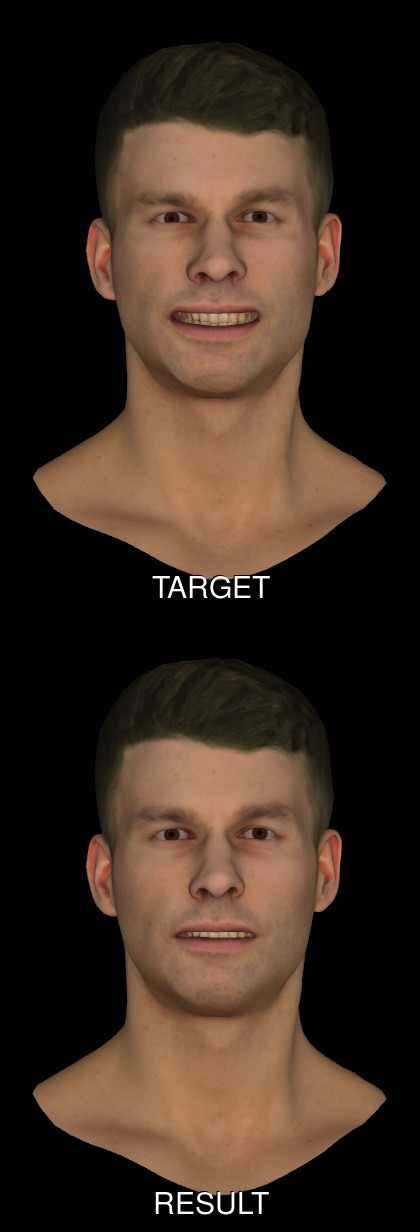}
&\rotatebox{90}{\parbox{3.5cm}{125-blendshape target}}  
&\includegraphics[scale=0.4, trim = 2mm 0mm 0mm 0mm, clip]{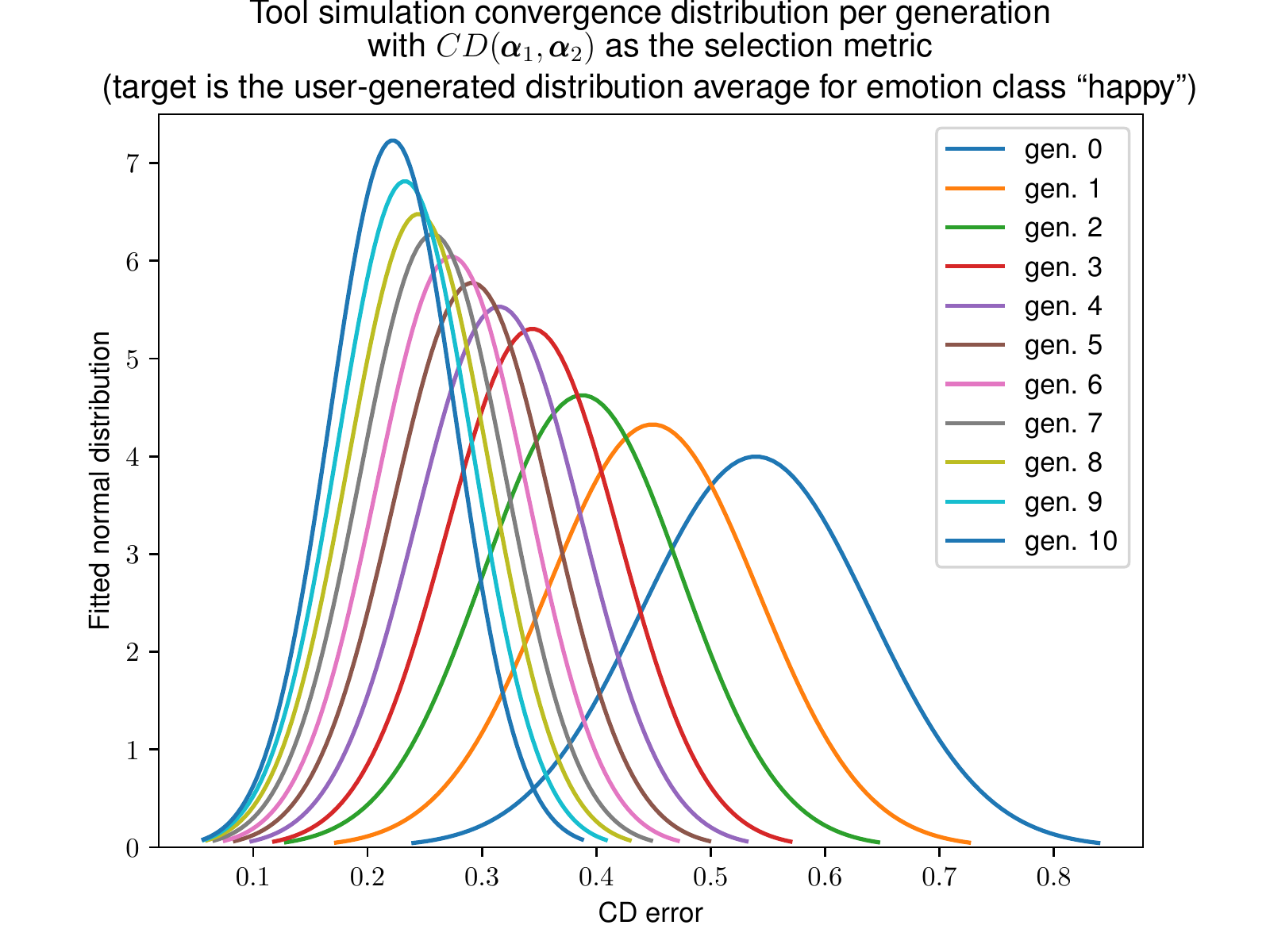} 
&\includegraphics[scale=0.11]{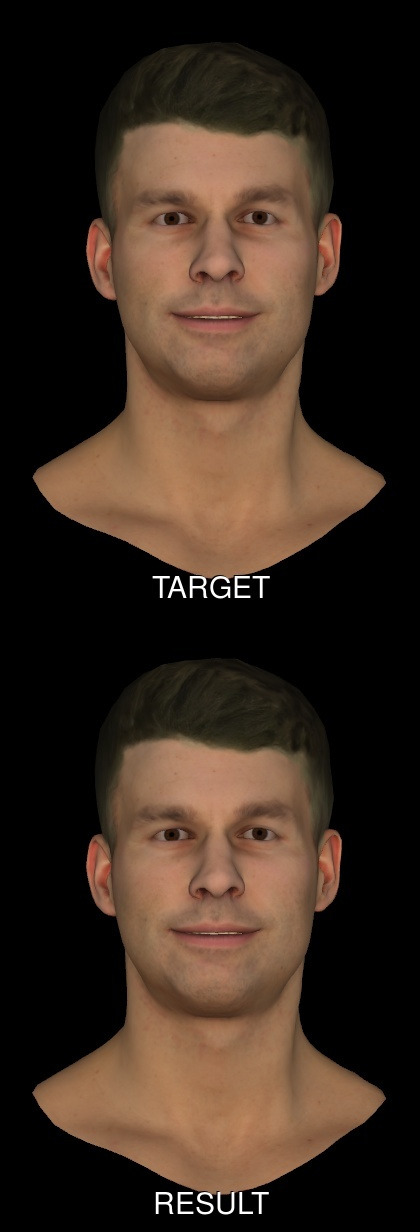} \\
\end{tabular}
\caption{Simulated EmoGen convergence using $CD(\boldsymbol{\alpha}_{1}, \boldsymbol{\alpha}_{2})$ as the selection metric for targets of increasing complexity defined by the number of non-zero weight blendshapes. Also shown for each target are the average faces converged to in the final generation.}
\label{convergence_complexity}
\end{figure*}
\begin{table}[h!]
\centering
\begin{tabular}{ccc}
target &  $\mu$ & $\sigma$ \\
\hline
1-blendshape   & 0.426  & 0.1643\\  \hline
3-blendshape   & 0.267  & 0.0997 \\ \hline
12-blendshape  & 0.237  & 0.0760 \\ \hline
125-blendshape & 0.222  & 0.0551 \\ \hline
\end{tabular}
\vspace{0.1cm}
\caption{Mean $\mu$ and standard deviation $\sigma$ of the $CD(\boldsymbol{\alpha}_{1}, \boldsymbol{\alpha}_{2})$ error in the final generation simulated distributions of Figure~\ref{convergence_complexity}.}
\label{convergence_complexity_numbers}
\end{table}
\par \textbf{Genetic algorithm noise.} In the EmoGen methodology, the \textit{uniform random distribution} is sampled in the initialisation by protocol and to drive the core processes (mutation, cross-breeding and averaging) of the GA-based space search for the subsequent expression evolution. Given a consistent target, any variance in the final evolved result due to the stochastic nature of the search mechanism is the \textit{genetic algorithm noise}. The noise is important to quantify because it gives an indication of \textit{significant} variances in user-generated distributions i.e. facial differences reflecting human perception beyond the noise threshold of the generation mechanism. One way to quantify the noise is by the mean and standard deviation ($\mu\pm\sigma$) of the cosine distance error in simulated convergence. We present and analyse convergence profiles for targets of increasing complexity in terms of the number of blendshapes. Specifically, the targets  are the three from Figure~\ref{targets} (targets 1, 3 and 2 have respectively one, three and twelve active blendshapes) and 
the user-generated distribution average for emotion class ``happy'' with 125 active blendshapes, 32 of which have a weight $\geq 0.1$. Figure~\ref{convergence_complexity} illustrates the targets.
\par From Figure~\ref{convergence_complexity} and Table~\ref{convergence_complexity_numbers}, we can observe that  the GA-noise decreases with increasing target complexity, tapering off to approximately $0.2\pm0.065$ cosine distance error (averaging the results of the 12- and 125-blendshape targets). Targets with a larger number of active blendshapes converge better because they are more likely to find options matching at least some of their blendshapes earlier in the process (see error distributions at the earlier generations) and hence have more generations left for refinement via the GA. 
\par We conclude $0.2\pm0.065$ cosine distance error to be the sought GA noise threshold that needs to be exceeded in any user-generated distribution to signify a sufficient signal-to-noise ratio for human-behaviour-related trend extraction. Specifically, we choose the noise level corresponding to the higher complexity targets as the threshold as it is more likely to reflect internalised emotional representations (we assume participants are less likely to search very specifically for a single-blendshape target but rather opt for a more realistic facial expression, distributed in terms of blendshape activation). GA noise cannot be removed from user-generated distributions prior to data  analysis. Since the internalised targets in the user studies are latent, distribution means can be used as proxy targets to estimate convergence statistics to compare to the simulated GA noise threshold. It is possible to extract additional statistics from simulated distributions to assess significance in user data. One example is the comparison of the mean and standard deviation of the inter-sample distance in user data against that in simulated distributions with consistent targets. 
\par \textbf{Genetic algorithm convergence.} We can discuss convergence of EmoGen's GA based on the visual comparison of the means evolved by the final generation relative to the targets in Figure~\ref{convergence_complexity}. Recall that cosine distance used as the selection metric in simulations is 
agnostic to the absolute intensity of the expression optimising the blendshape composition: Figure~\ref{intensity} illustrates samples on the intensity scale of the same blendshape composition vector.
Similarly, we observe that, for all targets in Figure~\ref{convergence_complexity}, the resultant average of the distribution in the last generation visually approximates a face on the intensity scale of the target's blendshape composition vector. This is an indication of convergence to the level of cosine distance sensitivity. In the case of the 125-blendshape composition, even the intensity of the evolved expression is close to the target. Furthermore, in Table~\ref{gmm} we have already shown that the target distributions obtained for target 1, 2 and 3 are separable despite significant similarity between targets 1 and 3. Hence the simulation data confirms convergence of the GA employed in the EmoGen methodology.  
\par \textbf{Selection pressure.} The GA implemented in our EmoGen methodology is distinct by its operating without continuous fitness ranking of the selected population members beyond the choice of the elite. Hence the algorithm is characterised by a low selection pressure counteracted only by exercising elitism (guaranteed propagation of the best chosen face to the next generation). 
\par We have tested how convergence is affected by increased selection pressure, specifically by incorporating fitness ranking. By using the cosine distance similarity metric, we can establish a sample ranking at each generation, in terms of their similarity to the target. Next, instead of the random parent selection exercised by default, in this comparative simulation we only propagate the fittest parents (closest to the target) by allowing them to cross-breed and mutate to produce two children, replacing both parents in the next generation. Each pair of parents is allowed to breed only once. Simulations with this set-up have shown convergence accuracy to be unaffected by introducing the continuous ranking of the population samples. So the absence of such in our GA is not a performance bottleneck.
 
\section{Conclusion}
In this paper, we presented our EmoGen methodology that harnesses technological advances in facial animation to enable quantitative studies of internalised emotion representation in psychology. The methodology is complete with tools for both data generation and analysis. Our configurable genetic algorithm for expression space search provides a way for a non-expert to generate arbitrary facial configurations in 3D via a user-friendly interface. Unlike prior art, our space search shows a degree of robustness against implausible configurations through tailored corrective mechanisms. Equally important within the EmoGen methodology is the derived principled approach to data analysis: we have found expressions are best represented as in the blendshape vector domain and quantitatively compared using cosine distance as the similarity metric.  Further, we validate and characterise the GA-based expression sample generation, including its parameter sensitivity, by means of simulated convergence statistics. These analytical contributions of the methodology facilitate principled experimental data collection and analysis in psychology research.
\par In terms of expression data generation, the key contribution of the EmoGen methodology is the  access to and easy control of a complex production-level blendshape model we provide for a non-expert user. No user expertise is required as robustness is maintained automatically. Further, the generation process is customisable and produces an exhaustive output of quantitative data from all stages. 
\par In terms of the analysis tools within the methodology, we highlight the observation that, notwithstanding its known limitation of agnosticism to the absolute intensity, the use of cosine distance as a similarity metric aids robust trend extraction in data distributions in the blendshape domain. This conclusion is supported by the observed superior performance of cosine distance as the selection metric in convergence simulations. Equally of practical importance are our conclusions on the preferable behaviour of protocol-generated initialisation and the estimates of the genetic algorithm noise we generated by simulation.
\par As future work, we plan to further expand flexibility of sample generation within the EmoGen methodology. Currently, the framework is limited to just two facial identities. The goal is to upgrade this binary selection to a space of choices, enabling a degree of user freedom in authoring the identity  through an accessible interface. Further, we intend to incorporate facial 4D dynamics into the EmoGen methodology by facilitating genetic evolution of animation curves. We believe such an extension to 4D, with spatio-temporal non-linearities of dynamic behaviour, will open up new exciting avenues for research in experimental psychology.

\bibliographystyle{IEEEtran}
\bibliography{bib}
\end{document}